\documentclass[%
 reprint,
 amsmath,amssymb,
 aps,
 pre,
 longbibliography
]{revtex4-2}

\usepackage[pdftex]{graphicx} 
\usepackage{dcolumn}
\usepackage{bm}
\usepackage[T1]{fontenc}
\usepackage[pdftex]{color, hyperref}
\hypersetup{
colorlinks = true,
linkcolor = blue,
citecolor = blue
}
\usepackage{here} 

\usepackage{color}

\newcommand{\argmax}{\mathop{\rm arg~max}\limits}

\begin{document}


\title{Information-Thermodynamic Bound on Information Flow in Turbulent Cascade}

\author{Tomohiro Tanogami$^1$}
\author{Ryo Araki$^{2}$}
\affiliation{$^1$Department of Earth and Space Science, Osaka University, Osaka 560-0043, Japan\\
$^2$Department of Mechanical and Aerospace Engineering, Faculty of Science
and Technology, Tokyo University of Science, Yamazaki 2641, Noda-shi
278-8510, Japan}




\date{\today}
\begin{abstract}
We investigate the nature of information flow in turbulence from an information-thermodynamic viewpoint.
For the fully developed three-dimensional fluid turbulence described by the fluctuating Navier--Stokes equation, we prove that information of large-scale eddies is transferred to small scales along with the energy cascade.
We numerically illustrate our findings using a shell model and further show that in the inertial range, the intensity of the information flow is nearly constant and can be scaled by the large-eddy turnover time.
Our numerical results also suggest that the corresponding information-thermodynamic efficiency is quite low compared to other typical information processing systems such as Maxwell's demon.
These findings provide a new perspective on how universality and intermittency of turbulent fluctuations emerge at small scales.
\end{abstract}

\pacs{Valid PACS appear here}

\maketitle
\section{Introduction}
Turbulence is characterized by the interference of fluctuations between disparate space-time scales. 
Despite its seemingly complicated and unpredictable nature, universal laws are hidden behind the disordered fluid motion.
For example, in fully developed three-dimensional turbulence, the energy spectrum exhibits the Kolmogorov spectrum $E(k)\propto k^{-5/3}$ at scales much smaller than the energy injection scale~\cite{K41_a,K41_b,K41_c,Frisch,davidson2015turbulence}.
The Kolmogorov spectrum is universal in the sense that it is independent of the details of the large scales, such as the boundary conditions or the mechanism of the external stirring.
Furthermore, in addition to the energy spectrum, which corresponds to the second-order moments of the velocity field, the higher-order moments also exhibit universal scaling laws~\cite{Frisch,data}.
Such remarkable universality of turbulent fluctuations at small scales is believed to be induced by the energy cascade process, where the energy is transferred conservatively from large to small scales.
More specifically, there is a common intuitive picture that the universal statistical properties emerge at small scales because ``information'' about the details of the large scales is lost in the chaotic stepwise cascade process~\cite{davidson2015turbulence,Eyink_Sreenivasan,Eyink_lecture}.

Somewhat contrary to this intuitive picture, some numerical and experimental observations suggest that the small-scale eddies do not ``forget'' about the large scales.
For example, it is known that fluctuations of small-scale quantities (e.g., the energy dissipation rate) follow those of large-scale quantities (e.g., the energy injection rate) with a time lag along with the energy cascade~\cite{yasuda2014quasi,goto2017hierarchy,araki2023minimal}.
This time lag is on the order of the large-eddy turnover time, which is the characteristic time scale for the largest eddies to be stretched into smaller eddies.
Another example is the chaos synchronization of small-scale motions induced by the energy cascade, where the small-scale velocity field is slaved to the chaotic dynamics of the large-scale velocity field~\cite{pecora1990synchronization,boccaletti2002synchronization,yoshida2005regeneration,lalescu2013synchronization,vela2021synchronisation}.
Moreover, small-scale intermittency can also be regarded as such an example because it implies that the turbulent fluctuations grow in each cascade step and thus ``remember'' the large scales~\cite{Frisch,Eyink_lecture,Eyink_Sreenivasan}.
These phenomena suggest that information about the large-scale fluctuations is not lost in the cascade process, but rather is transferred to small scales.

In order to deepen our understanding of the generation mechanism of universality and intermittency of turbulent fluctuations, it is thus desirable to reveal the nature of the information transfer across scales associated with the energy cascade.
As a first step toward this end, here we aim to prove that information of turbulent fluctuations is transferred from large to small scales in fully developed three-dimensional fluid turbulence.
While turbulence has been studied in various contexts from information-theoretic perspectives in recent decades~\cite{betchov1964measure,ikeda1989information,cerbus2013information,materassi2014information,cerbus2016information,goldburg2016turbulence,granero2016scaling,granero2018kullback,lozano2020causality,shavit2020singular,vladimirova2021fibonacci,PhysRevResearch.4.023195}, no previous studies have theoretically shown that information flows across scales along with turbulent cascade.

For this purpose, we employ \textit{information thermodynamics}, which is a thermodynamic framework for information flow between interacting subsystems~\cite{parrondo2015thermodynamics,ehrich2023energy,shiraishi2023introduction}.
While information thermodynamics has its origins in the thought experiment of Maxwell's demon, it has recently been applied to information processing at the cellular level in biological systems~\cite{barato2014efficiency,sartori2014thermodynamic,ito2015maxwell,hartich2016sensory,amano2022insights} and even to deterministic chemical reaction networks~\cite{penocchio2022information}.
By applying information thermodynamics to turbulence, we can clearly define the concept of ``information'' as a quantity closely related to thermodynamic quantities and obtain universal constraints on the flow of the information.
Note that information thermodynamics requires us to use a thermodynamically consistent model that includes thermal fluctuations, i.e., fluctuating hydrodynamic equations~\cite{landau1959fluid,de2006hydrodynamic}.
To put it another way, this approach also enables us to investigate the effects of thermal fluctuations on turbulence dynamics, which has recently been intensively investigated~\cite{ruelle1979microscopic,komatsu2014glimpse,bandak2021thermal,bandak2022dissipation,eyink2022high,mcmullen2022navier,bell2022thermal}.

In this paper, we prove that information of turbulent fluctuations is transferred from large to small scales along with the energy cascade.
We emphasize that our main results, [Ineqs.~(\ref{main result 1}) and (\ref{main result 2})], are exact and universal relations, independent of the details of the flow under consideration.
While we derive these relations for the fluctuating Navier--Stokes equation, our results are valid for various turbulence models, including shell models.
We numerically illustrate our findings using the Sabra shell model and further show that in the inertial range, the intensity of the information flow is nearly constant and can be scaled by the large-eddy turnover time [Eq.~(\ref{IF_scaling})].
This observation suggests that the information of large-scale turbulent fluctuations is transferred to small scales with nearly constant intensity by the energy cascade process.
Thus, our results challenge the conventional intuitive picture of how universality emerges at small scales.
Moreover, our numerical results suggest that the corresponding \textit{information-thermodynamic efficiency} is quite low compared to other typical information processing systems such as Maxwell's demon.
This implies that transferring information from large to small scales involves enormous thermodynamic costs, indicating the poor performance of turbulence as an information processing system.

This paper is organized as follows.
In Sec.~\ref{Setup}, we introduce the fluctuating Navier--Stokes equation and its corresponding Fokker--Planck equation.
In Sec.~\ref{Basic properties}, we briefly review some basic properties of fully developed turbulence described by the fluctuating Navier--Stokes equation.
In Sec.~\ref{Information-theoretic quantities}, we introduce the two information-theoretic quantities that are important in describing our main result: \textit{mutual information} and \textit{information flow}.
Then, in Sec.~\ref{Information-thermodynamic bound on information flow in turbulence}, we explain our main result on the information flow in turbulence and its derivation.
In the derivation, we first formulate the second law of thermodynamics for the fluctuating Navier--Stokes equation and then derive the \textit{second law of information thermodynamics}.
Section \ref{Numerical simulation} presents a numerical demonstration of our main result.
By introducing the concept of information-thermodynamic efficiency, we show that it is quite low in turbulence compared to other typical information processing systems.
We further show that in the inertial range, the intensity of the information flow is nearly constant and can be scaled by the large-eddy turnover time.
In Sec.~\ref{Concluding remarks}, we summarize our findings with some remarks and future perspectives.
The appendices contain details of the derivations and numerical simulations.

\section{Setup\label{Setup}}
While our results are valid for various thermodynamically consistent turbulence models, we focus on the fluctuating Navier--Stokes equation except for Sec.~\ref{Numerical simulation}, where we numerically illustrate our main result by using a shell model.
We consider an incompressible fluid with constant mass density $\rho$, temperature $T$, and kinematic viscosity $\nu$, confined in a cube with periodic boundary conditions $\Omega=L\mathbb{T}^3$.
Let ${\bm u}({\bm x},t)=(u^x({\bm x},t),u^y({\bm x},t),u^z({\bm x},t))$ be the fluid velocity at position ${\bm x}\in\Omega$ and time $t\in\mathbb{R}$.
Hereafter, we often omit the argument $t$ to simplify the notation.
The time evolution of the velocity field ${\bm u}$ is described by the fluctuating Navier--Stokes equation~\cite{landau1959fluid,de2006hydrodynamic,bandak2021thermal,bandak2022dissipation}:
\begin{align}
\partial_t{\bm u}+{\bm u}\cdot\nabla{\bm u}=-\nabla p+\nu\nabla^2{\bm u}+{\bm f}+\nabla\cdot\mathsf{s}
\label{fluctuating NS}
\end{align}
with the incompressibility condition $\nabla\cdot{\bm u}=0$, where $p$ denotes the kinematic pressure, ${\bm f}$ represents the external force per unit mass, which, without loss of generality, is assumed to be divergence-free.
We further assume that ${\bm f}$ acts only at large scales, i.e., it is supported in Fourier space at low wave numbers $\sim k_f$.
In the last term on the right-hand side of (\ref{fluctuating NS}), $\mathsf{s}$ denotes a thermal fluctuating stress prescribed as a zero-mean Gaussian random field that satisfies
\begin{align}
\langle\mathsf{s}^{ab}({\bm x},t)\mathsf{s}^{cd}({\bm x}',t')\rangle&=\dfrac{2\nu k_{\mathrm{B}}T}{\rho}\left(\delta^{ac}\delta^{bd}+\delta^{ad}\delta^{bc}-\dfrac{2}{3}\delta^{ab}\delta^{cd}\right)\notag\\
&\qquad\times\delta^3({\bm x}-{\bm x}')\delta(t-t'),
\end{align}
where $a,b,c,d\in\{x,y,z\}$ and $\delta^{ab}$ denotes the Kronecker delta, which is $1$ if $a=b$, and zero otherwise.
Here, the prefactor $2\nu k_{\mathrm{B}}T/\rho$, where $k_{\mathrm{B}}$ denotes the Boltzmann constant, is chosen according to the fluctuation-dissipation relation of the second kind~\cite{maes2021local,tanogami2022violation} so that the model (\ref{fluctuating NS}) is thermodynamically consistent~\cite{peliti2021stochastic}.

Let $\hat{\bm u}_{\bm k}$ be the Fourier mode of the velocity field with wave vector ${\bm k}\in(2\pi/L)\mathbb{Z}^3$ defined as
\begin{align}
\hat{{\bm u}}_{\bm k}=\dfrac{1}{V}\int_\Omega d^3{\bm x}e^{-i{\bm k}\cdot{\bm x}}{\bm u}({\bm x}),
\end{align}
where $V:=L^3$ denotes the volume of the fluid.
Here, we note that the fluctuating hydrodynamics describes fluid motions at the mesoscopic level~\cite{landau1959fluid,de2006hydrodynamic}.
In other words, there is a cutoff wave number $\Lambda$ such that $\ell^{-1}_{\mathrm{macro}}\ll\Lambda\ll\ell^{-1}_{\mathrm{micro}}$, where $\ell_{\mathrm{macro}}$ denotes the macroscopic length scale characterizing the macroscopic behaviors and $\ell_{\mathrm{micro}}$ denotes the microscopic length scale, such as the molecular size, the interaction length, or the mean free path~\cite{bandak2022dissipation}.
In the following, we shall assume that the summation over wave vectors $\sum_{\bm k}$ means the summation up to the cutoff wave number $\sum_{|{\bm k}|<\Lambda}$.
Because of the reality condition $\hat{\bm u}^*_{\bm k}=\hat{\bm u}_{-{\bm k}}$, only the modes ${\bm u}_{\bm k}$ whose wave vector lies in the half-set
\begin{align}
\mathcal{K}^+:=
\begin{Bmatrix}
  & k^x>0 & \text{or} \\
 {\bm k}: & k^y>0 & \text{if}\quad k^x=0\quad\text{or}\\
  & k^z\ge0 & \text{if}\quad k^x=k^y=0
\end{Bmatrix}
\label{independent Fourier modes}
\end{align}
are independent (for a schematic of this set, see Fig.~\ref{fig:Fourier_modes_division} in Sec.~\ref{Information-theoretic quantities}).
Then, the fluctuating Navier--stokes equation (\ref{fluctuating NS}) can be rewritten as stochastic differential equations for $\hat{\bm u}:=\{\hat{\bm u}_{\bm k}|{\bm k}\in\mathcal{K}^+\}$ and the complex-conjugate variables $\hat{\bm u}^*:=\{\hat{\bm u}^*_{\bm k}|{\bm k}\in\mathcal{K}^+\}$ (see Appendix~\ref{Detailed calculation of the Fourier transform of the fluctuating Navier--Stokes equation} for the derivation):
\begin{align}
\partial_t\hat{\bm u}_{\bm k}&={\bm B}_{\bm k}(\hat{\bm u},\hat{\bm u}^*)-\nu k^2\hat{\bm u}_{\bm k}+\hat{\bm f}_{\bm k}+\sqrt{\dfrac{2\nu k^2k_{\mathrm{B}}T}{\rho}}\hat{\bm \xi}_{\bm k},\label{fNS_k}\\
\partial_t\hat{\bm u}^*_{\bm k}&={\bm B}^*_{\bm k}(\hat{\bm u},\hat{\bm u}^*)-\nu k^2\hat{\bm u}^*_{\bm k}+\hat{\bm f}^*_{\bm k}+\sqrt{\dfrac{2\nu k^2k_{\mathrm{B}}T}{\rho}}\hat{\bm \xi}^*_{\bm k}.\label{fNS_k_*}
\end{align}
Here, ${\bm B}_{\bm k}(\hat{\bm u},\hat{\bm u}^*)$ denotes the nonlinear term 
\begin{align}
B^a_{\bm k}(\hat{\bm u},\hat{\bm u}^*):=-ik^c\left(\delta^{ab}-\dfrac{k^ak^b}{k^2}\right)\sum_{{\bm p}+{\bm q}={\bm k}}\hat{u}^b_{\bm p}\hat{u}^c_{\bm q},
\label{def:B}
\end{align}
where $k:=|{\bm k}|$, and $\hat{\bm \xi}_{\bm k}$ denotes the zero-mean white Gaussian noise that satisfies $\hat{\bm \xi}^*_{\bm k}=\hat{\bm \xi}_{-{\bm k}}$, ${\bm k}\cdot\hat{\bm \xi}_{\bm k}=0$, and
\begin{align}
\langle\hat{\xi}^a_{\bm k}(t)\hat{\xi}^{b*}_{{\bm k}'}(t')\rangle=\dfrac{1}{V}\left(\delta^{ab}-\dfrac{k^ak^b}{k^2}\right)\delta_{{\bm k},{\bm k}'}\delta(t-t').
\label{xi_xi_corelation}
\end{align}
Note that the wave vectors ${\bm p}$ and ${\bm q}$, which are summed over in (\ref{def:B}), may not belong to the half-set $\mathcal{K}^+$.
If ${\bm p}\notin\mathcal{K}^+$, then $\hat{\bm u}_{\bm p}$ should be interpreted instead as $\hat{\bm u}^*_{-{\bm p}}$.
We also remark that the noise intensity $2\nu k^2k_{\mathrm{B}}T/\rho$ becomes large in the higher wave number region to balance the viscous damping.

Let $p_t(\hat{\bm u},\hat{\bm u}^*)$ be the probability density of the total independent Fourier modes $\{\hat{\bm u},\hat{\bm u}^*\}$ at time $t$.
The time evolution of $p_t(\hat{\bm u},\hat{\bm u}^*)$ is governed by the following Fokker--Planck equation equivalent to the stochastic differential equations (\ref{fNS_k}) and (\ref{fNS_k_*}):
\begin{align}
&\partial_tp_t(\hat{\bm u},\hat{\bm u}^*)\notag\\
&=\sum_{{\bm k}\in\mathcal{K}^+}\left[-\dfrac{\partial}{\partial \hat{\bm u}_{\bm k}}\cdot {\bm J}_{\bm k}(\hat{\bm u},\hat{\bm u}^*)-\dfrac{\partial}{\partial \hat{\bm u}^*_{\bm k}}\cdot {\bm J}^*_{\bm k}(\hat{\bm u},\hat{\bm u}^*)\right],
\label{FP-fNS}
\end{align}
where we note that the summation over ${\bm k}$ is restricted to the half-set $\mathcal{K}^+$.
Here, ${\bm J}_{\bm k}(\hat{\bm u},\hat{\bm u}^*)$ denotes the probability current associated with a Fourier mode $\hat{\bm u}_{\bm k}$:
\begin{align}
{\bm J}_{\bm k}(\hat{\bm u},\hat{\bm u}^*)&:={\bm A}_{\bm k}(\hat{\bm u},\hat{\bm u}^*)p_t(\hat{\bm u},\hat{\bm u}^*)\notag\\
&\quad-\dfrac{1}{V}\dfrac{\nu k^2k_{\mathrm{B}}T}{\rho}\left(\mathsf{I}-\dfrac{{\bm k}{\bm k}}{k^2}\right)\cdot\dfrac{\partial}{\partial \hat{\bm u}^*_{\bm k}}p_t(\hat{\bm u},\hat{\bm u}^*),
\label{probability current}
\end{align}
where ${\bm A}_{\bm k}(\hat{\bm u},\hat{\bm u}^*)$ denotes a drift vector defined by
\begin{align}
{\bm A}_{\bm k}(\hat{\bm u},\hat{\bm u}^*):={\bm B}_{\bm k}(\hat{\bm u},\hat{\bm u}^*)-\nu k^2\hat{\bm u}_{\bm k}+\hat{\bm f}_{\bm k},
\label{A_k}
\end{align}
and $\mathsf{I}$ denotes the identity matrix.
Note that, from the incompressibility condition $\nabla\cdot{\bm u}=0$ and the definition of ${\bm f}$ and ${\bm B}_{\bm k}(\hat{\bm u},\hat{\bm u}^*)$, we have $\hat{\bm u}_{\bm k}\cdot{\bm k}=\hat{\bm f}_{\bm k}\cdot{\bm k}={\bm B}_{\bm k}(\hat{\bm u},\hat{\bm u}^*)\cdot{\bm k}=0$, and thus ${\bm J}_{\bm k}(\hat{\bm u},\hat{\bm u}^*)\cdot{\bm k}=0$.
Below, we assume that the system eventually reaches a statistically steady state with a stationary distribution $p_{\mathrm{ss}}(\hat{\bm u},\hat{\bm u}^*)$ after a sufficiently long time.

\section{Basic properties\label{Basic properties}}
In this section, we briefly review some basic properties of fully developed turbulence described by the fluctuating Navier--Stokes equation (\ref{fluctuating NS}).

\subsection{Energy balance}
We first consider the time evolution of the mean kinetic energy per unit mass $\langle|{\bm u}|^2\rangle/2$, where $\langle\cdot\rangle$ denotes the average with respect to $p_t(\hat{\bm u},\hat{\bm u}^*)$ (hereafter, we omit ``per unit mass'' for brevity).
Importantly, the nonlinear term ${\bm B}_{\bm k}(\hat{\bm u},\hat{\bm u}^*)$ satisfies the relation
\begin{align}
\sum_{\bm k}\left[{\bm B}_{\bm k}(\hat{\bm u},\hat{\bm u}^*)\cdot\hat{\bm u}^*_{\bm k}+{\bm B}^*_{\bm k}(\hat{\bm u},\hat{\bm u}^*)\cdot\hat{\bm u}_{\bm k}\right]=0.
\label{energy conservation property}
\end{align}
Then, from the Fokker--Planck equation (\ref{FP-fNS}), or equivalently from the stochastic differential equations (\ref{fNS_k}) and (\ref{fNS_k_*}) with stochastic calculus~\cite{gardiner1985handbook}, we find that the energy balance equation reads
\begin{align}
\dfrac{d}{dt}\sum_{\bm k}\dfrac{1}{2}\langle|\hat{\bm u}_{\bm k}|^2\rangle&=-\varepsilon+\sum_{\bm k}\dfrac{1}{2}\langle\hat{\bm f}_{\bm k}\cdot\hat{\bm u}^*_{\bm k}+\hat{\bm f}^*_{\bm k}\cdot\hat{\bm u}_{\bm k}\rangle\notag\\
&\quad+\sum_{\bm k}\dfrac{2\nu k^2k_{\mathrm{B}}T}{\rho V},
\label{energy balance}
\end{align}
where we have introduced the energy dissipation rate:
\begin{align}
\varepsilon:=\sum_{\bm k}\nu k^2\langle|\hat{\bm u}_{\bm k}|^2\rangle.
\end{align}
The second and third terms on the right-hand side of (\ref{energy balance}) denote the energy injection rate due to the external force and the internal thermal noise, respectively.
In the steady state, the energy dissipation rate balances the injection rate
\begin{align}
\varepsilon=\sum_{\bm k}\dfrac{1}{2}\langle\hat{\bm f}_{\bm k}\cdot\hat{\bm u}^*_{\bm k}+\hat{\bm f}^*_{\bm k}\cdot\hat{\bm u}_{\bm k}\rangle.
\label{dissipation = injection}
\end{align}
Here, we have ignored the energy injection due to the thermal noise by noting that the kinetic energy is much larger than the thermal energy over a wide range of scales in standard cases~\cite{bandak2021thermal,bandak2022dissipation}.

\subsection{Energy cascade}
While the nonlinear term ${\bm B}_{\bm k}(\hat{\bm u},\hat{\bm u}^*)$ does not contribute to the energy balance equation (\ref{energy balance}), it redistributes the energy over a wide range of scales.
To see this, we investigate the energy exchange across scales by considering the time evolution equation of the large-scale energy, which is defined as the total energy up to an arbitrary wave number $K$:
\begin{align}
\dfrac{d}{dt}\sum_{k\le K}\dfrac{1}{2}\langle|\hat{\bm u}_{\bm k}|^2\rangle&=-\Pi_K-\sum_{k\le K}\nu k^2\langle|\hat{\bm u}_{\bm k}|^2\rangle+\sum_{k\le K}\dfrac{2\nu k^2k_{\mathrm{B}}T}{\rho V}\notag\\
&\quad+\sum_{k\le K}\dfrac{1}{2}\langle\hat{\bm f}_{\bm k}\cdot\hat{\bm u}^*_{\bm k}+\hat{\bm f}^*_{\bm k}\cdot\hat{\bm u}_{\bm k}\rangle,
\label{large-scale energy balance}
\end{align}
where $\sum_{k\le K}$ denotes the summation over all ${\bm k}$ that satisfies $k\le K$.
Here, $\Pi_K$ denotes the scale-to-scale energy flux from large to small scales:
\begin{align}
\Pi_K:=-\sum_{k\le K}\dfrac{1}{2}\langle{\bm B}_{\bm k}(\hat{\bm u},\hat{\bm u}^*)\cdot\hat{\bm u}^*_{\bm k}+{\bm B}^*_{\bm k}(\hat{\bm u},\hat{\bm u}^*)\cdot\hat{\bm u}_{\bm k}\rangle.
\label{energy flux}
\end{align}

Now, we suppose that the system reaches a steady state and that the energy dissipation rate $\varepsilon$ remains finite in the inviscid limit $\nu\rightarrow0$~\cite{Frisch}.
Because the viscous dissipation is negligible at scales much larger than the Kolmogorov dissipation scale $\eta\equiv k_\nu^{-1}:=\nu^{3/4}\varepsilon^{-1/4}$, the second and third terms on the right-hand side of (\ref{large-scale energy balance}) can be ignored in the range $K\ll k_\nu$.
Similarly, by noting the relation (\ref{dissipation = injection}), the last term on the right-hand side of (\ref{large-scale energy balance}) can be approximated as the energy dissipation rate $\varepsilon$ in the range $K\gg k_f$.
Therefore, we obtain
\begin{align}
\Pi_K=\varepsilon
\label{cascade condition}
\end{align}
in the inertial range $k_f\ll K\ll k_\nu$.
The energy is thus transferred conservatively from large to small scales within the inertial range.
This \textit{energy cascade} process underlies various unique properties of turbulence~\cite{Frisch,davidson2015turbulence,Eyink_lecture} and is also essential for deriving our main results.

\section{Information-theoretic quantities\label{Information-theoretic quantities}}
\begin{figure}[t]
\includegraphics[width=7.5cm]{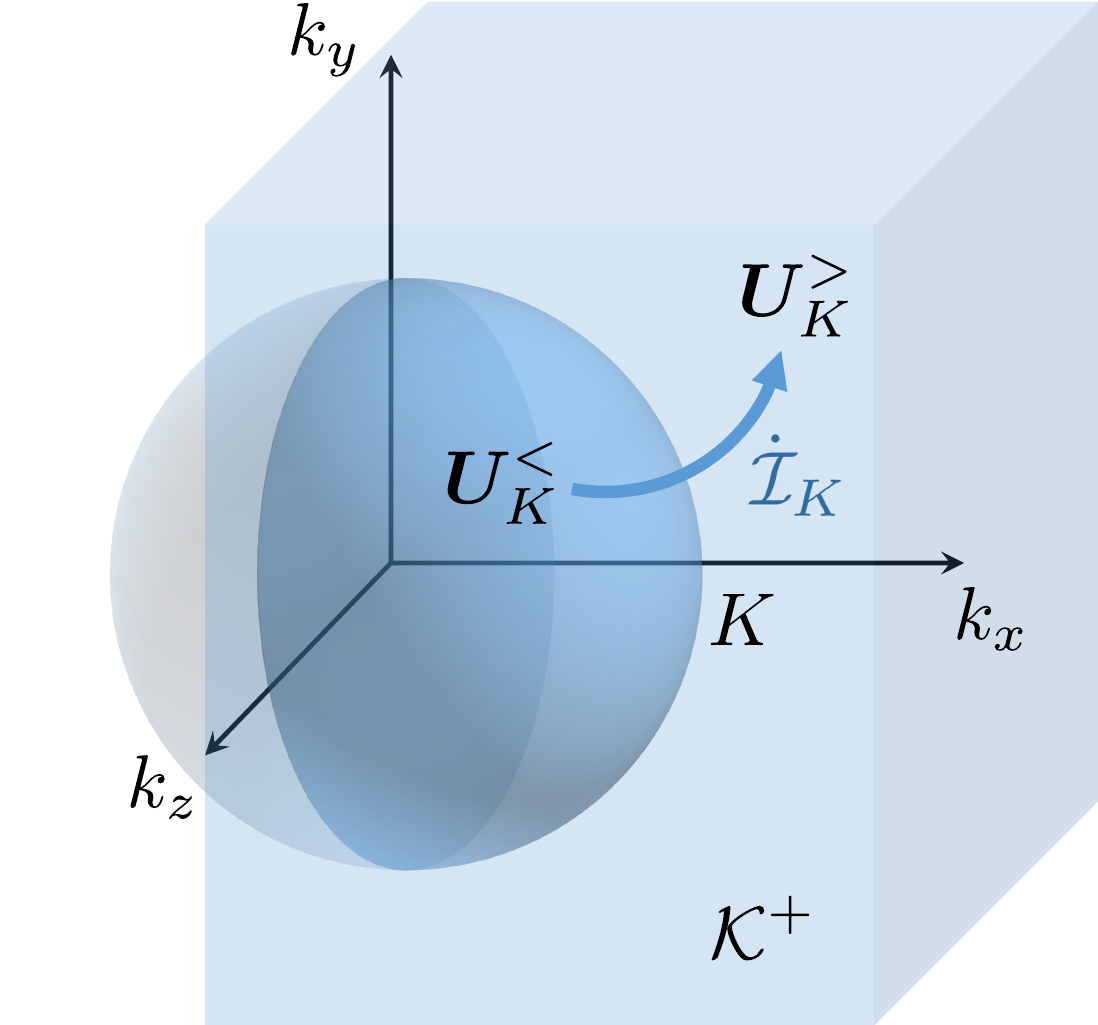}
\caption{Schematic of the information flow across scales.
The light blue shaded half-space represents the set of the wave vectors $\mathcal{K}^+$, defined by (\ref{independent Fourier modes}), associated with the independent Fourier modes $\{\hat{\bm u},\hat{\bm u}^*\}$.
We divide the independent Fourier modes into two parts at an arbitrary wave number $K$: $\{\hat{\bm u},\hat{\bm u}^*\}={\bm U}^<_K\cup{\bm U}^>_K$. 
The dark blue shaded hemisphere denotes the set of the wave vectors associated with the large-scale modes ${\bm U}^<_K$.
If the information flow $\dot{\mathcal{I}}_K$ is positive, then it means that the small-scale modes ${\bm U}^>_K$ are gaining information about the large-scale modes ${\bm U}^<_K$, as shown by the thick blue arrow pointing outward from the hemisphere.}
\label{fig:Fourier_modes_division}
\end{figure}

In this section, we introduce the two information-theoretic quantities that are important in describing our main result: \textit{mutual information} and \textit{information flow}.
Since we are interested in the information transfer across scales, we first divide the set of independent Fourier modes $\{\hat{\bm u},\hat{\bm u}^*\}$ into two parts at an arbitrary intermediate scale $K$ (see Fig.~\ref{fig:Fourier_modes_division}):
\begin{align}
\{\hat{\bm u},\hat{\bm u}^*\}={\bm U}^<_K\cup{\bm U}^>_K,
\end{align}
where ${\bm U}^<_K:=\{\hat{\bm u}_{\bm k},\hat{\bm u}^*_{\bm k}|{\bm k}\in\mathcal{K}^+, k\le K\}$ and ${\bm U}^>_K:=\{\hat{\bm u}_{\bm k},\hat{\bm u}^*_{\bm k}|{\bm k}\in\mathcal{K}^+, k>K\}$ denote the large-scale and small-scale modes, respectively.

The strength of the correlation between the large-scale modes ${\bm U}^<_K$ and small-scale modes ${\bm U}^>_K$ at time $t$ is quantified by the \textit{mutual information}~\cite{cover1999elements}:
\begin{align}
I[{\bm U}^<_K(t):{\bm U}^>_K(t)]:=\left\langle\ln\dfrac{p_t({\bm U}^<_K,{\bm U}^>_K)}{p^<_t({\bm U}^<_K)p^>_t({\bm U}^>_K)}\right\rangle,
\label{def: MI}
\end{align}
where $\langle\cdot\rangle$ denotes the average with respect to the joint probability distribution $p_t({\bm U}^<_K,{\bm U}^>_K)$, and $p^<_t({\bm U}^<_K)$ and $p^>_t({\bm U}^>_K)$ are the marginal distributions for the large-scale and small-scale modes, respectively.
Note that the joint probability distribution $p_t({\bm U}^<_K,{\bm U}^>_K)$ is nothing but the probability density for the total independent Fourier modes $p_t(\hat{\bm u},\hat{\bm u}^*)$ governed by the Fokker--Planck equation (\ref{FP-fNS}).
The mutual information is nonnegative and is equal to zero if and only if ${\bm U}^<_K$ and ${\bm U}^>_K$ are statistically independent.

Because the mutual information is symmetric between the two variables, it cannot quantify the directional flow of information from one variable to the other.
The directional flow of information can be quantified in terms of the \textit{information flow}, which is also called the \textit{learning rate}~\cite{hartich2014stochastic,barato2014efficiency,hartich2016sensory,matsumoto2018role}.
The information flow that characterizes the rate at which ${\bm U}^<_K$ acquires information about ${\bm U}^>_K$ is defined as
\begin{align}
\dot{I}^<_K:=\lim_{dt\rightarrow0^+}\dfrac{I[{\bm U}^<_K(t+dt):{\bm U}^>_K(t)]-I[{\bm U}^<_K(t):{\bm U}^>_K(t)]}{dt}.
\label{IF_def-1}
\end{align}
Similarly, the information flow associated with ${\bm U}^>_K$ is defined by
\begin{align}
\dot{I}^>_K:=\lim_{dt\rightarrow0^+}\dfrac{I[{\bm U}^<_K(t):{\bm U}^>_K(t+dt)]-I[{\bm U}^<_K(t):{\bm U}^>_K(t)]}{dt}.
\label{IF_def-2}
\end{align}
It follows from these definitions that the sum of these two information flows gives the time derivative of the mutual information:
\begin{align}
d_tI[{\bm U}^<_K:{\bm U}^>_K]=\dot{I}^<_K+\dot{I}^>_K.
\label{dtI}
\end{align}
See Appendix~\ref{Derivation of Eqs} for the derivation.
Therefore, in the steady state, there is only one information flow
\begin{align}
\dot{\mathcal{I}}_K:=\dot{I}^>_K=-\dot{I}^<_K
\label{single_IF}
\end{align}
because $d_tI[{\bm U}^<_K:{\bm U}^>_K]=0$.
The important point here is that we can detect the direction of the flow of information from the sign of $\dot{\mathcal{I}}_K$.
If $\dot{\mathcal{I}}_K>0$ ($\dot{\mathcal{I}}_K<0$), then it means that the small-scale modes ${\bm U}^>_K$ are gaining (destroying) information about the large-scale modes ${\bm U}^<_K$.
In other words, the positivity (negativity) of the information flow indicates that information about the large-scale (small-scale) modes is being transferred to small (large) scales (see Fig.~\ref{fig:Fourier_modes_division}).

We finally provide a few remarks on our definition of the mutual information and information flow.
Since the small-scale modes ${\bm U}^>_K$ include Fourier modes significantly affected by the viscous damping and thermal noise, the mutual information (\ref{def: MI}) and information flows (\ref{IF_def-1}) and (\ref{IF_def-2}) can possibly depend on the viscosity $\nu$ and the temperature $T$ even for $K$ within the inertial range.
Similarly, since the large-scale modes ${\bm U}^<_K$ include Fourier modes directly affected by the external force ${\bm f}$, these information-theoretic quantities can also depend on ${\bm f}$.
These points will be discussed in Sec.~\ref{Numerical simulation}, where we present some numerical results suggesting that these dependencies are weak in the inertial range.

\section{Information-thermodynamic bound on information flow in turbulence\label{Information-thermodynamic bound on information flow in turbulence}}
In this section, we first present our main result on the information flow in turbulence in Sec.~\ref{Main result}.
Then, we provide a detailed derivation of this result in Sec.~\ref{Derivation of the main result}.

\subsection{Main result\label{Main result}}
\begin{figure}[t]
\includegraphics[width=8.6cm]{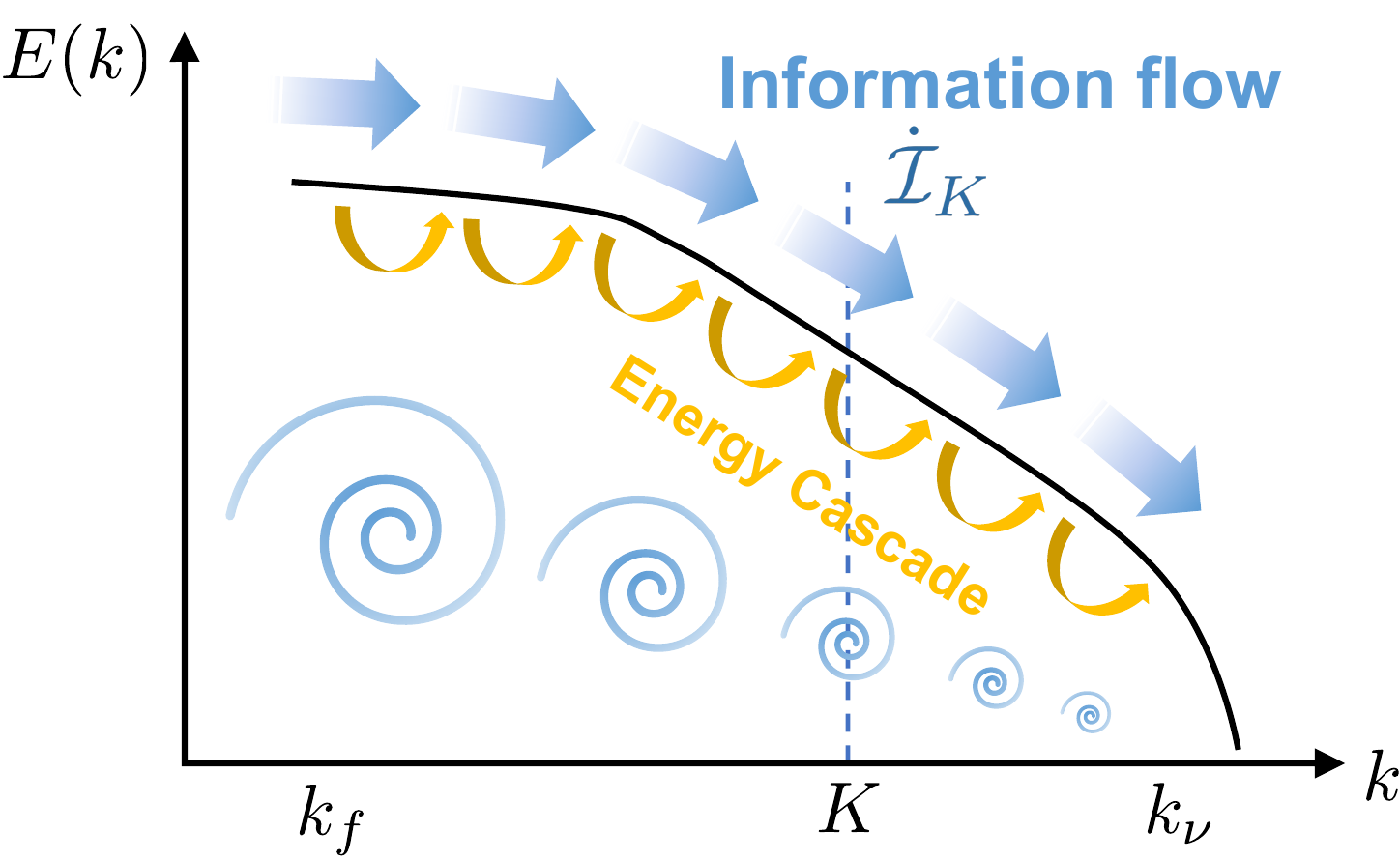}
\caption{Schematic of information flow in the energy cascade.}
\label{fig:main_result}
\end{figure}

We now state our first main result:
in the steady state, for any $K$ within the inertial range $k_f\ll K\ll k_\nu$, the information flow (\ref{single_IF}) is always nonnegative:
\begin{align}
\dot{\mathcal{I}}_K\ge0.
\label{main result 1}
\end{align}
This inequality states that information of large-scale eddies is transferred to small scales along with the energy cascade (see Fig.~\ref{fig:main_result}).
In other words, small-scale modes ${\bm U}^>_K$ are ``learning'' about the large-scale modes ${\bm U}^<_K$ while receiving the kinetic energy from large scales.
Furthermore, there is an upper bound on the information flow determined by the energy dissipation rate and the temperature of the fluid:
\begin{align}
\dfrac{\rho V\varepsilon}{k_{\mathrm{B}}T}&\ge\dot{\mathcal{I}}_K,
\label{main result 2}
\end{align}
which is the second main result of this paper.

Before proving these relations, here we provide several remarks.
First, no \textit{ad hoc} assumptions are used in deriving the inequalities (\ref{main result 1}) and (\ref{main result 2}).
These relations are based on the second law of information thermodynamics and the property of the energy cascade (\ref{cascade condition}).
Second, these inequalities hold independent of the details of the flow under consideration, such as the mechanism of the external forcing.
That is, (\ref{main result 1}) and (\ref{main result 2}) are universal relations, which are valid for all types of flow described by the fluctuating Navier--Stokes equation exhibiting the energy cascade (\ref{cascade condition}).
Furthermore, we can also prove the same relations even for other turbulence models such as shell models.
Here, we note that the information flow $\dot{\mathcal{I}}_K$ itself may not be universal, even if $K$ lies in the inertial range, as pointed out at the end of the previous section.
Nevertheless, our numerical simulation result suggests that the magnitude of the information flow is also universal in the inertial range (see Eq.~(\ref{IF_scaling})).
Third, these relations hold for arbitrary temperature $T$, including the limit $T\rightarrow0$, which formally corresponds to the deterministic case.
While the second inequality (\ref{main result 2}) becomes a trivial inequality in the limit $T\rightarrow0$, the first inequality (\ref{main result 1}) still provides a meaningful bound on the information flow.

\subsection{Derivation of the main result\label{Derivation of the main result}}
The derivation of the main result is based on the second law of information thermodynamics for bipartite systems~\cite{horowitz2014thermodynamics}.
Below, we first formulate the second law of thermodynamics [Ineq.~(\ref{total entropy production})] and then derive the second law of information thermodynamics [Ineqs.~(\ref{2nd law of information thermodynamics_large-scale}) and (\ref{2nd law of information thermodynamics_small-scale})].
Finally, from the inequalities (\ref{2nd law of information thermodynamics_large-scale}) and (\ref{2nd law of information thermodynamics_small-scale}), we derive the main result.

\subsubsection{Formulation of the second law of stochastic thermodynamics}
First, we formulate the standard second law of thermodynamics.
From a thermodynamic point of view, the fluctuating Navier--Stokes equation (\ref{fNS_k}) and (\ref{fNS_k_*}) consists of two parts: \textit{system} and \textit{thermal environment}~\cite{sekimoto2010stochastic}.
Here, by \textit{system}, we mean the independent Fourier modes $\{\hat{\bm u},\hat{\bm u}^*\}$, and by \textit{thermal environment}, we mean the fast degrees of freedom associated with the microscopic molecular motion, which induce the viscous damping and thermal noise.
In this paper, we treat entropy as dimensionless by dividing it by the Boltzmann constant $k_{\mathrm{B}}$.

Let $S[\hat{\bm u},\hat{\bm u}^*]$ be the entropy of the system, identified with the Shannon entropy
\begin{align}
S[\hat{\bm u},\hat{\bm u}^*]:=-\int d\hat{\bm u}d\hat{\bm u}^*p_t(\hat{\bm u},\hat{\bm u}^*)\ln p_t(\hat{\bm u},\hat{\bm u}^*),
\end{align}
where $d\hat{\bm u}d\hat{\bm u}^*:=\prod_{{\bm k}\in\mathcal{K}^+}\prod_{c\in\{x,y,z\}}d\mathrm{Re}[\hat{u}^c_{\bm k}]d\mathrm{Im}[\hat{u}^c_{\bm k}]$.
Here, we use the notation $S[\hat{\bm u},\hat{\bm u}^*]$ to indicate the relevant random variables $\hat{\bm u}$ and $\hat{\bm u}^*$ although $S[\hat{\bm u},\hat{\bm u}^*]$ is not a function of $\hat{\bm u}$ and $\hat{\bm u}^*$.
The average rate of change of the system entropy is given by
\begin{align}
&\quad\dfrac{d}{dt}S[\hat{\bm u},\hat{\bm u}^*]\notag\\
&=-\dfrac{d}{dt}\int d\hat{\bm u}d\hat{\bm u}^*p_t(\hat{\bm u},\hat{\bm u}^*)\ln p_t(\hat{\bm u},\hat{\bm u}^*)\notag\\
&=-\int d\hat{\bm u}d\hat{\bm u}^*(\partial_tp_t(\hat{\bm u},\hat{\bm u}^*))\ln p_t(\hat{\bm u},\hat{\bm u}^*)\notag\\
&\qquad-\int d\hat{\bm u}d\hat{\bm u}^*\partial_tp_t(\hat{\bm u},\hat{\bm u}^*)\notag\\
&=\int d\hat{\bm u}d\hat{\bm u}^*\sum_{{\bm k}\in\mathcal{K}^+}\left[\dfrac{\partial}{\partial \hat{\bm u}_{\bm k}}\cdot {\bm J}_{\bm k}(\hat{\bm u},\hat{\bm u}^*)+\dfrac{\partial}{\partial \hat{\bm u}^*_{\bm k}}\cdot {\bm J}^*_{\bm k}(\hat{\bm u},\hat{\bm u}^*)\right]\notag\\
&\qquad\times\ln p_t(\hat{\bm u},\hat{\bm u}^*)\notag\\
&=\sum_{{\bm k}\in\mathcal{K}^+}\dot{S}_{\bm k}[\hat{\bm u},\hat{\bm u}^*].
\label{shannon entropy change}
\end{align}
Here, in the third equality, we have used the Fokker--Planck equation (\ref{FP-fNS}) and the fact that $\int d\hat{\bm u}d\hat{\bm u}^*\partial_tp_t(\hat{\bm u},\hat{\bm u}^*)=d_t\int d\hat{\bm u}d\hat{\bm u}^*p_t(\hat{\bm u},\hat{\bm u}^*)=0$.
In the last equality, we have introduced $\dot{S}_{\bm k}[\hat{\bm u},\hat{\bm u}^*]$, which is given by
\begin{align}
\dot{S}_{\bm k}[\hat{\bm u},\hat{\bm u}^*]&:=-\int d\hat{\bm u}d\hat{\bm u}^*\left[{\bm J}_{\bm k}(\hat{\bm u},\hat{\bm u}^*)\cdot\dfrac{\partial}{\partial \hat{\bm u}_{\bm k}}\ln p_t(\hat{\bm u},\hat{\bm u}^*)\right.\notag\\
&\quad\left.+{\bm J}^*_{\bm k}(\hat{\bm u},\hat{\bm u}^*)\cdot\dfrac{\partial}{\partial \hat{\bm u}^*_{\bm k}}\ln p_t(\hat{\bm u},\hat{\bm u}^*)\right],
\end{align}
where the over-dot denotes the rates of change of observables that are not a time derivative of a state function.

We identify the entropy change in the environment according to Sekimoto's argument~\cite{sekimoto2010stochastic}.
Since the model satisfies the fluctuation-dissipation relation of the second kind, the thermal environment is ensured to be always in equilibrium at temperature $T$.
Then, by noting that $-\nu k^2\hat{\bm u}_{\bm k}+\sqrt{2\nu k^2k_{\mathrm{B}}T/\rho}\hat{\bm \xi}_{\bm k}$ can be interpreted as a force exerted by the environment on the system, the entropy change in the environment is identified as the work done by the system on the environment per unit time divided by $k_{\mathrm{B}}T$:
\begin{align}
\dot{S}^{\mathrm{env}}&=\sum_{{\bm k}}\dfrac{\rho V}{2k_{\mathrm{B}}T}\left\langle \hat{\bm u}^*_{\bm k}\circ\left[\nu k^2\hat{\bm u}_{\bm k}-\sqrt{\dfrac{2\nu k^2k_{\mathrm{B}}T}{\rho}}\hat{\bm \xi}_{\bm k}\right]+\mathrm{c.c.}\right\rangle\notag\\
&=\sum_{{\bm k}\in\mathcal{K}^+}\dot{S}^{\mathrm{env}}_{\bm k},
\label{medium entropy change}
\end{align}
where $\mathrm{c.c.}$ denotes the complex conjugate term and $\dot{S}^{\mathrm{env}}_{\bm k}$ denotes the entropy change in the environment associated with a wave vector ${\bm k}\in\mathcal{K}^+$,
\begin{align}
\dot{S}^{\mathrm{env}}_{\bm k}&:=\dfrac{\rho V}{k_{\mathrm{B}}T}\left\langle \hat{\bm u}^*_{\bm k}\circ\left[\nu k^2\hat{\bm u}_{\bm k}-\sqrt{\dfrac{2\nu k^2k_{\mathrm{B}}T}{\rho}}\hat{\bm \xi}_{\bm k}\right]+\mathrm{c.c.}\right\rangle.
\end{align}
Here, the symbol ``$\circ$'' denotes the multiplication in the sense of Stratonovich~\cite{gardiner1985handbook}.
We remark that this identification is consistent with the \textit{local detailed balance} (see Appendix~\ref{Local detailed balance}).

The total entropy production rate, which we denote by $\dot{\sigma}$, is identified as the sum of the average rate of change of the system entropy (\ref{shannon entropy change}) and the entropy change in the environment (\ref{medium entropy change}):
\begin{align}
\dot{\sigma}&:=\dfrac{d}{dt}S[\hat{\bm u},\hat{\bm u}^*]+\dot{S}^{\mathrm{env}}\notag\\
&=\sum_{{\bm k}\in\mathcal{K}^+}\left(\dot{S}_{\bm k}[\hat{\bm u},\hat{\bm u}^*]+\dot{S}^{\mathrm{env}}_{\bm k}\right).
\end{align}
We now show that $\dot{\sigma}\ge0$, which is a manifestation of the second law of thermodynamics and is sometimes called the \textit{second law of stochastic thermodynamics}~\cite{peliti2021stochastic}.
To this end, it is convenient to decompose the probability current (\ref{probability current}) into two parts: ${\bm J}_{\bm k}(\hat{\bm u},\hat{\bm u}^*)={\bm J}^{\mathrm{ir}}_{\bm k}(\hat{\bm u},\hat{\bm u}^*)+{\bm J}^{\mathrm{rev}}_{\bm k}(\hat{\bm u},\hat{\bm u}^*)$, where ${\bm J}^{\mathrm{ir}}_{\bm k}(\hat{\bm u},\hat{\bm u}^*)$ denotes the irreversible probability current, defined by
\begin{align}
{\bm J}^{\mathrm{ir}}_{\bm k}(\hat{\bm u},\hat{\bm u}^*)&:=\dfrac{1}{2}\left[{\bm J}_{\bm k}(\hat{\bm u},\hat{\bm u}^*)-{\bm J}_{\bm k}(-\hat{\bm u},-\hat{\bm u}^*)\right]\notag\\
&=-\nu k^2\hat{\bm u}_{\bm k}p_t(\hat{\bm u},\hat{\bm u}^*)\notag\\
&\quad-\dfrac{1}{V}\dfrac{\nu k^2k_{\mathrm{B}}T}{\rho}\left(\mathsf{I}-\dfrac{{\bm k}{\bm k}}{k^2}\right)\cdot\dfrac{\partial}{\partial \hat{\bm u}^*_{\bm k}}p_t(\hat{\bm u},\hat{\bm u}^*),
\end{align}
and ${\bm J}^{\mathrm{rev}}_{\bm k}(\hat{\bm u},\hat{\bm u}^*)$ denotes the reversible probability current, defined by
\begin{align}
{\bm J}^{\mathrm{rev}}_{\bm k}(\hat{\bm u},\hat{\bm u}^*)&:=\dfrac{1}{2}[{\bm J}_{\bm k}(\hat{\bm u},\hat{\bm u}^*)+{\bm J}_{\bm k}(-\hat{\bm u},-\hat{\bm u}^*)]\notag\\
&=\left({\bm B}_{\bm k}(\hat{\bm u},\hat{\bm u}^*)+\hat{\bm f}_{\bm k}\right)p_t(\hat{\bm u},\hat{\bm u}^*).
\end{align}
We then rewrite the average rate of change of the system entropy (\ref{shannon entropy change}) as
\begin{widetext}
\begin{align}
\dfrac{d}{dt}S[\hat{\bm u},\hat{\bm u}^*]&=-\sum_{{\bm k}\in\mathcal{K}^+}\int d\hat{\bm u}d\hat{\bm u}^*\left[{\bm J}_{\bm k}(\hat{\bm u},\hat{\bm u}^*)\cdot\dfrac{\partial}{\partial \hat{\bm u}_{\bm k}}\ln p_t(\hat{\bm u},\hat{\bm u}^*)+\mathrm{c.c.}\right]\notag\\
&=\sum_{{\bm k}\in\mathcal{K}^+}\dfrac{\rho V}{\nu k^2k_{\mathrm{B}}T}\int d\hat{\bm u}d\hat{\bm u}^*\left[\dfrac{{\bm J}^{\mathrm{ir}}_{\bm k}(\hat{\bm u},\hat{\bm u}^*)}{p_t(\hat{\bm u},\hat{\bm u}^*)}\cdot\left\{-\dfrac{1}{V}\dfrac{\nu k^2k_{\mathrm{B}}T}{\rho}\left(\mathsf{I}-\dfrac{{\bm k}{\bm k}}{k^2}\right)\cdot\dfrac{\partial}{\partial \hat{\bm u}_{\bm k}}p_t(\hat{\bm u},\hat{\bm u}^*)\right\}+\mathrm{c.c.}\right],
\label{shannon entropy change_rewrite}
\end{align}
where we have used the fact that
\begin{align}
\dfrac{\partial}{\partial \hat{\bm u}_{\bm k}}\cdot\dfrac{{\bm J}^{\mathrm{rev}}_{\bm k}(\hat{\bm u},\hat{\bm u}^*)}{p_t(\hat{\bm u},\hat{\bm u}^*)}+\dfrac{\partial}{\partial \hat{\bm u}^*_{\bm k}}\cdot\dfrac{{\bm J}^{\mathrm{rev}*}_{\bm k}(\hat{\bm u},\hat{\bm u}^*)}{p_t(\hat{\bm u},\hat{\bm u}^*)}=-2i{\bm k}\cdot\hat{\bm u}_{\bm 0}+2i{\bm k}\cdot\hat{\bm u}_{\bm 0}=0
\end{align}
and that ${\bm J}^{\mathrm{ir}}_{\bm k}(\hat{\bm u},\hat{\bm u}^*)$ is orthogonal to ${\bm k}$, i.e., ${\bm J}^{\mathrm{ir}}_{\bm k}(\hat{\bm u},\hat{\bm u}^*)\cdot(\mathsf{I}-{\bm k}{\bm k}/k^2)={\bm J}^{\mathrm{ir}}_{\bm k}(\hat{\bm u},\hat{\bm u}^*)$.
We also rewrite the entropy change in the environment (\ref{medium entropy change}) as
\begin{align}
\dot{S}^{\mathrm{env}}&:=\sum_{{\bm k}\in\mathcal{K}^+}\dfrac{\rho V}{k_{\mathrm{B}}T}\left\langle \hat{\bm u}^*_{\bm k}\circ\left[\nu k^2\hat{\bm u}_{\bm k}-\sqrt{\dfrac{2\nu k^2k_{\mathrm{B}}T}{\rho}}\hat{\bm \xi}_{\bm k}\right]+\mathrm{c.c.}\right\rangle\notag\\
&=\sum_{{\bm k}\in\mathcal{K}^+}\dfrac{\rho V}{k_{\mathrm{B}}T}\left\langle\hat{\bm u}^*_{\bm k}\cdot\left[\nu k^2\hat{\bm u}_{\bm k}-\sqrt{\dfrac{2\nu k^2k_{\mathrm{B}}T}{\rho}}\hat{\bm \xi}_{\bm k}\right]-\dfrac{1}{2}\dfrac{\partial}{\partial\hat{\bm u}^*_{\bm k}}\hat{\bm u}^*_{\bm k}:\left(\mathsf{I}-\dfrac{{\bm k}{\bm k}}{k^2}\right)\dfrac{2\nu k^2k_{\mathrm{B}}T}{\rho V}+\mathrm{c.c.}\right\rangle\notag\\
&=\sum_{{\bm k}\in\mathcal{K}^+}\dfrac{\rho V}{\nu k^2k_{\mathrm{B}}T}\int d\hat{\bm u}d\hat{\bm u}^*\left[\dfrac{-\nu k^2\hat{\bm u}^*_{\bm k}p_t(\hat{\bm u},\hat{\bm u}^*)}{p_t(\hat{\bm u},\hat{\bm u}^*)}\cdot{\bm J}^{\mathrm{ir}}_{\bm k}(\hat{\bm u},\hat{\bm u}^*)+\mathrm{c.c.}\right].
\label{medium entropy change_rewrite}
\end{align}
\end{widetext}
Here, in the second line, we have used the relation between the Stratonovich and the Ito integral~\cite{gardiner1985handbook}, so that the inner product ``$\cdot$'' here should be interpreted as the multiplication in the sense of Ito.
Then, by combining (\ref{shannon entropy change_rewrite}) and (\ref{medium entropy change_rewrite}), we can confirm that the total entropy production rate is nonnegative:
\begin{align}
\dot{\sigma}&=\sum_{{\bm k}\in\mathcal{K}^+}\left(\dot{S}_{\bm k}[\hat{\bm u},\hat{\bm u}^*]+\dot{S}^{\mathrm{env}}_{\bm k}\right)\notag\\
&=\sum_{{\bm k}\in\mathcal{K}^+}\int d\hat{\bm u}d\hat{\bm u}^*\dfrac{2\rho V}{\nu k^2k_{\mathrm{B}}T}\dfrac{|{\bm J}^{\mathrm{ir}}_{\bm k}(\hat{\bm u},\hat{\bm u}^*)|^2}{p_t(\hat{\bm u},\hat{\bm u}^*)}\ge0.
\label{total entropy production}
\end{align}

\subsubsection{Derivation of the second law of information thermodynamics}
Now, we derive the second law of information thermodynamics.
Let $\dot{\sigma}_{\bm k}:=\dot{S}_{\bm k}[\hat{\bm u},\hat{\bm u}^*]+\dot{S}^{\mathrm{env}}_{\bm k}$ be the \textit{partial entropy production rate}~\cite{shiraishi2015fluctuation} associated with a wave vector ${\bm k}\in\mathcal{K}^+$, so that $\dot{\sigma}=\sum_{{\bm k}\in\mathcal{K}^+}\dot{\sigma}_{\bm k}$.
As is clear from the expression (\ref{total entropy production}), $\dot{\sigma}_{\bm k}$ is also nonnegative for each wave vector ${\bm k}$:
\begin{align}
\dot{\sigma}_{\bm k}=\int d\hat{\bm u}d\hat{\bm u}^*\dfrac{2\rho V}{\nu k^2k_{\mathrm{B}}T}\dfrac{|{\bm J}^{\mathrm{ir}}_{\bm k}(\hat{\bm u},\hat{\bm u}^*)|^2}{p_t(\hat{\bm u},\hat{\bm u}^*)}\ge0.
\label{partial entropy production per k}
\end{align}
From this relation, we can derive the second law of information thermodynamics for the two sets of Fourier modes ${\bm U}^<_K$ and ${\bm U}^>_K$.
We first note that the information flow $\dot{I}^<_K$ associated with the large-scale modes ${\bm U}^<_K$ can be rewritten as
\begin{widetext}
\begin{align}
\dot{I}^<_K&:=\lim_{dt\rightarrow0^+}\dfrac{I[{\bm U}^<_K(t+dt):{\bm U}^>_K(t)]-I[{\bm U}^<_K(t):{\bm U}^>_K(t)]}{dt}\notag\\
&=\sum_{{\bm k}\in\mathcal{K}^+,k\le K}\int d\hat{\bm u}d\hat{\bm u}^*\left[{\bm J}_{\bm k}(\hat{\bm u},\hat{\bm u}^*)\cdot\dfrac{\partial}{\partial \hat{\bm u}_{\bm k}}\ln\dfrac{p_t({\bm U}^<_K,{\bm U}^>_K)}{p^<_t({\bm U}^<_K)p^>_t({\bm U}^>_K)}+{\bm J}^*_{\bm k}(\hat{\bm u},\hat{\bm u}^*)\cdot\dfrac{\partial}{\partial \hat{\bm u}^*_{\bm k}}\ln\dfrac{p_t({\bm U}^<_K,{\bm U}^>_K)}{p^<_t({\bm U}^<_K)p^>_t({\bm U}^>_K)}\right].
\label{IF_rewrite}
\end{align}
For the derivation, see Appendix~\ref{Derivation of Eqs}.
By using this relation, we obtain
\begin{align}
\dot{I}^<_K&=-\sum_{{\bm k}\in\mathcal{K}^+,k\le K}\int d\hat{\bm u}d\hat{\bm u}^*\left[{\bm J}_{\bm k}(\hat{\bm u},\hat{\bm u}^*)\cdot\dfrac{\partial}{\partial \hat{\bm u}_{\bm k}}\ln p^<_t({\bm U}^<_K)+{\bm J}^*_{\bm k}(\hat{\bm u},\hat{\bm u}^*)\cdot\dfrac{\partial}{\partial \hat{\bm u}^*_{\bm k}}\ln p^<_t({\bm U}^<_K)\right]\notag\\
&\quad+\sum_{{\bm k}\in\mathcal{K}^+,k\le K}\int d\hat{\bm u}d\hat{\bm u}^*\left[{\bm J}_{\bm k}(\hat{\bm u},\hat{\bm u}^*)\cdot\dfrac{\partial}{\partial \hat{\bm u}_{\bm k}}\ln p_t(\hat{\bm u},\hat{\bm u}^*)+{\bm J}^*_{\bm k}(\hat{\bm u},\hat{\bm u}^*)\cdot\dfrac{\partial}{\partial \hat{\bm u}^*_{\bm k}}\ln p_t(\hat{\bm u},\hat{\bm u}^*)\right]\notag\\
&=\dfrac{d}{dt}S[{\bm U}^<_K]-\sum_{{\bm k}\in\mathcal{K}^+,k\le K}\dot{S}_{\bm k}[\hat{\bm u},\hat{\bm u}^*],
\label{information flow and conditional Shannon entropy}
\end{align}
\end{widetext}
where $S[{\bm U}^<_K]:=-\int d{\bm U}^<_Kp^<_t({\bm U}^<_K)\ln p^<_t({\bm U}^<_K)$ denotes the Shannon entropy of the large-scale modes ${\bm U}^<_K$.
Then, by summing (\ref{partial entropy production per k}) over all ${\bm k}$ that satisfies ${\bm k}\in\mathcal{K}^+$ and $k\le K$ and by using (\ref{information flow and conditional Shannon entropy}), we obtain the second law of information thermodynamics for the large-scale modes:
\begin{align}
\sum_{{\bm k}\in\mathcal{K}^+,k\le K}\dot{\sigma}_{\bm k}=\dfrac{d}{dt}S[{\bm U}^<_K]+\dot{S}^<_{\mathrm{env}}-\dot{I}^<_K\ge0,
\label{2nd law of information thermodynamics_large-scale}
\end{align}
where $\dot{S}^<_{\mathrm{env}}:=\sum_{{\bm k}\in\mathcal{K}^+,k\le K}\dot{S}^{\mathrm{env}}_{\bm k}$ denotes the entropy change in the environment due to the large-scale modes.
Similarly, we can obtain the second law of information thermodynamics for the small-scale modes:
\begin{align}
\sum_{{\bm k}\in\mathcal{K}^+,k> K}\dot{\sigma}_{\bm k}=\dfrac{d}{dt}S[{\bm U}^>_K]+\dot{S}^>_{\mathrm{env}}-\dot{I}^>_K\ge0.
\label{2nd law of information thermodynamics_small-scale}
\end{align}
Note that the first two terms on the right-hand side of (\ref{2nd law of information thermodynamics_large-scale}) and (\ref{2nd law of information thermodynamics_small-scale}) can be interpreted as the total entropy production rate associated with the large-scale and small-scale modes, respectively.
Then, (\ref{2nd law of information thermodynamics_large-scale}) and (\ref{2nd law of information thermodynamics_small-scale}) state that the total entropy production associated with each mode is not necessarily nonnegative but is bounded by the information flow.
In particular, if ${\bm U}^<_K$ and ${\bm U}^>_K$ are statistically independent, then $\dot{I}^<_K=\dot{I}^>_K=0$ and the standard second law of thermodynamics holds for each mode.
In contrast, if they are correlated, then the inequalities (\ref{2nd law of information thermodynamics_large-scale}) and (\ref{2nd law of information thermodynamics_small-scale}) give nontrivial bounds on the information flow in terms of the entropy production.

\subsubsection{Derivation of the main result}
We now derive the main results (\ref{main result 1}) and (\ref{main result 2}) from the second law of information thermodynamics (\ref{2nd law of information thermodynamics_large-scale}) and (\ref{2nd law of information thermodynamics_small-scale}).
We assume that the system is in the steady state.
Then, by noting that $\dot{\mathcal{I}}_K=\dot{I}^>_K=-\dot{I}^<_K$, (\ref{2nd law of information thermodynamics_large-scale}) and (\ref{2nd law of information thermodynamics_small-scale}) can be rewritten as
\begin{align}
\dot{S}^<_{\mathrm{env}}+\dot{\mathcal{I}}_K\ge0,\label{2nd law of information thermodynamics_large-scale_NESS}\\
\dot{S}^>_{\mathrm{env}}-\dot{\mathcal{I}}_K\ge0,\label{2nd law of information thermodynamics_small-scale_NESS}
\end{align}
respectively.
We set $K$ to be within the inertial range $k_f\ll K\ll k_\nu$.
Then, $\dot{S}^<_{\mathrm{env}}$ can be expressed in terms of the energy flux (\ref{energy flux}) as
\begin{align}
\dot{S}^<_{\mathrm{env}}
&=\sum_{{\bm k}\in\mathcal{K}^+,k\le K}\dfrac{\rho V}{k_{\mathrm{B}}T}\notag\\
&\quad\times\left\langle \hat{\bm u}^*_{\bm k}\circ\left[\nu k^2\hat{\bm u}_{\bm k}-\sqrt{\dfrac{2\nu k^2k_{\mathrm{B}}T}{\rho}}\hat{\bm \xi}_{\bm k}\right]+\mathrm{c.c.}\right\rangle\notag\\
&=\sum_{k\le K}\dfrac{1}{2}\dfrac{\rho V}{k_{\mathrm{B}}T}\notag\\
&\quad\times\left\langle \hat{\bm u}^*_{\bm k}\circ\left[{\bm B}_{\bm k}(\hat{\bm u},\hat{\bm u}^*)+\hat{\bm f}_{\bm k}-\partial_t\hat{\bm u}_{\bm k}\right]+\mathrm{c.c.}\right\rangle\notag\\
&=\dfrac{\rho V}{k_{\mathrm{B}}T}\left(\varepsilon-\Pi_K\right),
\end{align}
where we have used $\sum_{k\le K}\langle\hat{\bm f}_{\bm k}\cdot\hat{\bm u}^*_{\bm k}+\hat{\bm f}^*_{\bm k}\cdot\hat{\bm u}_{\bm k}\rangle/2=\varepsilon$ for $K$ within the inertial range $k_f\ll K\ll k_\nu$ and $\langle \hat{\bm u}^*_{\bm k}\circ\partial_t\hat{\bm u}_{\bm k}+\hat{\bm u}_{\bm k}\circ\partial_t\hat{\bm u}^*_{\bm k}\rangle=d_t\langle|\hat{\bm u}_{\bm k}|^2\rangle=0$ in the steady state.
Similarly, $\dot{S}^>_{\mathrm{env}}$ can be expressed as
\begin{align}
\dot{S}^>_{\mathrm{env}}&=\dfrac{\rho V}{k_{\mathrm{B}}T}\Pi_K,
\end{align}
where we have used the property of the nonlinear term (\ref{energy conservation property}).
By substituting these expressions into (\ref{2nd law of information thermodynamics_large-scale_NESS}) and (\ref{2nd law of information thermodynamics_small-scale_NESS}) and by noting that $\Pi_K\rightarrow\varepsilon$ as $K/k_\nu\rightarrow0$, we arrive at the main result (\ref{main result 1}) and (\ref{main result 2}).

\section{Numerical simulation\label{Numerical simulation}}
We here numerically illustrate the main result by estimating the information flow $\dot{\mathcal{I}}_K$.
Since the estimation of the information flow for the fluctuating Navier--Stokes equation requires an enormous computational cost, we instead use a fluctuating shell model, which is a simplified caricature of the fluctuating Navier--Stokes equation in wave number space.
Even for the fluctuating shell model, we can easily confirm that the main results (\ref{main result 1}) and (\ref{main result 2}) are valid.
In the following, we first introduce the fluctuating shell model in Sec.~\ref{Model_numerical simulation}.
Next, we explain the setup of the numerical simulation in Sec.~\ref{Setup of the numerical simulation}.
The numerical simulation results are presented in Sec.~\ref{Results_numerical simulation}.

\subsection{Model\label{Model_numerical simulation}}
We consider the Sabra shell model with thermal noise~\cite{l1998improved,bandak2021thermal,bandak2022dissipation}.
Let $u_n(t)\in\mathbb{C}$ be the ``velocity'' at time $t$ with the wave number $k_n=k_02^n$ ($n=0,1,\cdots,N$).
The time evolution of the complex shell variables $u:=\{u_n\}$ is given by the following Langevin equation:
\begin{align}
\partial_t{u}_n=B_n(u,u^*)-\nu k^2_nu_n+\sqrt{\dfrac{2\nu k^2_nk_{\mathrm{B}}T}{\rho}}\xi_n+f_n
\label{sabra shell model}
\end{align}
with the scale-local nonlinear interactions given by
\begin{align}
B_n(u,u^*)&:=i\biggl(k_{n+1}u_{n+2}u^*_{n+1}\notag\\
&\quad-\dfrac{1}{2}k_nu_{n+1}u^*_{n-1}+\dfrac{1}{2}k_{n-1}u_{n-1}u_{n-2}\biggr),
\end{align}
where we set $u_{-1}=u_{-2}=u_{N+1}=u_{N+2}=0$.
Here, $\nu>0$ represents the kinematic viscosity, $f_n\in\mathbb{C}$ denotes the external body force that acts only at large scales, i.e., $f_n=0$ for $n>n_f$, and $\xi_n\in\mathbb{C}$ is the zero-mean white Gaussian noise that satisfies $\langle\xi_n(t)\xi^*_{n'}(t')\rangle=2\delta_{nn'}\delta(t-t')$.
The specific form of the thermal noise term satisfies the fluctuation-dissipation relation of the second kind, where $T$ denotes the absolute temperature, $k_{\mathrm{B}}$ the Boltzmann constant, and $\rho$ the mass ``density''.

Although the shell model has a much simpler form than the Navier--Stokes equation, it exhibits rich temporal and multiscale statistics that are similar to those observed in real turbulent flow~\cite{biferale2003shell,bohr1998dynamical}.
In particular, the energy cascade property (\ref{cascade condition}) is satisfied even for this model.
Then, we can easily confirm that the main results (\ref{main result 1}) and (\ref{main result 2}) are valid:
\begin{align}
\dot{\mathcal{I}}_K&\ge0,\\
\dfrac{\rho\varepsilon}{k_{\mathrm{B}}T}&\ge\dot{\mathcal{I}}_K,\label{main_result_shell model}
\end{align}
for $K$ within the inertial range $k_f\ll K\ll k_\nu$, where $k_f:=k_{n_f}$ and $k_\nu\equiv\eta^{-1}:=\nu^{-3/4}\varepsilon^{1/4}$.
Note that, in contrast to (\ref{main result 2}), the volume of the fluid $V$ does not appear in (\ref{main_result_shell model}) because $\rho$ has units of mass in the shell model.

\subsection{Setup of the numerical simulation\label{Setup of the numerical simulation}}
\renewcommand{\arraystretch}{1.2} 
\begin{table}[b]
\caption{\label{table: three cases}%
The largest shell number $N$, the achieved Reynolds number $\mathrm{Re}$, and the dimensionless temperature $\theta_\eta$ of the three different cases.
}
\begin{ruledtabular}
\begin{tabular}{lccc}
\textrm{Case}&
$N$&
$\mathrm{Re}$&
$\theta_\eta$\\
\colrule
I & $22$ & $1.46\times10^6$ & $2.328\times10^{-8}$\\
II & $19$ & $9.25\times10^4$ & $2.328\times10^{-8}$\\
III & $22$ & $1.46\times10^6$ & $0$\\
\end{tabular}
\end{ruledtabular}
\end{table}
To investigate the Reynolds number and temperature dependence of the information flow, we consider three different cases as listed in Table \ref{table: three cases}.
In Case I, we set $N=22$ and $n_f=1$ to ensure that the external force acts only on the 0th and 1st shells of the total 23 shells.
In choosing the parameter values, we note that the presence of thermal fluctuations introduces another dimensionless quantity $\theta_\eta:=k_{\mathrm{B}}T/\rho u^2_\eta$ in addition to the Reynolds number $\mathrm{Re}$.
Here, $u_\eta:=(\varepsilon\nu)^{1/4}$ denotes the characteristic velocity at the Kolmogorov dissipation scale, and thus the dimensionless temperature $\theta_\eta$ is the ratio of the thermal energy to the kinetic energy at the Kolmogorov dissipation scale.
Then, the values of the external force and the other parameters are chosen following Refs.~\cite{bandak2021thermal,bandak2022dissipation} so that the achieved Reynolds number $\mathrm{Re}$ and the dimensionless temperature $\theta_\eta$ are both comparable to the typical values in the atmospheric boundary layer, i.e., $\mathrm{Re}\sim10^6$ and $\theta_\eta\sim10^{-8}$.
In Case II, we set $N=19$ so that the achieved Reynolds number is lowered to $\mathrm{Re}\sim10^5$ while leaving the other parameter values unchanged.
In Case III, we consider the standard deterministic case by setting $T=0$ ($\theta_\eta=0$) while leaving the other parameter values unchanged.
In all three cases, we have used $N_{\mathrm{samp}}=3\times10^5$ samples in the following averaging and estimation.

In the estimation of the mutual information, we first note that the naive binning approach is not feasible because it requires estimation of the $2(N+1)$-dimensional probability density $p_t({\bm U}^<_K,{\bm U}^>_K)$.
Instead, we use the so-called Kraskov-St\"ogbauer-Grassberger (KSG) estimator~\cite{kraskov2004estimating,khan2007relative,holmes2019estimation}, which has the advantage that it does not require estimation of the underlying probability density.
The KSG estimator uses the distances to the $\kappa$-th nearest neighbors of the sample points in the data to detect the structures of the underlying probability distribution.
While we set $\kappa=4$ here, following Ref.~\cite{kraskov2004estimating}, essentially the same result can be obtained for other values of $\kappa$.
Because the KSG estimator is based on the local uniformity assumption of the probability density, the estimated value approaches the true value as $N_{\mathrm{samp}}\rightarrow\infty$ when this assumption is satisfied.
In the following, we denote by $\hat{I}^{(\kappa)}_{\mathrm{KSG}}[{\bm U}^<_K(t):{\bm U}^>_K(t)]$ the KSG estimator of the mutual information $I[{\bm U}^<_K(t):{\bm U}^>_K(t)]$.

The information flow $\dot{\mathcal{I}}_K$ can be estimated by using the KSG estimator.
Note that this procedure requires high accuracy in the estimation of the mutual information because the information flow is defined through infinitesimal increments in the mutual information.
Because it is not feasible to increase the number of samples, we instead take the approach of using the largest possible time increment $\Delta t$.
That is, we define the estimated information flow $\hat{\dot{\mathcal{I}}}_K$ by
\begin{align}
\hat{\dot{\mathcal{I}}}_K:=-\dfrac{\hat{I}^{(\kappa)}_{\mathrm{KSG}}[{\bm U}^<_K(t+\Delta t):{\bm U}^>_K(t)]-\hat{I}^{(\kappa)}_{\mathrm{KSG}}[{\bm U}^<_K(t):{\bm U}^>_K(t)]}{\Delta t}.
\end{align}
Because we are interested in $K$ within the inertial range, we choose $\Delta t$ such that it is smaller than the smallest time scale in the inertial range. 
Therefore, we set $\Delta t=0.1\tau_\eta$, where $\tau_\eta:=\eta/u_\eta$ denotes the typical time scale at the Kolmogorov dissipation scale.
Note that $\Delta t$ is different from the time step $\delta t:=10^{-5}\tau_\eta$ used in solving (\ref{sabra shell model}) numerically.

Further details of the numerical simulation are given in Appendix~\ref{Details of the numerical simulation}, including the details of the KSG estimator.

\subsection{Results\label{Results_numerical simulation}}
\subsubsection{Energy spectrum}
\begin{figure}[t]
\includegraphics[width=8.6cm]{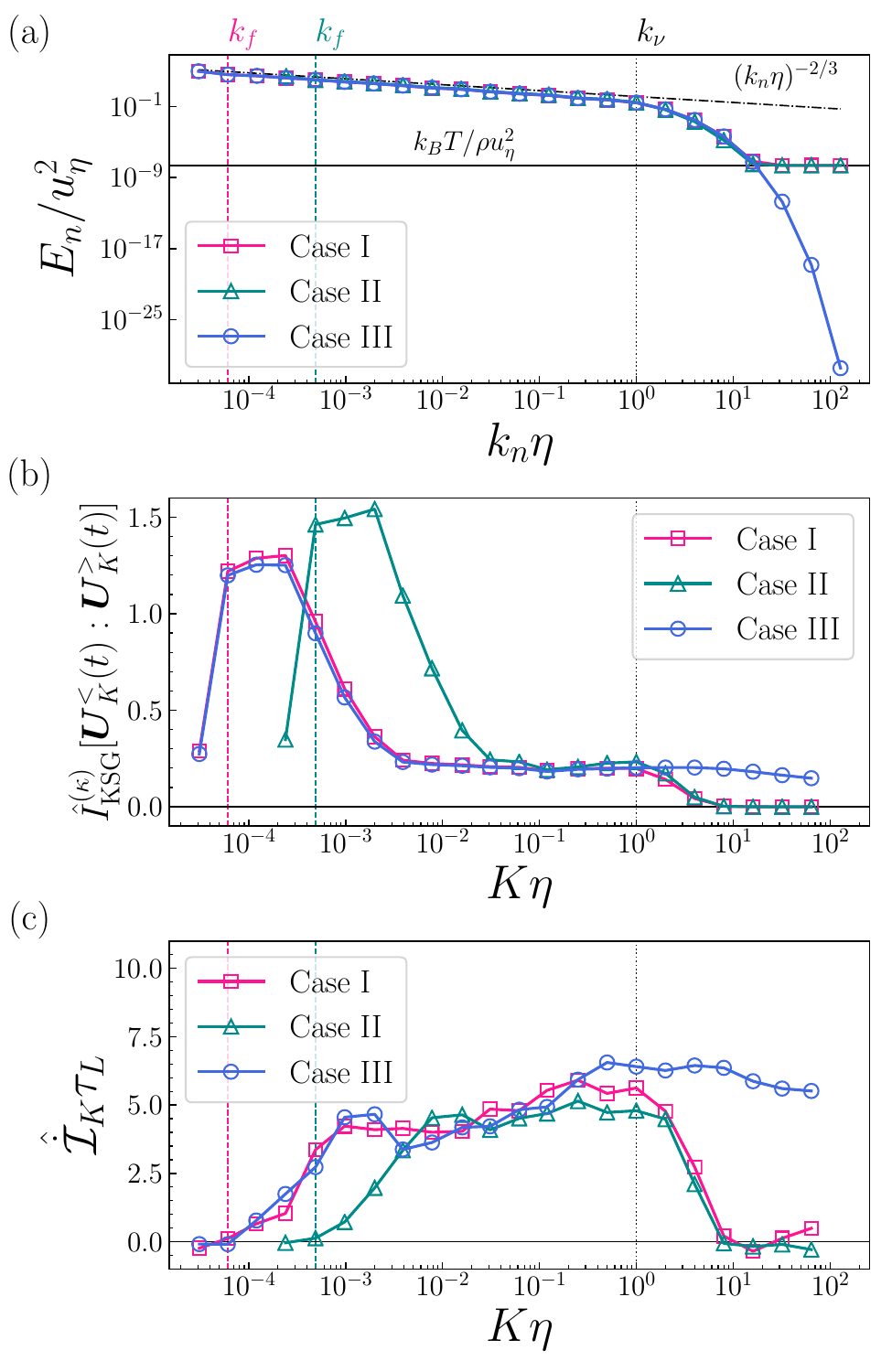}
\caption{(a) Scale dependence of the energy spectrum $E_n=\langle|u_n|^2\rangle_{\mathrm{ss}}/2$. The dash-dotted line represents $\varepsilon^{2/3}k^{-2/3}_n$. The solid line represents the thermal equipartition value $k_{\mathrm{B}}T/\rho$. (b) Scale dependence of the estimated mutual information $\hat{I}^{(\kappa)}_{\mathrm{KSG}}[{\bm U}^<_K(t):{\bm U}^>_K(t)]$ with $\kappa=4$. The error bars are within the marker size. (c) Scale dependence of the estimated information flow $\hat{\dot{\mathcal{I}}}_K$. Note that it is plotted in units of the inverse of $\tau_L$.
In all panels, the dotted and dashed lines represent the Kolmogorov dissipation scale $k_\nu=1/\eta$ and injection scale $k_f$, respectively.}
\label{fig:deterministic+noise_psd_MI_LR_n=-15_-12_07_thetaK=2.328e-08_0.000e+00_Re=1.000e+00_DeltaT=1.0e-01_Nens=3000_Nt=100_k=4}
\end{figure}

Figure~\ref{fig:deterministic+noise_psd_MI_LR_n=-15_-12_07_thetaK=2.328e-08_0.000e+00_Re=1.000e+00_DeltaT=1.0e-01_Nens=3000_Nt=100_k=4}(a) shows the energy spectrum $E_n:=\langle|u_n|^2\rangle_{\mathrm{ss}}/2$ in the steady state.
The achieved Reynolds numbers are $\mathrm{Re}\simeq9.25\times10^4$ for Case II and $1.46\times10^6$ for Case I and III.
In all three cases, we can see that the spectrum is consistent with the Kolmogorov spectrum in the inertial range, $E_n\propto k^{-2/3}_n$.
In the dissipation range, the spectrum exhibits the stretched-exponential decay in Case III, while it exhibits the equipartition of energy $E_n=k_{\mathrm{B}}T/\rho$ ($E_n/u^2_\eta=\theta_\eta$ in the dimensionless form) in Case I and II.
We also note that while the thermal fluctuation effects are negligible in the inertial range, they become relevant already at the length scale $k\eta\sim10$, as pointed out in Ref.~\cite{bandak2021thermal,bandak2022dissipation}.

\subsubsection{Mutual information}
Figure~\ref{fig:deterministic+noise_psd_MI_LR_n=-15_-12_07_thetaK=2.328e-08_0.000e+00_Re=1.000e+00_DeltaT=1.0e-01_Nens=3000_Nt=100_k=4}(b) shows the scale dependence of the estimated mutual information $\hat{I}^{(\kappa)}_{\mathrm{KSG}}[{\bm U}^<_K(t):{\bm U}^>_K(t)]$.
Its standard deviation is also estimated to be $\sim10^{-3}$ by subsampling~\cite{holmes2019estimation} (see Appendix~\ref{Estimation of the variance of the KSG estimator}), which lies within the marker size.
Notably, the mutual information is almost independent of $K$ in the inertial range.
Furthermore, Fig.~\ref{fig:deterministic+noise_psd_MI_LR_n=-15_-12_07_thetaK=2.328e-08_0.000e+00_Re=1.000e+00_DeltaT=1.0e-01_Nens=3000_Nt=100_k=4}(b) implies that the mutual information is also independent of $\mathrm{Re}$ and $T$ in the inertial range.
In other words, if we divide the total shell variables into the large-scale and small-scale modes at an arbitrary wave number $K$ within the inertial range, then the correlation between the large-scale and small-scale modes is not affected by $\mathrm{Re}$ and $T$.
In the energy injection and dissipation scale range, however, the mutual information significantly depends on $\mathrm{Re}$ and $T$.
In particular, the mutual information becomes zero in the dissipation range for Case I and II, while it remains finite for Case III.
This is because the thermal fluctuation destroys the correlation, and the large-scale and small-scale modes become statistically independent.

\subsubsection{Information flow}
In Fig.~\ref{fig:deterministic+noise_psd_MI_LR_n=-15_-12_07_thetaK=2.328e-08_0.000e+00_Re=1.000e+00_DeltaT=1.0e-01_Nens=3000_Nt=100_k=4}(c), we show the estimated information flow $\hat{\dot{\mathcal{I}}}_K$ in units of the inverse of the large-eddy turnover time $\tau_L:=1/k_0u_{\mathrm{rms}}$, where $u^2_{\mathrm{rms}}:=\sum^N_{n=0}\langle|u_n|^2\rangle_{\mathrm{ss}}$.
We find that $\tau_L\simeq181\tau_\eta$ for Case II, while $\tau_L\simeq734\tau_\eta$ for Case I and III.
From this figure, we can see that the information flow takes positive values for $K$ within the inertial range, consistent with the main result (\ref{main result 1}).
Thus, information of large-scale turbulent fluctuations is indeed transferred to small scales.

For the second inequality (\ref{main result 2}), we note that the upper bound in units of $\tau^{-1}_L$ reads $(\rho\varepsilon/k_{\mathrm{B}}T)\tau_L\simeq7.79\times10^9$ for Case II and $3.15\times10^{10}$ for Case I and III.
Therefore, the inequality (\ref{main result 2}) is also satisfied, although it is a rather loose bound.
This result can be reinterpreted using the concept of \textit{information-thermodynamic efficiency}~\cite{horowitz2014thermodynamics}, defined by
\begin{align}
\eta_{\mathrm{eff}}:=\dfrac{\dot{\mathcal{I}}_K}{\dot{S}^>_{\mathrm{env}}}=\dfrac{\dot{\mathcal{I}}_K}{\rho\varepsilon/k_{\mathrm{B}}T}.
\end{align}
From the main result (\ref{main result 1}) and the second law of information thermodynamics in the steady state (\ref{2nd law of information thermodynamics_small-scale_NESS}), it immediately follows that $0\le\eta_{\mathrm{eff}}\le1$.
This efficiency quantifies how efficiently the small-scale modes ${\bm U}^>_K$ gain information about the large-scale modes ${\bm U}^<_K$ relative to the energy dissipation or thermodynamic cost.
Then, the previous result states that $\eta_{\mathrm{eff}}\ll1$, which suggests that the small-scale eddies acquire information about the large-scale eddies at a relatively high thermodynamic cost.
This property is in contrast to other typical information processing systems such as Maxwell's demon~\cite{horowitz2014thermodynamics,hartich2016sensory,matsumoto2018role} and thus characterizes turbulence dynamics.

Furthermore, Fig.~\ref{fig:deterministic+noise_psd_MI_LR_n=-15_-12_07_thetaK=2.328e-08_0.000e+00_Re=1.000e+00_DeltaT=1.0e-01_Nens=3000_Nt=100_k=4}(c) suggests that the information flow may be scaled as
\begin{align}
\dot{\mathcal{I}}_K\sim \dfrac{C}{\tau_L}
\label{IF_scaling}
\end{align}
in the inertial range, where $C$ is a dimensionless constant that is almost independent of $\mathrm{Re}$, $K$, and $T$.
By noting that $\tau_L$ can be interpreted as the characteristic time scale for the largest eddies to be stretched into smaller eddies, this result implies that the information of large-scale eddies is transferred to small scales by the energy cascade process with nearly constant intensity.
This result also implies that, although thermal fluctuations are crucial in deriving the main results (\ref{main result 1}) and (\ref{main result 2}), the information flow itself is mainly governed by the large-scale dynamics rather than by the thermal fluctuations.

\section{Concluding remarks\label{Concluding remarks}}
In summary, we have proved that, in fully developed three-dimensional fluid turbulence, information of turbulent fluctuations is transferred from large to small scales along with the energy cascade.
Our main results (\ref{main result 1}) and (\ref{main result 2}) are a direct consequence of the second law of information thermodynamics, and thus they are exact and universal relations, independent of the details of the flow under consideration.
Furthermore, our numerical simulation using a shell model suggests that the intensity of the information flow is nearly constant in the inertial range and that the rate of information transfer is characterized by the large-eddy turnover time [Eq.~(\ref{IF_scaling})].
This observation states that the information of large-scale turbulent fluctuations is transferred to small scales by the energy cascade process with nearly constant intensity.
Thus, our results challenge the conventional intuitive picture that the universal statistical properties emerge at small scales because information about the details of the large scales is lost in the cascade process.
Moreover, we have found that the information-thermodynamic efficiency is quite low compared to other typical information processing systems such as Maxwell's demon.
This implies that transferring information from large to small scales involves enormous thermodynamic costs, indicating the poor performance of turbulence as an information processing system.

We now provide some technical remarks on the estimation of the information flow.
Although the KSG estimator used here is asymptotically unbiased for $N_{\mathrm{samp}}\rightarrow\infty$, there are a sample-size-dependent bias and a $\kappa$-dependent bias for a finite $N_{\mathrm{samp}}$ in general~\cite{holmes2019estimation}.
In our case, we have found that the magnitude of $\hat{I}^{(\kappa)}_{\mathrm{KSG}}[{\bm U}^<_K:{\bm U}^>_K]$ depends on $\kappa$.
This may be because the probability distribution $p_t({\bm U}^<_K,{\bm U}^>_K)$ is skewed and has heavy tails, thus violating the local uniformity condition~\cite{holmes2019estimation}.
Nevertheless, we have confirmed that the sign of $\hat{\dot{\mathcal{I}}}_K$ does not depend on the choice of $\kappa$.
See Appendix~\ref{Details of the numerical simulation} for more details on these subtle points.
It should also be noted that the number of samples $N_{\mathrm{samp}}$ used here is not sufficient for high accurate estimation of the information flow because the standard deviation of the estimated mutual information is comparable to its increment.
In other words, if we naively estimate the error bar of the information flow $\hat{\dot{\mathcal{I}}}_K$ by using the estimated standard deviation of the mutual information, it is of the same order as $\hat{\dot{\mathcal{I}}}_K$ itself.
It is therefore desirable to perform the numerical calculations with higher accuracy while taking the bias into account.

Our study opens several possible directions for future research.
The first direction concerns the origin of the universality of turbulent fluctuations at small scales.
As we have mentioned above, our result here is somewhat contrary to the common intuitive picture of how universality emerges at small scales.
Then, it seems natural to ask how the universality emerges at small scales under the influence of the information flow from large scales.
We conjecture that the coexistence of the universality and information flow can be explained by the stepwise ``information cascade'' process where ``irrelevant information'' is ``deamplified'' as the cascade develops.
The role of various energy cascade mechanisms~\cite{goto2017hierarchy, Carbone2020_is_vortex} in this process would be an interesting question to be investigated.
Note that this cascade picture is analogous to that proposed by Wilson in the context of the critical phenomena~\cite{wilson1975renormalization}.
The second direction concerns the intermittency.
Since there is an information flow of turbulent fluctuations from large to small scales, the small-scale intermittency must be affected by the information flow.
Indeed, the intermittency implies that the turbulent fluctuations grow in each cascade step and thus ``remember'' the large scales~\cite{Frisch,Eyink_lecture,Eyink_Sreenivasan}.
Therefore, we guess that there are universal relations between the intermittency and the information flow that restrict the possible values of the structure function exponent $\zeta_p$.
Finally, because turbulent cascade is a ubiquitous phenomenon found in quantum fluids~\cite{tanogami2021theoretical,tanogami2022reply,krstulovic2022comment,skrbek2021phenomenology}, supercritical fluids near a critical point~\cite{tanogami2021van}, elastic bodies~\cite{Nazarenko_2011,zakharov1992kolmogorov}, and even spin systems~\cite{tanogami2022xy,tsubota2013spin,rodriguez2021turbulent}, it would be interesting research direction to investigate the nature of the information flow in these various systems.
We hope that our work opens up a new research area, ``\textit{information hydrodynamics},'' which would provide a theoretical framework to elucidate and control the dynamics of complicated hydrodynamic phenomena.
\vspace{5mm}
\begin{acknowledgements}
We thank Masanobu Inubushi, Wouter J.\ T.\ Bos,  Susumu Goto, and Shin-ichi Sasa for fruitful discussions.
We also thank Dmytro Bandak and Gregory L.\ Eyink for their helpful comments on the numerical simulation.
T.T. was supported by JSPS KAKENHI Grant No. 20J20079, a Grant-in-Aid for JSPS Fellows.
R.A. was supported by the Takenaka Scholarship Foundation.
\end{acknowledgements}

\appendix
\section{Detailed calculation of the Fourier transform of the fluctuating Navier--Stokes equation\label{Detailed calculation of the Fourier transform of the fluctuating Navier--Stokes equation}}
By applying the Fourier transform to the fluctuating Navier--Stokes equation (\ref{fluctuating NS}), we obtain
\begin{align}
\partial_t\hat{\bm u}_{\bm k}&=-i{\bm k}\cdot\sum_{{\bm p}+{\bm q}={\bm k}}\hat{\bm u}_{\bm p}\hat{\bm u}_{\bm q}-i{\bm k}\hat{p}_{\bm k}-\nu k^2\hat{\bm u}_{\bm k}+\hat{\bm f}_{\bm k}+i{\bm k}\cdot\hat{\mathsf{s}}_{\bm k}.
\label{appendix: fNS_k}
\end{align}
Here, $\hat{\mathsf{s}}_{\bm k}$ denotes the zero-mean white Gaussian noise that satisfies
\begin{align}
&\quad\langle\hat{\mathsf{s}}^{ab}_{\bm k}(t)\hat{\mathsf{s}}^{cd*}_{{\bm k}'}(t')\rangle\notag\\
&=\dfrac{1}{V}\dfrac{2\nu k_{\mathrm{B}}T}{\rho}\left(\delta^{ac}\delta^{bd}+\delta^{ad}\delta^{bc}-\dfrac{2}{3}\delta^{ab}\delta^{cd}\right)\delta_{{\bm k},{\bm k}'}\delta(t-t').
\end{align}
We define $\hat{\bm \xi}_{\bm k}$ by $\sqrt{2\nu k^2k_{\mathrm{B}}T/\rho}\hat{\bm \xi}_{\bm k}=i{\bm k}\cdot\hat{\mathsf{s}}_{\bm k}$, which satisfies
\begin{align}
\langle\hat{\xi}^a_{\bm k}(t)\hat{\xi}^{b*}_{{\bm k}'}(t')\rangle=\dfrac{1}{V}\left(\delta^{ab}+\dfrac{1}{3}\dfrac{k^ak^b}{k^2}\right)\delta_{{\bm k},{\bm k}'}\delta(t-t').
\end{align}
We decompose $\hat{\bm \xi}_{\bm k}$ into components parallel and perpendicular to ${\bm k}$: $\hat{\bm \xi}_{\bm k}=\hat{\bm \xi}^{\parallel}_{\bm k}+\hat{\bm \xi}^{\perp}_{\bm k}$, where $\hat{\bm \xi}^{\perp}_{\bm k}=(\mathsf{1}-{\bm k}{\bm k}/k^2)\cdot\hat{\bm \xi}_{\bm k}$, which satisfies
\begin{align}
\langle\hat{\xi}^{\perp a}_{\bm k}(t)\hat{\xi}^{\perp b*}_{{\bm k}'}(t')\rangle=\dfrac{1}{V}\left(\delta^{ab}-\dfrac{k^ak^b}{k^2}\right)\delta_{{\bm k},{\bm k}'}\delta(t-t').
\end{align}
Then, by multiplying ${\bm k}$ to (\ref{appendix: fNS_k}) and by noting that ${\bm k}\cdot\hat{\bm u}_{\bm k}={\bm k}\cdot\hat{\bm f}_{\bm k}=0$, we obtain
\begin{align}
-i\hat{p}_{\bm k}+\sqrt{\dfrac{2\nu k_{\mathrm{B}}T}{\rho}}\hat{\xi}^{\parallel}_{\bm k}=i\dfrac{{\bm k}{\bm k}}{k^2}:\sum_{{\bm p}+{\bm q}={\bm k}}\hat{\bm u}_{\bm p}\hat{\bm u}_{\bm q},
\end{align}
where $\hat{\xi}^{\parallel}_{\bm k}=({\bm k}/k)\cdot\hat{\bm \xi}^{\parallel}_{\bm k}$.
By substituting this relation into (\ref{appendix: fNS_k}), we arrive at (\ref{fNS_k}).
Note that, in (\ref{fNS_k}), $\hat{\bm \xi}^{\perp}_{\bm k}$ is rewritten as $\hat{\bm \xi}_{\bm k}$ for notational simplicity.
\vspace{7mm}

\section{Derivation of Eqs.~(\ref{dtI}) and (\ref{IF_rewrite})\label{Derivation of Eqs}}
We first derive the expression (\ref{IF_rewrite}).
Let $p(\tilde{\bm U}^<_K,t+h;{\bm U}^>_K,t)$ be the two-point probability density with $\tilde{\bm U}^<_K={\bm U}^<_K(t+h)$ and ${\bm U}^>_K={\bm U}^>_K(t)$.
When $h=0$, it corresponds to the joint probability density at the same time $t$: $p({\bm U}^<_K,t;{\bm U}^>_K,t)=p_t({\bm U}^<_K,{\bm U}^>_K)$.
Then, from the definition of the information flow (\ref{IF_def-1}), we find that
\begin{widetext}
\begin{align}
\dot{I}^<_K&:=\lim_{dt\rightarrow0^+}\dfrac{I[{\bm U}^<_K(t+dt):{\bm U}^>_K(t)]-I[{\bm U}^<_K(t):{\bm U}^>_K(t)]}{dt}\notag\\
&=\lim_{h\rightarrow0^+}\dfrac{1}{h}\left(\int d\tilde{\bm U}^<_Kd{\bm U}^>_Kp(\tilde{\bm U}^<_K,t+h;{\bm U}^>_K,t)\ln\dfrac{p(\tilde{\bm U}^<_K,t+h;{\bm U}^>_K,t)}{p^<_{t+h}(\tilde{\bm U}^<_K)p^>_t({\bm U}^>_K)}-\int d{\bm U}^<_Kd{\bm U}^>_Kp_t({\bm U}^<_K,{\bm U}^>_K)\ln\dfrac{p_t({\bm U}^<_K,{\bm U}^>_K)}{p^<_t({\bm U}^<_K)p^>_t({\bm U}^>_K)}\right)\notag\\
&=\int d\tilde{\bm U}^<_Kd\tilde{\bm U}^>_Kd{\bm U}^<_Kd{\bm U}^>_K\left.\dfrac{\partial}{\partial h}p(\tilde{\bm U}^<_K,\tilde{\bm U}^>_K,t+h|{\bm U}^<_K,{\bm U}^>_K,t)\right|_{h=0}p_t({\bm U}^<_K,{\bm U}^>_K)\ln\dfrac{p_t(\tilde{\bm U}^<_K,{\bm U}^>_K)}{p^<_t(\tilde{\bm U}^<_K)p^>_t({\bm U}^>_K)},
\end{align}
where we have used $\int d\tilde{\bm U}^<_Kd{\bm U}^>_K\frac{\partial}{\partial h}p(\tilde{\bm U}^<_K,t+h;{\bm U}^>_K,t)=\frac{\partial}{\partial h}\int d\tilde{\bm U}^<_Kd{\bm U}^>_Kp(\tilde{\bm U}^<_K,t+h;{\bm U}^>_K,t)=0$ and
\begin{align}
p(\tilde{\bm U}^<_K,t+h;{\bm U}^>_K,t)&=\int d\tilde{\bm U}^>_Kd{\bm U}^<_Kp(\tilde{\bm U}^<_K,\tilde{\bm U}^>_K,t+h;{\bm U}^<_K,{\bm U}^>_K,t)\notag\\
&=\int d\tilde{\bm U}^>_Kd{\bm U}^<_Kp(\tilde{\bm U}^<_K,\tilde{\bm U}^>_K,t+h|{\bm U}^<_K,{\bm U}^>_K,t)p_t({\bm U}^<_K,{\bm U}^>_K),
\end{align}
where $p(\tilde{\bm U}^<_K,\tilde{\bm U}^>_K,t+h|{\bm U}^<_K,{\bm U}^>_K,t)=p(\tilde{\bm U}^<_K,\tilde{\bm U}^>_K,t+h;{\bm U}^<_K,{\bm U}^>_K,t)/p_t({\bm U}^<_K,{\bm U}^>_K)$ denotes the conditional probability density.
By noting that $p(\tilde{\bm U}^<_K,\tilde{\bm U}^>_K,t+h|{\bm U}^<_K,{\bm U}^>_K,t)$ obeys the Fokker--Planck equation (\ref{FP-fNS}) and that $p(\tilde{\bm U}^<_K,\tilde{\bm U}^>_K,t|{\bm U}^<_K,{\bm U}^>_K,t)=\delta(\tilde{\bm U}^<_K-{\bm U}^<_K)\delta(\tilde{\bm U}^>_K-{\bm U}^>_K)$, we arrive at (\ref{IF_rewrite}):
\begin{align}
\dot{I}^<_K&=\int d{\bm U}^<_Kd{\bm U}^>_K\sum_{{\bm k}\in\mathcal{K}^+,k\le K}\left[-\dfrac{\partial}{\partial \hat{\bm u}_{\bm k}}\cdot {\bm J}_{\bm k}(\hat{\bm u},\hat{\bm u}^*)-\dfrac{\partial}{\partial \hat{\bm u}^*_{\bm k}}\cdot {\bm J}^*_{\bm k}(\hat{\bm u},\hat{\bm u}^*)\right]\ln\dfrac{p_t({\bm U}^<_K,{\bm U}^>_K)}{p^<_t({\bm U}^<_K)p^>_t({\bm U}^>_K)}\notag\\
&=\sum_{{\bm k}\in\mathcal{K}^+,k\le K}\int d\hat{\bm u}d\hat{\bm u}^*\left[{\bm J}_{\bm k}(\hat{\bm u},\hat{\bm u}^*)\cdot\dfrac{\partial}{\partial \hat{\bm u}_{\bm k}}\ln\dfrac{p_t({\bm U}^<_K,{\bm U}^>_K)}{p^<_t({\bm U}^<_K)p^>_t({\bm U}^>_K)}+{\bm J}^*_{\bm k}(\hat{\bm u},\hat{\bm u}^*)\cdot\dfrac{\partial}{\partial \hat{\bm u}^*_{\bm k}}\ln\dfrac{p_t({\bm U}^<_K,{\bm U}^>_K)}{p^<_t({\bm U}^<_K)p^>_t({\bm U}^>_K)}\right].
\label{appendix: IF_rewrite_<}
\end{align}
Similarly, for the information flow associated with the small-scale modes $\dot{I}^>_K$, we can obtain
\begin{align}
\dot{I}^>_K=\sum_{{\bm k}\in\mathcal{K}^+,k>K}\int d\hat{\bm u}d\hat{\bm u}^*\left[{\bm J}_{\bm k}(\hat{\bm u},\hat{\bm u}^*)\cdot\dfrac{\partial}{\partial \hat{\bm u}_{\bm k}}\ln\dfrac{p_t({\bm U}^<_K,{\bm U}^>_K)}{p^<_t({\bm U}^<_K)p^>_t({\bm U}^>_K)}+{\bm J}^*_{\bm k}(\hat{\bm u},\hat{\bm u}^*)\cdot\dfrac{\partial}{\partial \hat{\bm u}^*_{\bm k}}\ln\dfrac{p_t({\bm U}^<_K,{\bm U}^>_K)}{p^<_t({\bm U}^<_K)p^>_t({\bm U}^>_K)}\right].
\label{appendix: IF_rewrite_>}
\end{align}
By combining (\ref{appendix: IF_rewrite_<}) and (\ref{appendix: IF_rewrite_>}), we can easily confirm that (\ref{dtI}) holds: $d_tI[{\bm U}^<_K:{\bm U}^>_K]=\dot{I}^<_K+\dot{I}^>_K$.

\section{Local detailed balance\label{Local detailed balance}}
Here, we show that the identification of the entropy change in the environment (\ref{medium entropy change}) is consistent with the local detailed balance (LDB).
Note that the LDB is essentially equivalent to the fluctuation-dissipation relation of the second kind~\cite{harada2006energy,maes2021local}.
Therefore, if the expression (\ref{medium entropy change}) is thermodynamically consistent, then it should also be consistent with the LDB.

Let $p(\hat{\bm u}',\hat{\bm u}'^*,t+dt|\hat{\bm u},\hat{\bm u}^*,t)$ be the transition probability density with $\hat{\bm u}'=\hat{\bm u}(t+dt)$ and $\hat{\bm u}=\hat{\bm u}(t)$.
Then, in the Ito scheme, it can be expressed as~\cite{risken1996fokker}
\begin{align}
p(\hat{\bm u}',\hat{\bm u}'^*,t+dt|\hat{\bm u},\hat{\bm u}^*,t)&=\prod_{{\bm k}\in\mathcal{K}^+}\left(\dfrac{\rho V}{2\pi\nu k^2k_{\mathrm{B}}Tdt}\right)^2\exp\left(-\dfrac{\rho V}{2\nu k^2k_{\mathrm{B}}Tdt}\left|d\hat{\bm u}_{\bm k}-{\bm A}_{\bm k}(\hat{\bm u},\hat{\bm u}^*)dt\right|^2\right),
\label{transition probability density}
\end{align}
where $d\hat{\bm u}_{\bm k}=\hat{\bm u}_{\bm k}'-\hat{\bm u}_{\bm k}$ and ${\bm A}_{\bm k}(\hat{\bm u},\hat{\bm u}^*)$ is given by (\ref{A_k}).
Similarly, the transition probability density for the backward transition $p(-\hat{\bm u},-\hat{\bm u}^*,t+dt|-\hat{\bm u}',-\hat{\bm u}'^*,t)$ is given by
\begin{align}
&p(-\hat{\bm u},-\hat{\bm u}^*,t+dt|-\hat{\bm u}',-\hat{\bm u}'^*,t)\notag\\
&=\prod_{{\bm k}\in\mathcal{K}^+}\left(\dfrac{\rho V}{2\pi\nu k^2k_{\mathrm{B}}Tdt}\right)^2\exp\left(-\dfrac{\rho V}{2\nu k^2k_{\mathrm{B}}Tdt}\left|d\hat{\bm u}_{\bm k}-\left[-{\bm A}^{\mathrm{ir}}_{\bm k}(\hat{\bm u},\hat{\bm u}^*)+{\bm A}^{\mathrm{rev}}_{\bm k}(\hat{\bm u},\hat{\bm u}^*)\right]dt\right|^2\right.\notag\\
&\quad\left.+\left(\mathsf{I}-\dfrac{{\bm k}{\bm k}}{k^2}\right):\left[-\dfrac{\partial}{\partial \hat{\bm u}_{\bm k}}{\bm A}^{\mathrm{ir}}_{\bm k}(\hat{\bm u},\hat{\bm u}^*)-\dfrac{\partial}{\partial \hat{\bm u}^*_{\bm k}}{\bm A}^{\mathrm{ir}*}_{\bm k}(\hat{\bm u},\hat{\bm u}^*)+\dfrac{\partial}{\partial \hat{\bm u}_{\bm k}}{\bm A}^{\mathrm{rev}}_{\bm k}(\hat{\bm u},\hat{\bm u}^*)+\dfrac{\partial}{\partial \hat{\bm u}^*_{\bm k}}{\bm A}^{\mathrm{rev}*}_{\bm k}(\hat{\bm u},\hat{\bm u}^*)\right]dt\right),
\label{backward transition probability density}
\end{align}
where ${\bm A}^{\mathrm{ir}}_{\bm k}(\hat{\bm u},\hat{\bm u}^*)$ and ${\bm A}^{\mathrm{rev}}_{\bm k}(\hat{\bm u},\hat{\bm u}^*)$ denote the irreversible and reversible parts of ${\bm A}_{\bm k}(\hat{\bm u},\hat{\bm u}^*)$, respectively:
\begin{align}
{\bm A}^{\mathrm{ir}}_{\bm k}(\hat{\bm u},\hat{\bm u}^*)&:=\dfrac{1}{2}[{\bm A}_{\bm k}(\hat{\bm u},\hat{\bm u}^*)-{\bm A}_{\bm k}(-\hat{\bm u},-\hat{\bm u}^*)]\notag\\
&=-\nu k^2\hat{\bm u}_{\bm k},\\
{\bm A}^{\mathrm{rev}}_{\bm k}(\hat{\bm u},\hat{\bm u}^*)&:=\dfrac{1}{2}[{\bm A}_{\bm k}(\hat{\bm u},\hat{\bm u}^*)+{\bm A}_{\bm k}(-\hat{\bm u},-\hat{\bm u}^*)]\notag\\
&={\bm B}_{\bm k}(\hat{\bm u},\hat{\bm u}^*)+\hat{\bm f}_{\bm k}.
\end{align}

Then, by imposing the local detailed balance, the entropy change in the environment during the time interval $dt$ can be identified as
\begin{align}
\dot{S}^{\mathrm{env}}dt=\left\langle\ln\dfrac{p(\hat{\bm u}',\hat{\bm u}'^*,t+dt|\hat{\bm u},\hat{\bm u}^*,t)}{p(-\hat{\bm u},-\hat{\bm u}^*,t+dt|-\hat{\bm u}',-\hat{\bm u}'^*,t)}\right\rangle,
\end{align}
where $\langle\cdot\rangle$ denotes the average with respect to the two-point probability density $p(\hat{\bm u}',\hat{\bm u}'^*,t+dt;\hat{\bm u},\hat{\bm u}^*,t)$.
By substituting (\ref{transition probability density}) and (\ref{backward transition probability density}), we obtain
\begin{align}
\dot{S}^{\mathrm{env}}dt&=\sum_{{\bm k}\in\mathcal{K}^+}\left\langle\dfrac{\rho V}{\nu k^2k_{\mathrm{B}}T}\left[{\bm A}^{\mathrm{ir}}_{\bm k}(\hat{\bm u},\hat{\bm u}^*)\cdot\left(d\hat{\bm u}^*_{\bm k}-{\bm A}^{\mathrm{rev}*}_{\bm k}(\hat{\bm u},\hat{\bm u}^*)dt\right)+{\bm A}^{\mathrm{ir}*}_{\bm k}(\hat{\bm u},\hat{\bm u}^*)\cdot\left(d\hat{\bm u}_{\bm k}-{\bm A}^{\mathrm{rev}}_{\bm k}(\hat{\bm u},\hat{\bm u}^*)dt\right)\right]\right.\notag\\
&\quad\left.-\left(\mathsf{I}-\dfrac{{\bm k}{\bm k}}{k^2}\right):\left[-\dfrac{\partial}{\partial \hat{\bm u}_{\bm k}}{\bm A}^{\mathrm{ir}}_{\bm k}(\hat{\bm u},\hat{\bm u}^*)-\dfrac{\partial}{\partial \hat{\bm u}^*_{\bm k}}{\bm A}^{\mathrm{ir}*}_{\bm k}(\hat{\bm u},\hat{\bm u}^*)+\dfrac{\partial}{\partial \hat{\bm u}_{\bm k}}{\bm A}^{\mathrm{rev}}_{\bm k}(\hat{\bm u},\hat{\bm u}^*)+\dfrac{\partial}{\partial \hat{\bm u}^*_{\bm k}}{\bm A}^{\mathrm{rev}*}_{\bm k}(\hat{\bm u},\hat{\bm u}^*)\right]dt\right\rangle,
\end{align}
where the inner product ``$\cdot$'' between ${\bm A}^{\mathrm{ir}}_{\bm k}(\hat{\bm u},\hat{\bm u}^*)$ and $d\hat{\bm u}^*_{\bm k}$ (similarly, ${\bm A}^{\mathrm{ir}*}_{\bm k}(\hat{\bm u},\hat{\bm u}^*)$ and $d\hat{\bm u}_{\bm k}$) should be interpreted as the multiplication in the sense of Ito.
Note that
\begin{align}
\dfrac{\partial}{\partial \hat{\bm u}_{\bm k}}{\bm A}^{\mathrm{rev}}_{\bm k}(\hat{\bm u},\hat{\bm u}^*)+\dfrac{\partial}{\partial \hat{\bm u}^*_{\bm k}}{\bm A}^{\mathrm{rev}*}_{\bm k}(\hat{\bm u},\hat{\bm u}^*)&=\dfrac{\partial}{\partial \hat{\bm u}_{\bm k}}{\bm B}_{\bm k}(\hat{\bm u},\hat{\bm u}^*)+\dfrac{\partial}{\partial \hat{\bm u}^*_{\bm k}}{\bm B}^*_{\bm k}(\hat{\bm u},\hat{\bm u}^*)\notag\\
&=-i{\bm k}\cdot\hat{\bm u}_{\bm 0}\left(\mathsf{I}-\dfrac{{\bm k}{\bm k}}{k^2}\right)-i{\bm k}\left(\mathsf{I}-\dfrac{{\bm k}{\bm k}}{k^2}\right)\cdot\hat{\bm u}_{\bm 0}+\mathrm{c.c.}\notag\\
&=0,
\end{align}
and that the Ito integral is related to the Stratonovich integral as
\begin{align}
\dfrac{\rho V}{\nu k^2k_{\mathrm{B}}T}{\bm A}^{\mathrm{ir}}_{\bm k}(\hat{\bm u},\hat{\bm u}^*)\cdot d\hat{\bm u}^*_{\bm k}+\left(\mathsf{I}-\dfrac{{\bm k}{\bm k}}{k^2}\right):\dfrac{\partial}{\partial \hat{\bm u}_{\bm k}}{\bm A}^{\mathrm{ir}}_{\bm k}(\hat{\bm u},\hat{\bm u}^*)=\dfrac{\rho V}{\nu k^2k_{\mathrm{B}}T}{\bm A}^{\mathrm{ir}}_{\bm k}(\hat{\bm u},\hat{\bm u}^*)\circ d\hat{\bm u}^*_{\bm k}.
\end{align}
Therefore, we obtain
\begin{align}
\dot{S}^{\mathrm{env}}dt&=\sum_{{\bm k}\in\mathcal{K}^+}\dfrac{\rho V}{\nu k^2k_{\mathrm{B}}T}\left\langle{\bm A}^{\mathrm{ir}}_{\bm k}(\hat{\bm u},\hat{\bm u}^*)\circ\left(d\hat{\bm u}^*_{\bm k}-{\bm A}^{\mathrm{rev}*}_{\bm k}(\hat{\bm u},\hat{\bm u}^*)dt\right)+{\bm A}^{\mathrm{ir}*}_{\bm k}(\hat{\bm u},\hat{\bm u}^*)\circ\left(d\hat{\bm u}_{\bm k}-{\bm A}^{\mathrm{rev}}_{\bm k}(\hat{\bm u},\hat{\bm u}^*)dt\right)\right\rangle\notag\\
&=\sum_{{\bm k}\in\mathcal{K}^+}\dfrac{\rho V}{k_{\mathrm{B}}T}\left\langle\hat{\bm u}^*_{\bm k}\circ\left[{\bm B}_{\bm k}(\hat{\bm u},\hat{\bm u}^*)+\hat{\bm f}_{\bm k}-\partial_t\hat{\bm u}_{\bm k}\right]+\hat{\bm u}_{\bm k}\circ\left[{\bm B}^*_{\bm k}(\hat{\bm u},\hat{\bm u}^*)+\hat{\bm f}^*_{\bm k}-\partial_t\hat{\bm u}^*_{\bm k}\right]\right\rangle dt,
\end{align}
which corresponds to the expression (\ref{medium entropy change}).
\end{widetext}

\section{Details of the numerical simulation\label{Details of the numerical simulation}}
In this section, we explain the details of the numerical simulation.
After describing the setup, the details of the KSG estimator are explained.
In particular, we provide a detailed explanation of the method used to estimate the variance and bias of the KSG estimator.

\subsection{Setup\label{appendix: Setup}}
To evaluate the inertial range straightforwardly, we first nondimensionalize the equation (\ref{sabra shell model}) with the Kolmogorov dissipation scale $\eta$ and the velocity scale $u_\eta:=(\epsilon\nu)^{1/4}$ by setting
\begin{align}
\hat{u}_n:=u_n/u_\eta,\quad \hat{k}_n:=\eta k_n,\quad \hat{t}:=t/\tau_\eta,\notag\\
\hat{\xi}_n:=(u_\eta/\eta)^{-1/2}\xi_n,\quad \hat{f}_n:=f_n/F,
\end{align}
where $\tau_\eta:=\eta/u_\eta$ denotes the typical time scale at the Kolmogorov dissipation scale, and $F$ denotes the typical magnitude of the force per mass.
The nondimensionalized equation (\ref{sabra shell model}) reads
\begin{align}
\partial_{\hat{t}}{\hat{u}}_n&=\hat{B}_n(\hat{u},\hat{u}^*)-\hat{k}^2_n\hat{u}_n+(2\theta_\eta)^{1/2}\hat{k}_n\hat{\xi}_n+\mathcal{F}_\eta \hat{f}_n,
\label{sabra shell model_dimensionless}
\end{align}
where
\begin{align}
\hat{B}_n(\hat{u},\hat{u}^*)&:=i\biggl(\hat{k}_{n+1}\hat{u}_{n+2}\hat{u}^*_{n+1}-\dfrac{1}{2}\hat{k}_n\hat{u}_{n+1}\hat{u}^*_{n-1}\notag\\
&\qquad+\dfrac{1}{2}\hat{k}_{n-1}\hat{u}_{n-1}\hat{u}_{n-2}\biggr).
\end{align}
Here, $\theta_\eta:=k_{\mathrm{B}}T/\rho u^2_\eta$ denotes the ratio of the thermal energy to the kinetic energy at the Kolmogorov dissipation scale, and $\mathcal{F}_\eta:=F\eta/u^2_\eta$ denotes the nondimensionalized magnitude of the force.
Correspondingly, we set the shell index to be $n=M,\cdots, R$ with $M=-[(3/4)\log_2(\mathrm{Re})]$ and $R=N-M$, so that $k_0=1$ corresponds to the Kolmogorov dissipation scale.

We use a slaved $3/2$-strong-order Ito-Taylor scheme~\cite{lord2004numerical} with the time-step $\delta \hat{t}:=10^{-5}$, which is smaller than the viscous time scale at the highest wave number $\hat{\tau}_{\mathrm{vis}}:=1/\hat{k}^2_R\sim10^{-4}$.
We consider three different cases as listed in Table \ref{table: three cases} in Sec.~\ref{Numerical simulation}.
In Case I, the parameter values are set to the same values used in Ref.~\cite{bandak2022dissipation,bandak2021thermal}, which are consistent with the typical values in the atmospheric boundary layer.
Specifically, the range of shell numbers is chosen as $n=-15,\cdots,7$ so that the achieved Reynolds number is comparable to the typical value in the atmospheric boundary layer of $\mathrm{Re}\sim10^6$.
Similarly, the dimensionless temperature is chosen as $\theta_\eta=2.328\times10^{-8}$.
For the external force, we set $n_f=-14$ to ensure that the external force acts only on the 0th and 1st shells of the total 23 shells.
The values of the external forces are adjusted such that $\hat{u}_{\mathrm{rms}}:=\sqrt{\sum^R_{n=M}\langle|\hat{u}_n|^2\rangle}\sim10^2$ and $\hat{\epsilon}:=\sum^R_{n=M}\hat{k}^2_n\langle|\hat{u}_n|^2\rangle\simeq1$~\cite{bandak2022dissipation,bandak2021thermal}:
\begin{align}
\mathcal{F}_\eta \hat{f}_{-15}&=-0.008900918232183095\notag\\
&\qquad- 0.0305497603210104i,\\
\mathcal{F}_\eta \hat{f}_{-14}&=0.005116337459331228\notag\\
&\qquad- 0.018175040700335127i.
\end{align}
In Case II, we set $n=-12,\cdots,7$ so that the achieved Reynolds number is lowered to $\mathrm{Re}\sim10^5$.
The dimensionless temperature is the same as in Case I.
For the external force, we set $n_f=-11$ and
\begin{align}
\mathcal{F}_\eta \hat{f}_{-12}&=-0.017415685046854878\notag\\
&\qquad- 0.05977417049893835i,\\
\mathcal{F}_\eta \hat{f}_{-11}&=0.010010711194151034\notag\\
&\qquad- 0.03556158772544649i.
\end{align}
In Case III, we set $\theta_\eta=0$ while the other parameter values are the same as in Case I.

In the averaging of the energy spectrum and the estimation of the mutual information, we use $N_{\mathrm{samp}}=3\times10^5$ samples.
For Case I and III, these samples are obtained by sampling $100$ snapshots at time $\hat{t}=1000i$ ($i=1,2,\cdots,100$) for each of the $3000$ noise realizations.
That is, for each of the $3000$ independent runs, we sample $100$ snapshots.
Here, the time interval of the sampling, $1000$, is chosen to be larger than one large-eddy turnover time $\tau_L/\tau_\eta\simeq734<1000$.
Similarly, for Case II, $N_{\mathrm{samp}}=3\times10^5$ samples are obtained by sampling $100$ snapshots at time $\hat{t}=500i$ ($i=1,2,\cdots,100$) for each of the $3000$ noise realizations, where the time interval, $500$, is chosen to be larger than one large-eddy turnover time $\tau_L/\tau_\eta\simeq181<500$.

\subsection{The KSG estimator}
The KSG estimator for the mutual information $I[X:Y]$ (either or both of the random variables $X$ and $Y$ can be multidimensional) is defined as follows~\cite{kraskov2004estimating}:
\begin{align}
\hat{I}^{(\kappa)}_{\mathrm{KSG}}[X:Y]&:=\psi(\kappa)-1/\kappa+\psi(N_{\mathrm{samp}})\notag\\
&\quad-\dfrac{1}{N_{\mathrm{samp}}}\sum^{N_{\mathrm{samp}}}_{i=1}\left[\psi(n_x(i))+\psi(n_y(i))\right],
\label{KSG estimator}
\end{align}
where $\kappa\in\mathbb{N}$ denotes the parameter of the KSG estimator, $\psi$ is the digamma function, $N_{\mathrm{samp}}$ denotes the total number of samples, and $n^{(\kappa)}_\alpha(i)$ ($\alpha=x,y$) is the number of samples such that $\|\alpha_j-\alpha_i\|\le\epsilon^{(\kappa)}_\alpha(i)/2$.
Here, $\epsilon^{(\kappa)}_\alpha(i)$ denotes the $\alpha$ extent of the smallest hyper-rectangle in the $(x,y)$ space centered at the $i$-th sample $(x_i,y_i)$ that contains $\kappa$ of its neighboring samples.
While any norms can be used for $\|\alpha_j-\alpha_i\|$, we use the standard Euclidean norm here.

Note that $\kappa$ is the only free parameter of the KSG estimator.
By varying $\kappa$, we can detect the structure of the underlying probability distribution in different spatial resolutions.
To choose the optimal $\kappa$ (if it exists), we must estimate both the standard deviation and the bias of the KSG estimator~\cite{holmes2019estimation}.

\subsection{Estimation of the variance of the KSG estimator\label{Estimation of the variance of the KSG estimator}}
In this section, we explain the method used to estimate the variance of the KSG estimator based on the subsampling approach proposed by Holmes and Nemenman~\cite{holmes2019estimation}.
This method is based on the fact that the variance of any function that is an average of $N$ i.i.d. random variables scales as $1/N$. 
Therefore, we write the variance of the KSG estimator as
\begin{align}
\mathrm{Var}_{N_{\mathrm{samp}}}[\hat{I}^{(\kappa)}_{\mathrm{KSG}}]=\dfrac{B^{(\kappa)}}{N_{\mathrm{samp}}}.
\label{variance of the KSG estimator}
\end{align}
We estimate $B^{(\kappa)}$ via a subsampling approach.
Specifically, we first partition the $N_\mathrm{samp}=N$ samples into $n$ nonoverlapping subsets of equal size as much as possible.
Let $\hat{I}^{(\kappa)}_{\mathrm{KSG},i}(N/n)$ be the $i$-th realization of $\hat{I}^{(\kappa)}_{\mathrm{KSG}}[X:Y]$ with $N_{\mathrm{samp}}=N/n$ ($i=1,2,\cdots,n$).
Then, we calculate the unbiased sample variance of these $n$ values of $\hat{I}^{(\kappa)}_{\mathrm{KSG},i}(N/n)$:
\begin{align}
\sigma^2_{\kappa,N/n}:=\dfrac{1}{n-1}\sum^{n}_{i=1}\left(\hat{I}^{(\kappa)}_{\mathrm{KSG},i}(N/n)-\dfrac{1}{n}\sum^n_{i=1}\hat{I}^{(\kappa)}_{\mathrm{KSG},i}(N/n)\right)^2.
\label{unbiased sample variance}
\end{align}
This is our estimate of $\mathrm{Var}_{N/n}[\hat{I}^{(\kappa)}_{\mathrm{KSG}}]=nB^{(\kappa)}/N$.
Finally, we estimate $B^{(\kappa)}$ by using maximum likelihood estimation.
In doing so, we first calculate $\sigma^2_{\kappa,N/n_\ell}$ for various $n_\ell$ ($\ell=1,2,\cdots,L$).
Then, from Cochran's theorem, $(n_\ell-1)\sigma^2_{\kappa,N/n_\ell}/\mathrm{Var}_{N/n_\ell}[\hat{I}^{(\kappa)}_{\mathrm{KSG}}]$ follows the $\chi^2$-distribution with $n_\ell-1$ degrees of freedom:
\begin{align}
P^{(\chi^2)}_{n_\ell-1}(x):=\dfrac{1}{2^{(n_\ell-1)/2}\Gamma(\frac{n_\ell-1}{2})}x^{\frac{n_\ell-1}{2}-1}e^{-x/2}.
\end{align}
By assuming independence of $\{\sigma^2_{\kappa,N/n_\ell}\}^L_{\ell=1}$, a likelihood function for $B^{(\kappa)}$ is
\begin{align}
\prod^L_{\ell=1}P^{(\chi^2)}_{n_\ell-1}\left(\dfrac{N(n_\ell-1)\sigma^2_{\kappa,N/n_\ell}}{B^{(\kappa)}n_\ell}\right).
\end{align}
We then obtain the maximum likelihood estimator:
\begin{align}
\hat{B}^{(\kappa)}&:=\argmax_{B^{(\kappa)}}\prod^L_{\ell=1}P^{(\chi^2)}_{n_\ell-1}\left(\dfrac{N(n_\ell-1)\sigma^2_{\kappa,N/n_\ell}}{B^{(\kappa)}n_\ell}\right)\notag\\
&=\dfrac{\sum^L_{\ell=1}\frac{n_\ell-1}{n_\ell}N\sigma^2_{\kappa,N/n_\ell}}{\sum^L_{\ell=1}(n_\ell-3)}.
\label{maximum liklihood estimator}
\end{align}
By combining (\ref{variance of the KSG estimator}) and (\ref{maximum liklihood estimator}), we can estimate the variance of the KSG estimator to be $\hat{B}^{(\kappa)}/N_{\mathrm{samp}}$.

\newpage
\subsection{Estimation bias of the KSG estimator}
Although the KSG estimator is asymptotically unbiased for sufficiently regular probability distributions as $N_{\mathrm{samp}}\rightarrow\infty$, both sample-size-dependent bias and $\kappa$-dependent bias generally exist for a finite $N_{\mathrm{samp}}$~\cite{holmes2019estimation}.
The sample-size-dependent (resp.~$\kappa$-dependent) bias can be detected by comparing the sample-size-dependence (resp.~$\kappa$-dependence) of the estimated mutual information with its standard deviation.
If the sample-size-dependence (resp.~$\kappa$-dependence) of the estimated mutual information is much larger than its standard deviation, then a sample-size-dependent (resp.~$\kappa$-dependent) bias may be present.

Note that $\kappa$ is related to the spatial resolution in detecting the structure of the underlying probability distribution.
For large $\kappa$, because the fine structure of the probability distribution cannot be detected, we would expect the mutual information to be underestimated.
At the same time, because $n_x(i)$ and $n_y(i)$ both increase with increasing $\kappa$, the standard deviation of the estimated mutual information will be smaller for large $\kappa$.
If there is no $\kappa$-dependent bias, we can choose the optimal $\kappa$ such that there is no sample-size-dependence compared to the standard deviation and the standard deviation is the smallest.

\begin{figure}[H]
\includegraphics[width=8.6cm]{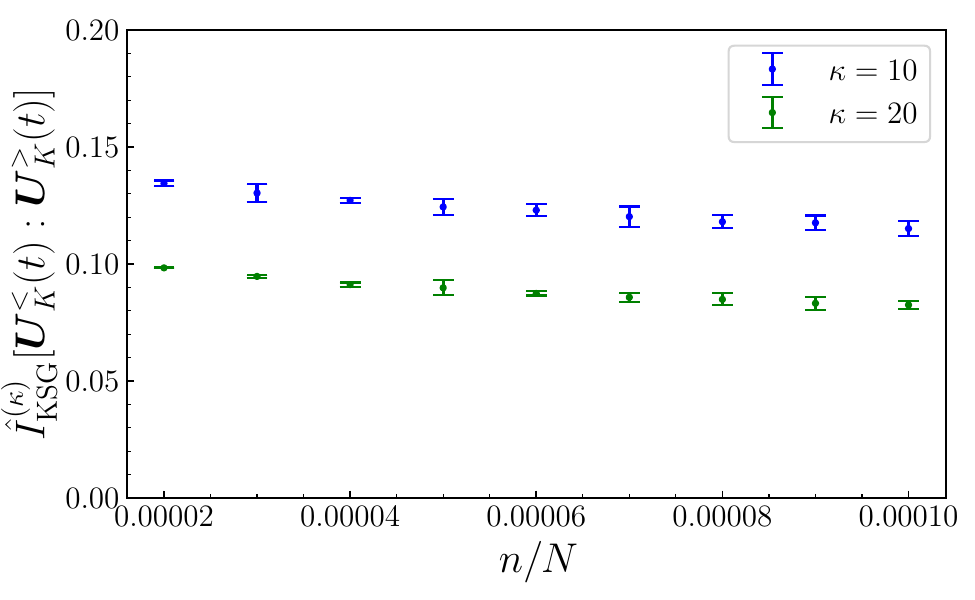}
\includegraphics[width=8.6cm]{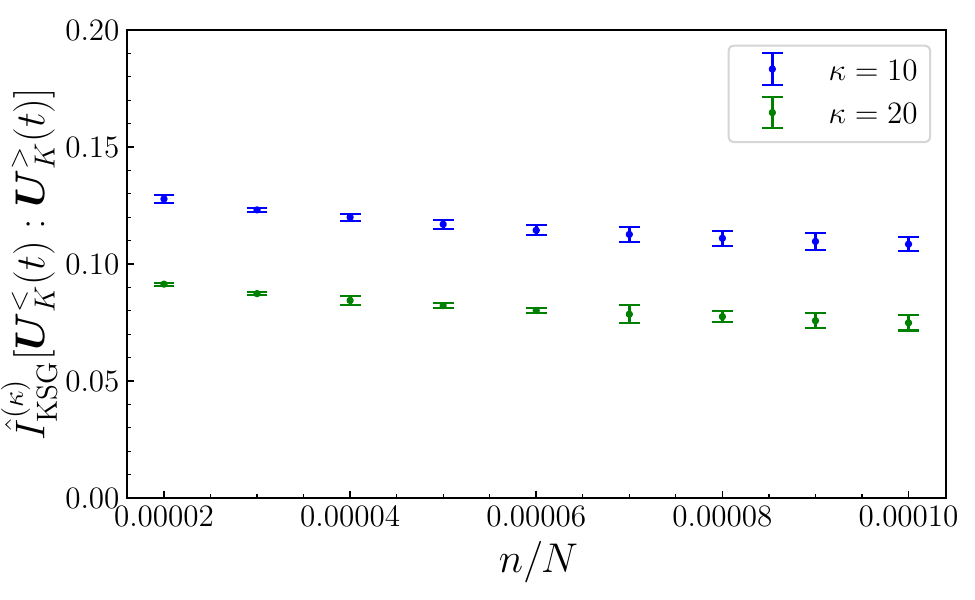}
\caption{Bias of the KSG estimator $\hat{I}^{(\kappa)}_{\mathrm{KSG}}[{\bm U}^<_K(t):{\bm U}^>_K(t)]$ as a function of $1/N_{\mathrm{samp}}=n/N$ with $n=2,3,\cdots,10$ and $N=10^5$. $K=k_{10}$ (top) and $K=k_{15}$ (bottom).}
\label{fig:KSGbias_n=-15_07_thetaK=2.328e-08_Re=1.000e+00_Nens=1000_Nt=100}
\includegraphics[width=8.6cm]{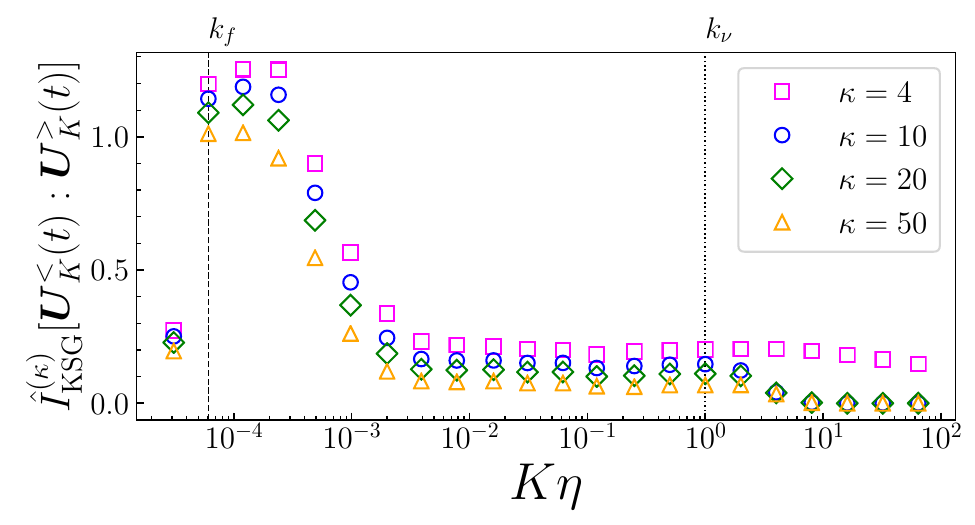}
\includegraphics[width=8.6cm]{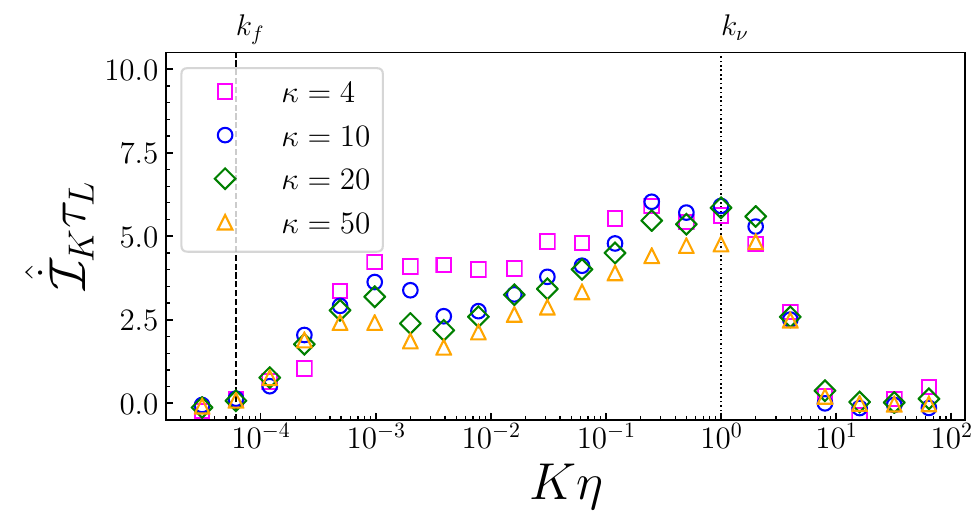}
\caption{Scale dependences of the estimated mutual information $\hat{I}^{(\kappa)}_{\mathrm{KSG}}[{\bm U}^<_K(t):{\bm U}^>_K(t)]$ (top) and information flow $\hat{\dot{\mathcal{I}}}_K$ (bottom). Note that the information flow is plotted in units of the inverse of $\tau_L$.}
\label{fig:Euclidean_MI_n=-15_07_thetaK=2.328e-08_Re=1.000e+00_ΔT=1.0e-01_N=200000_k=4_10_20_50}
\end{figure}

Figure~\ref{fig:KSGbias_n=-15_07_thetaK=2.328e-08_Re=1.000e+00_Nens=1000_Nt=100} shows bias of the KSG estimator $\hat{I}^{(\kappa)}_{\mathrm{KSG}}[{\bm U}^<_K(t):{\bm U}^>_K(t)]$ as a function of $1/N_{\mathrm{samp}}=n/N$ with $N=10^5$ in the case of $\mathrm{Re}\sim10^6$.
Here, we use $n=2,3,\cdots,10$, following Ref.~\cite{holmes2019estimation}.
The wave number $K$ is within the inertial range, $K=k_{10}$ (top), and at the Kolmogorov dissipation scale, $K=k_{15}$ (bottom).
The error bars are estimated by using the unbiased sample variance (\ref{unbiased sample variance}).
From (\ref{variance of the KSG estimator}) and (\ref{maximum liklihood estimator}), the standard deviation of $\hat{I}^{(\kappa)}_{\mathrm{KSG}}[{\bm U}^<_K(t):{\bm U}^>_K(t)]$ is estimated to be $\sim10^{-3}$ for $N_{\mathrm{samp}}\sim10^5$.
It can be seen from Fig.~\ref{fig:KSGbias_n=-15_07_thetaK=2.328e-08_Re=1.000e+00_Nens=1000_Nt=100} that, while there is no significant sample-size-dependent bias, a $\kappa$-dependent bias does exist.
In particular, the estimated mutual information is underestimated as $\kappa$ is increased.

Figure~\ref{fig:Euclidean_MI_n=-15_07_thetaK=2.328e-08_Re=1.000e+00_ΔT=1.0e-01_N=200000_k=4_10_20_50} shows the scale dependences of the estimated mutual information and information flow for $\kappa=4,10,20,50$ with $N_{\mathrm{samp}}=2\times10^5$ in the case of $\mathrm{Re}\sim10^6$.
Here, $\kappa=4$ is chosen because $\kappa=2,3,4$ are recommended in~\cite{kraskov2004estimating}.
These results clearly show that the estimated mutual information and information flow are underestimated as $\kappa$ is increased.
Therefore, it is difficult to choose the optimal $\kappa$ in this case.
The important point here is that the estimated information flow is positive for $K$ within the inertial range for all $\kappa$.
As mentioned in the main text, we remark that the error bar of the information flow $\hat{\dot{\mathcal{I}}}_K$ is of the same order as $\hat{\dot{\mathcal{I}}}_K$ itself if we naively estimate it by using the estimated standard deviation of the mutual information flow.

\bibliography{information_cascade}

\begin{thebibliography}{77}%
\makeatletter
\providecommand \@ifxundefined [1]{%
 \@ifx{#1\undefined}
}%
\providecommand \@ifnum [1]{%
 \ifnum #1\expandafter \@firstoftwo
 \else \expandafter \@secondoftwo
 \fi
}%
\providecommand \@ifx [1]{%
 \ifx #1\expandafter \@firstoftwo
 \else \expandafter \@secondoftwo
 \fi
}%
\providecommand \natexlab [1]{#1}%
\providecommand \enquote  [1]{``#1''}%
\providecommand \bibnamefont  [1]{#1}%
\providecommand \bibfnamefont [1]{#1}%
\providecommand \citenamefont [1]{#1}%
\providecommand \href@noop [0]{\@secondoftwo}%
\providecommand \href [0]{\begingroup \@sanitize@url \@href}%
\providecommand \@href[1]{\@@startlink{#1}\@@href}%
\providecommand \@@href[1]{\endgroup#1\@@endlink}%
\providecommand \@sanitize@url [0]{\catcode `\\12\catcode `\$12\catcode
  `\&12\catcode `\#12\catcode `\^12\catcode `\_12\catcode `\%12\relax}%
\providecommand \@@startlink[1]{}%
\providecommand \@@endlink[0]{}%
\providecommand \url  [0]{\begingroup\@sanitize@url \@url }%
\providecommand \@url [1]{\endgroup\@href {#1}{\urlprefix }}%
\providecommand \urlprefix  [0]{URL }%
\providecommand \Eprint [0]{\href }%
\providecommand \doibase [0]{https://doi.org/}%
\providecommand \selectlanguage [0]{\@gobble}%
\providecommand \bibinfo  [0]{\@secondoftwo}%
\providecommand \bibfield  [0]{\@secondoftwo}%
\providecommand \translation [1]{[#1]}%
\providecommand \BibitemOpen [0]{}%
\providecommand \bibitemStop [0]{}%
\providecommand \bibitemNoStop [0]{.\EOS\space}%
\providecommand \EOS [0]{\spacefactor3000\relax}%
\providecommand \BibitemShut  [1]{\csname bibitem#1\endcsname}%
\let\auto@bib@innerbib\@empty
\bibitem [{\citenamefont {Kolmogorov}(1941{\natexlab{a}})}]{K41_a}%
  \BibitemOpen
  \bibfield  {author} {\bibinfo {author} {\bibfnamefont {A.~N.}\ \bibnamefont
  {Kolmogorov}},\ }\bibfield  {title} {\bibinfo {title} {The local structure of
  turbulence in incompressible viscous fluid for very large {R}eynolds
  numbers},\ }\href@noop {} {\bibfield  {journal} {\bibinfo  {journal} {Dokl.
  Akad. Nauk SSSR}\ }\textbf {\bibinfo {volume} {30}},\ \bibinfo {pages} {9}
  (\bibinfo {year} {1941}{\natexlab{a}})}\BibitemShut {NoStop}%
\bibitem [{\citenamefont {Kolmogorov}(1941{\natexlab{b}})}]{K41_b}%
  \BibitemOpen
  \bibfield  {author} {\bibinfo {author} {\bibfnamefont {A.~N.}\ \bibnamefont
  {Kolmogorov}},\ }\bibfield  {title} {\bibinfo {title} {On degeneration of
  isotropic turbulence in an incompressible viscous liquid},\ }\href@noop {}
  {\bibfield  {journal} {\bibinfo  {journal} {Dokl. Akad. Nauk SSSR}\ }\textbf
  {\bibinfo {volume} {31}},\ \bibinfo {pages} {538} (\bibinfo {year}
  {1941}{\natexlab{b}})}\BibitemShut {NoStop}%
\bibitem [{\citenamefont {Kolmogorov}(1941{\natexlab{c}})}]{K41_c}%
  \BibitemOpen
  \bibfield  {author} {\bibinfo {author} {\bibfnamefont {A.~N.}\ \bibnamefont
  {Kolmogorov}},\ }\bibfield  {title} {\bibinfo {title} {Dissipation of energy
  in locally isotropic turbulence},\ }\href@noop {} {\bibfield  {journal}
  {\bibinfo  {journal} {Dokl. Akad. Nauk SSSR}\ }\textbf {\bibinfo {volume}
  {32}},\ \bibinfo {pages} {16} (\bibinfo {year}
  {1941}{\natexlab{c}})}\BibitemShut {NoStop}%
\bibitem [{\citenamefont {Frisch}(1995)}]{Frisch}%
  \BibitemOpen
  \bibfield  {author} {\bibinfo {author} {\bibfnamefont {U.}~\bibnamefont
  {Frisch}},\ }\href@noop {} {\emph {\bibinfo {title} {Turbulence}}}\ (\bibinfo
   {publisher} {Cambridge university press},\ \bibinfo {year}
  {1995})\BibitemShut {NoStop}%
\bibitem [{\citenamefont {Davidson}(2015)}]{davidson2015turbulence}%
  \BibitemOpen
  \bibfield  {author} {\bibinfo {author} {\bibfnamefont {P.~A.}\ \bibnamefont
  {Davidson}},\ }\href@noop {} {\emph {\bibinfo {title} {Turbulence: {A}n
  {I}ntroduction for {S}cientists and {E}ngineers}}},\ \bibinfo {edition}
  {2nd}\ ed.\ (\bibinfo  {publisher} {Oxford University Press},\ \bibinfo
  {year} {2015})\BibitemShut {NoStop}%
\bibitem [{\citenamefont {Anselmet}\ \emph {et~al.}(1984)\citenamefont
  {Anselmet}, \citenamefont {Gagne}, \citenamefont {Hopfinger},\ and\
  \citenamefont {Antonia}}]{data}%
  \BibitemOpen
  \bibfield  {author} {\bibinfo {author} {\bibfnamefont {F.}~\bibnamefont
  {Anselmet}}, \bibinfo {author} {\bibfnamefont {Y.~l.}\ \bibnamefont {Gagne}},
  \bibinfo {author} {\bibfnamefont {E.~J.}\ \bibnamefont {Hopfinger}},\ and\
  \bibinfo {author} {\bibfnamefont {R.~A.}\ \bibnamefont {Antonia}},\
  }\bibfield  {title} {\bibinfo {title} {High-order velocity structure
  functions in turbulent shear flows},\ }\href@noop {} {\bibfield  {journal}
  {\bibinfo  {journal} {J. Fluid Mech.}\ }\textbf {\bibinfo {volume} {140}},\
  \bibinfo {pages} {63} (\bibinfo {year} {1984})}\BibitemShut {NoStop}%
\bibitem [{\citenamefont {Eyink}\ and\ \citenamefont
  {Sreenivasan}(2006)}]{Eyink_Sreenivasan}%
  \BibitemOpen
  \bibfield  {author} {\bibinfo {author} {\bibfnamefont {G.~L.}\ \bibnamefont
  {Eyink}}\ and\ \bibinfo {author} {\bibfnamefont {K.~R.}\ \bibnamefont
  {Sreenivasan}},\ }\bibfield  {title} {\bibinfo {title} {Onsager and the
  theory of hydrodynamic turbulence},\ }\href@noop {} {\bibfield  {journal}
  {\bibinfo  {journal} {Rev. Mod. Phys.}\ }\textbf {\bibinfo {volume} {78}},\
  \bibinfo {pages} {87} (\bibinfo {year} {2006})}\BibitemShut {NoStop}%
\bibitem [{\citenamefont {Eyink}()}]{Eyink_lecture}%
  \BibitemOpen
  \bibfield  {author} {\bibinfo {author} {\bibfnamefont {G.~L.}\ \bibnamefont
  {Eyink}},\ }\href@noop {} {\bibinfo {title} {Turbulence {T}heory, {C}ourse
  {N}otes}},\ \bibinfo {note}
  {\url{http://www.ams.jhu.edu/~eyink/Turbulence/notes/}}\BibitemShut {NoStop}%
\bibitem [{\citenamefont {Yasuda}\ \emph {et~al.}(2014)\citenamefont {Yasuda},
  \citenamefont {Goto},\ and\ \citenamefont {Kawahara}}]{yasuda2014quasi}%
  \BibitemOpen
  \bibfield  {author} {\bibinfo {author} {\bibfnamefont {T.}~\bibnamefont
  {Yasuda}}, \bibinfo {author} {\bibfnamefont {S.}~\bibnamefont {Goto}},\ and\
  \bibinfo {author} {\bibfnamefont {G.}~\bibnamefont {Kawahara}},\ }\bibfield
  {title} {\bibinfo {title} {Quasi-cyclic evolution of turbulence driven by a
  steady force in a periodic cube},\ }\href@noop {} {\bibfield  {journal}
  {\bibinfo  {journal} {Fluid Dyn. Res.}\ }\textbf {\bibinfo {volume} {46}},\
  \bibinfo {pages} {061413} (\bibinfo {year} {2014})}\BibitemShut {NoStop}%
\bibitem [{\citenamefont {Goto}\ \emph {et~al.}(2017)\citenamefont {Goto},
  \citenamefont {Saito},\ and\ \citenamefont {Kawahara}}]{goto2017hierarchy}%
  \BibitemOpen
  \bibfield  {author} {\bibinfo {author} {\bibfnamefont {S.}~\bibnamefont
  {Goto}}, \bibinfo {author} {\bibfnamefont {Y.}~\bibnamefont {Saito}},\ and\
  \bibinfo {author} {\bibfnamefont {G.}~\bibnamefont {Kawahara}},\ }\bibfield
  {title} {\bibinfo {title} {Hierarchy of antiparallel vortex tubes in
  spatially periodic turbulence at high {R}eynolds numbers},\ }\href@noop {}
  {\bibfield  {journal} {\bibinfo  {journal} {Phys. Rev. Fluids}\ }\textbf
  {\bibinfo {volume} {2}},\ \bibinfo {pages} {064603} (\bibinfo {year}
  {2017})}\BibitemShut {NoStop}%
\bibitem [{\citenamefont {Araki}\ \emph {et~al.}(2023)\citenamefont {Araki},
  \citenamefont {Bos},\ and\ \citenamefont {Goto}}]{araki2023minimal}%
  \BibitemOpen
  \bibfield  {author} {\bibinfo {author} {\bibfnamefont {R.}~\bibnamefont
  {Araki}}, \bibinfo {author} {\bibfnamefont {W.}~\bibnamefont {Bos}},\ and\
  \bibinfo {author} {\bibfnamefont {S.}~\bibnamefont {Goto}},\ }\bibfield
  {title} {\bibinfo {title} {Minimal model of quasi-cyclic behaviour in
  turbulence driven by taylor--green forcing},\ }\href@noop {} {\bibfield
  {journal} {\bibinfo  {journal} {Fluid Dyn. Res.}\ }\textbf {\bibinfo {volume}
  {55}},\ \bibinfo {pages} {035507} (\bibinfo {year} {2023})}\BibitemShut
  {NoStop}%
\bibitem [{\citenamefont {Pecora}\ and\ \citenamefont
  {Carroll}(1990)}]{pecora1990synchronization}%
  \BibitemOpen
  \bibfield  {author} {\bibinfo {author} {\bibfnamefont {L.~M.}\ \bibnamefont
  {Pecora}}\ and\ \bibinfo {author} {\bibfnamefont {T.~L.}\ \bibnamefont
  {Carroll}},\ }\bibfield  {title} {\bibinfo {title} {Synchronization in
  chaotic systems},\ }\href@noop {} {\bibfield  {journal} {\bibinfo  {journal}
  {Phys. Rev. Lett.}\ }\textbf {\bibinfo {volume} {64}},\ \bibinfo {pages}
  {821} (\bibinfo {year} {1990})}\BibitemShut {NoStop}%
\bibitem [{\citenamefont {Boccaletti}\ \emph {et~al.}(2002)\citenamefont
  {Boccaletti}, \citenamefont {Kurths}, \citenamefont {Osipov}, \citenamefont
  {Valladares},\ and\ \citenamefont {Zhou}}]{boccaletti2002synchronization}%
  \BibitemOpen
  \bibfield  {author} {\bibinfo {author} {\bibfnamefont {S.}~\bibnamefont
  {Boccaletti}}, \bibinfo {author} {\bibfnamefont {J.}~\bibnamefont {Kurths}},
  \bibinfo {author} {\bibfnamefont {G.}~\bibnamefont {Osipov}}, \bibinfo
  {author} {\bibfnamefont {D.}~\bibnamefont {Valladares}},\ and\ \bibinfo
  {author} {\bibfnamefont {C.}~\bibnamefont {Zhou}},\ }\bibfield  {title}
  {\bibinfo {title} {The synchronization of chaotic systems},\ }\href@noop {}
  {\bibfield  {journal} {\bibinfo  {journal} {Phys. Rep.}\ }\textbf {\bibinfo
  {volume} {366}},\ \bibinfo {pages} {1} (\bibinfo {year} {2002})}\BibitemShut
  {NoStop}%
\bibitem [{\citenamefont {Yoshida}\ \emph {et~al.}(2005)\citenamefont
  {Yoshida}, \citenamefont {Yamaguchi},\ and\ \citenamefont
  {Kaneda}}]{yoshida2005regeneration}%
  \BibitemOpen
  \bibfield  {author} {\bibinfo {author} {\bibfnamefont {K.}~\bibnamefont
  {Yoshida}}, \bibinfo {author} {\bibfnamefont {J.}~\bibnamefont {Yamaguchi}},\
  and\ \bibinfo {author} {\bibfnamefont {Y.}~\bibnamefont {Kaneda}},\
  }\bibfield  {title} {\bibinfo {title} {Regeneration of small eddies by data
  assimilation in turbulence},\ }\href@noop {} {\bibfield  {journal} {\bibinfo
  {journal} {Phys. Rev. Lett.}\ }\textbf {\bibinfo {volume} {94}},\ \bibinfo
  {pages} {014501} (\bibinfo {year} {2005})}\BibitemShut {NoStop}%
\bibitem [{\citenamefont {Lalescu}\ \emph {et~al.}(2013)\citenamefont
  {Lalescu}, \citenamefont {Meneveau},\ and\ \citenamefont
  {Eyink}}]{lalescu2013synchronization}%
  \BibitemOpen
  \bibfield  {author} {\bibinfo {author} {\bibfnamefont {C.~C.}\ \bibnamefont
  {Lalescu}}, \bibinfo {author} {\bibfnamefont {C.}~\bibnamefont {Meneveau}},\
  and\ \bibinfo {author} {\bibfnamefont {G.~L.}\ \bibnamefont {Eyink}},\
  }\bibfield  {title} {\bibinfo {title} {Synchronization of chaos in fully
  developed turbulence},\ }\href@noop {} {\bibfield  {journal} {\bibinfo
  {journal} {Phys. Rev. Lett.}\ }\textbf {\bibinfo {volume} {110}},\ \bibinfo
  {pages} {084102} (\bibinfo {year} {2013})}\BibitemShut {NoStop}%
\bibitem [{\citenamefont {Vela-Mart{\'\i}n}(2021)}]{vela2021synchronisation}%
  \BibitemOpen
  \bibfield  {author} {\bibinfo {author} {\bibfnamefont {A.}~\bibnamefont
  {Vela-Mart{\'\i}n}},\ }\bibfield  {title} {\bibinfo {title} {The
  synchronisation of intense vorticity in isotropic turbulence},\ }\href@noop
  {} {\bibfield  {journal} {\bibinfo  {journal} {J. Fluid Mech.}\ }\textbf
  {\bibinfo {volume} {913}} (\bibinfo {year} {2021})}\BibitemShut {NoStop}%
\bibitem [{\citenamefont {Betchov}(1964)}]{betchov1964measure}%
  \BibitemOpen
  \bibfield  {author} {\bibinfo {author} {\bibfnamefont {R.}~\bibnamefont
  {Betchov}},\ }\bibfield  {title} {\bibinfo {title} {Measure of the intricacy
  of turbulence},\ }\href@noop {} {\bibfield  {journal} {\bibinfo  {journal}
  {Phys. Fluids}\ }\textbf {\bibinfo {volume} {7}},\ \bibinfo {pages} {1160}
  (\bibinfo {year} {1964})}\BibitemShut {NoStop}%
\bibitem [{\citenamefont {Ikeda}\ and\ \citenamefont
  {Matsumoto}(1989)}]{ikeda1989information}%
  \BibitemOpen
  \bibfield  {author} {\bibinfo {author} {\bibfnamefont {K.}~\bibnamefont
  {Ikeda}}\ and\ \bibinfo {author} {\bibfnamefont {K.}~\bibnamefont
  {Matsumoto}},\ }\bibfield  {title} {\bibinfo {title} {Information theoretical
  characterization of turbulence},\ }\href@noop {} {\bibfield  {journal}
  {\bibinfo  {journal} {Phys. Rev. Lett.}\ }\textbf {\bibinfo {volume} {62}},\
  \bibinfo {pages} {2265} (\bibinfo {year} {1989})}\BibitemShut {NoStop}%
\bibitem [{\citenamefont {Cerbus}\ and\ \citenamefont
  {Goldburg}(2013)}]{cerbus2013information}%
  \BibitemOpen
  \bibfield  {author} {\bibinfo {author} {\bibfnamefont {R.~T.}\ \bibnamefont
  {Cerbus}}\ and\ \bibinfo {author} {\bibfnamefont {W.~I.}\ \bibnamefont
  {Goldburg}},\ }\bibfield  {title} {\bibinfo {title} {Information content of
  turbulence},\ }\href@noop {} {\bibfield  {journal} {\bibinfo  {journal}
  {Phys. Rev. E}\ }\textbf {\bibinfo {volume} {88}},\ \bibinfo {pages} {053012}
  (\bibinfo {year} {2013})}\BibitemShut {NoStop}%
\bibitem [{\citenamefont {Materassi}\ \emph {et~al.}(2014)\citenamefont
  {Materassi}, \citenamefont {Consolini}, \citenamefont {Smith},\ and\
  \citenamefont {De~Marco}}]{materassi2014information}%
  \BibitemOpen
  \bibfield  {author} {\bibinfo {author} {\bibfnamefont {M.}~\bibnamefont
  {Materassi}}, \bibinfo {author} {\bibfnamefont {G.}~\bibnamefont
  {Consolini}}, \bibinfo {author} {\bibfnamefont {N.}~\bibnamefont {Smith}},\
  and\ \bibinfo {author} {\bibfnamefont {R.}~\bibnamefont {De~Marco}},\
  }\bibfield  {title} {\bibinfo {title} {Information theory analysis of
  cascading process in a synthetic model of fluid turbulence},\ }\href@noop {}
  {\bibfield  {journal} {\bibinfo  {journal} {Entropy}\ }\textbf {\bibinfo
  {volume} {16}},\ \bibinfo {pages} {1272} (\bibinfo {year}
  {2014})}\BibitemShut {NoStop}%
\bibitem [{\citenamefont {Cerbus}\ and\ \citenamefont
  {Goldburg}(2016)}]{cerbus2016information}%
  \BibitemOpen
  \bibfield  {author} {\bibinfo {author} {\bibfnamefont {R.~T.}\ \bibnamefont
  {Cerbus}}\ and\ \bibinfo {author} {\bibfnamefont {W.~I.}\ \bibnamefont
  {Goldburg}},\ }\bibfield  {title} {\bibinfo {title} {Information theory
  demonstration of the {R}ichardson cascade},\ }\href@noop {} {\bibfield
  {journal} {\bibinfo  {journal} {arXiv preprint arXiv:1602.02980}\ } (\bibinfo
  {year} {2016})}\BibitemShut {NoStop}%
\bibitem [{\citenamefont {Goldburg}\ and\ \citenamefont
  {Cerbus}(2016)}]{goldburg2016turbulence}%
  \BibitemOpen
  \bibfield  {author} {\bibinfo {author} {\bibfnamefont {W.~I.}\ \bibnamefont
  {Goldburg}}\ and\ \bibinfo {author} {\bibfnamefont {R.~T.}\ \bibnamefont
  {Cerbus}},\ }\bibfield  {title} {\bibinfo {title} {Turbulence as
  information},\ }\href@noop {} {\bibfield  {journal} {\bibinfo  {journal}
  {arXiv preprint arXiv:1609.00471}\ } (\bibinfo {year} {2016})}\BibitemShut
  {NoStop}%
\bibitem [{\citenamefont {Granero-Belinchon}\ \emph {et~al.}(2016)\citenamefont
  {Granero-Belinchon}, \citenamefont {Roux},\ and\ \citenamefont
  {Garnier}}]{granero2016scaling}%
  \BibitemOpen
  \bibfield  {author} {\bibinfo {author} {\bibfnamefont {C.}~\bibnamefont
  {Granero-Belinchon}}, \bibinfo {author} {\bibfnamefont {S.~G.}\ \bibnamefont
  {Roux}},\ and\ \bibinfo {author} {\bibfnamefont {N.~B.}\ \bibnamefont
  {Garnier}},\ }\bibfield  {title} {\bibinfo {title} {Scaling of information in
  turbulence},\ }\href@noop {} {\bibfield  {journal} {\bibinfo  {journal}
  {Europhys. Lett.}\ }\textbf {\bibinfo {volume} {115}},\ \bibinfo {pages}
  {58003} (\bibinfo {year} {2016})}\BibitemShut {NoStop}%
\bibitem [{\citenamefont {Granero-Belinch{\'o}n}\ \emph
  {et~al.}(2018)\citenamefont {Granero-Belinch{\'o}n}, \citenamefont {Roux},\
  and\ \citenamefont {Garnier}}]{granero2018kullback}%
  \BibitemOpen
  \bibfield  {author} {\bibinfo {author} {\bibfnamefont {C.}~\bibnamefont
  {Granero-Belinch{\'o}n}}, \bibinfo {author} {\bibfnamefont {S.~G.}\
  \bibnamefont {Roux}},\ and\ \bibinfo {author} {\bibfnamefont {N.~B.}\
  \bibnamefont {Garnier}},\ }\bibfield  {title} {\bibinfo {title}
  {Kullback-{L}eibler divergence measure of intermittency: {A}pplication to
  turbulence},\ }\href@noop {} {\bibfield  {journal} {\bibinfo  {journal}
  {Phys. Rev. E}\ }\textbf {\bibinfo {volume} {97}},\ \bibinfo {pages} {013107}
  (\bibinfo {year} {2018})}\BibitemShut {NoStop}%
\bibitem [{\citenamefont {Lozano-Dur{\'a}n}\ \emph {et~al.}(2020)\citenamefont
  {Lozano-Dur{\'a}n}, \citenamefont {Bae},\ and\ \citenamefont
  {Encinar}}]{lozano2020causality}%
  \BibitemOpen
  \bibfield  {author} {\bibinfo {author} {\bibfnamefont {A.}~\bibnamefont
  {Lozano-Dur{\'a}n}}, \bibinfo {author} {\bibfnamefont {H.~J.}\ \bibnamefont
  {Bae}},\ and\ \bibinfo {author} {\bibfnamefont {M.~P.}\ \bibnamefont
  {Encinar}},\ }\bibfield  {title} {\bibinfo {title} {Causality of
  energy-containing eddies in wall turbulence},\ }\href@noop {} {\bibfield
  {journal} {\bibinfo  {journal} {J. Fluid Mech.}\ }\textbf {\bibinfo {volume}
  {882}} (\bibinfo {year} {2020})}\BibitemShut {NoStop}%
\bibitem [{\citenamefont {Shavit}\ and\ \citenamefont
  {Falkovich}(2020)}]{shavit2020singular}%
  \BibitemOpen
  \bibfield  {author} {\bibinfo {author} {\bibfnamefont {M.}~\bibnamefont
  {Shavit}}\ and\ \bibinfo {author} {\bibfnamefont {G.}~\bibnamefont
  {Falkovich}},\ }\bibfield  {title} {\bibinfo {title} {Singular measures and
  information capacity of turbulent cascades},\ }\href@noop {} {\bibfield
  {journal} {\bibinfo  {journal} {Phys. Rev. Lett.}\ }\textbf {\bibinfo
  {volume} {125}},\ \bibinfo {pages} {104501} (\bibinfo {year}
  {2020})}\BibitemShut {NoStop}%
\bibitem [{\citenamefont {Vladimirova}\ \emph {et~al.}(2021)\citenamefont
  {Vladimirova}, \citenamefont {Shavit},\ and\ \citenamefont
  {Falkovich}}]{vladimirova2021fibonacci}%
  \BibitemOpen
  \bibfield  {author} {\bibinfo {author} {\bibfnamefont {N.}~\bibnamefont
  {Vladimirova}}, \bibinfo {author} {\bibfnamefont {M.}~\bibnamefont
  {Shavit}},\ and\ \bibinfo {author} {\bibfnamefont {G.}~\bibnamefont
  {Falkovich}},\ }\bibfield  {title} {\bibinfo {title} {Fibonacci turbulence},\
  }\href@noop {} {\bibfield  {journal} {\bibinfo  {journal} {Phys. Rev. X}\
  }\textbf {\bibinfo {volume} {11}},\ \bibinfo {pages} {021063} (\bibinfo
  {year} {2021})}\BibitemShut {NoStop}%
\bibitem [{\citenamefont {Lozano-Dur\'an}\ and\ \citenamefont
  {Arranz}(2022)}]{PhysRevResearch.4.023195}%
  \BibitemOpen
  \bibfield  {author} {\bibinfo {author} {\bibfnamefont {A.}~\bibnamefont
  {Lozano-Dur\'an}}\ and\ \bibinfo {author} {\bibfnamefont {G.}~\bibnamefont
  {Arranz}},\ }\bibfield  {title} {\bibinfo {title} {Information-theoretic
  formulation of dynamical systems: {C}ausality, modeling, and control},\
  }\href {https://doi.org/10.1103/PhysRevResearch.4.023195} {\bibfield
  {journal} {\bibinfo  {journal} {Phys. Rev. Research}\ }\textbf {\bibinfo
  {volume} {4}},\ \bibinfo {pages} {023195} (\bibinfo {year}
  {2022})}\BibitemShut {NoStop}%
\bibitem [{\citenamefont {Parrondo}\ \emph {et~al.}(2015)\citenamefont
  {Parrondo}, \citenamefont {Horowitz},\ and\ \citenamefont
  {Sagawa}}]{parrondo2015thermodynamics}%
  \BibitemOpen
  \bibfield  {author} {\bibinfo {author} {\bibfnamefont {J.~M.}\ \bibnamefont
  {Parrondo}}, \bibinfo {author} {\bibfnamefont {J.~M.}\ \bibnamefont
  {Horowitz}},\ and\ \bibinfo {author} {\bibfnamefont {T.}~\bibnamefont
  {Sagawa}},\ }\bibfield  {title} {\bibinfo {title} {Thermodynamics of
  information},\ }\href@noop {} {\bibfield  {journal} {\bibinfo  {journal}
  {Nat. Phys.}\ }\textbf {\bibinfo {volume} {11}},\ \bibinfo {pages} {131}
  (\bibinfo {year} {2015})}\BibitemShut {NoStop}%
\bibitem [{\citenamefont {Ehrich}\ and\ \citenamefont
  {Sivak}(2023)}]{ehrich2023energy}%
  \BibitemOpen
  \bibfield  {author} {\bibinfo {author} {\bibfnamefont {J.}~\bibnamefont
  {Ehrich}}\ and\ \bibinfo {author} {\bibfnamefont {D.~A.}\ \bibnamefont
  {Sivak}},\ }\bibfield  {title} {\bibinfo {title} {Energy and information
  flows in autonomous systems},\ }\href@noop {} {\bibfield  {journal} {\bibinfo
   {journal} {Front. Phys.}\ }\textbf {\bibinfo {volume} {11}},\ \bibinfo
  {pages} {155} (\bibinfo {year} {2023})}\BibitemShut {NoStop}%
\bibitem [{\citenamefont {Shiraishi}(2023)}]{shiraishi2023introduction}%
  \BibitemOpen
  \bibfield  {author} {\bibinfo {author} {\bibfnamefont {N.}~\bibnamefont
  {Shiraishi}},\ }\href@noop {} {\emph {\bibinfo {title} {{An Introduction to
  Stochastic Thermodynamics: From Basic to Advanced}}}},\ Vol.\ \bibinfo
  {volume} {212}\ (\bibinfo  {publisher} {Springer Nature},\ \bibinfo {year}
  {2023})\BibitemShut {NoStop}%
\bibitem [{\citenamefont {Barato}\ \emph {et~al.}(2014)\citenamefont {Barato},
  \citenamefont {Hartich},\ and\ \citenamefont
  {Seifert}}]{barato2014efficiency}%
  \BibitemOpen
  \bibfield  {author} {\bibinfo {author} {\bibfnamefont {A.~C.}\ \bibnamefont
  {Barato}}, \bibinfo {author} {\bibfnamefont {D.}~\bibnamefont {Hartich}},\
  and\ \bibinfo {author} {\bibfnamefont {U.}~\bibnamefont {Seifert}},\
  }\bibfield  {title} {\bibinfo {title} {Efficiency of cellular information
  processing},\ }\href@noop {} {\bibfield  {journal} {\bibinfo  {journal} {New
  J. Phys.}\ }\textbf {\bibinfo {volume} {16}},\ \bibinfo {pages} {103024}
  (\bibinfo {year} {2014})}\BibitemShut {NoStop}%
\bibitem [{\citenamefont {Sartori}\ \emph {et~al.}(2014)\citenamefont
  {Sartori}, \citenamefont {Granger}, \citenamefont {Lee},\ and\ \citenamefont
  {Horowitz}}]{sartori2014thermodynamic}%
  \BibitemOpen
  \bibfield  {author} {\bibinfo {author} {\bibfnamefont {P.}~\bibnamefont
  {Sartori}}, \bibinfo {author} {\bibfnamefont {L.}~\bibnamefont {Granger}},
  \bibinfo {author} {\bibfnamefont {C.~F.}\ \bibnamefont {Lee}},\ and\ \bibinfo
  {author} {\bibfnamefont {J.~M.}\ \bibnamefont {Horowitz}},\ }\bibfield
  {title} {\bibinfo {title} {Thermodynamic costs of information processing in
  sensory adaptation},\ }\href@noop {} {\bibfield  {journal} {\bibinfo
  {journal} {PLoS Comput. Biol.}\ }\textbf {\bibinfo {volume} {10}},\ \bibinfo
  {pages} {e1003974} (\bibinfo {year} {2014})}\BibitemShut {NoStop}%
\bibitem [{\citenamefont {Ito}\ and\ \citenamefont
  {Sagawa}(2015)}]{ito2015maxwell}%
  \BibitemOpen
  \bibfield  {author} {\bibinfo {author} {\bibfnamefont {S.}~\bibnamefont
  {Ito}}\ and\ \bibinfo {author} {\bibfnamefont {T.}~\bibnamefont {Sagawa}},\
  }\bibfield  {title} {\bibinfo {title} {Maxwell's demon in biochemical signal
  transduction with feedback loop},\ }\href@noop {} {\bibfield  {journal}
  {\bibinfo  {journal} {Nat. Commun.}\ }\textbf {\bibinfo {volume} {6}},\
  \bibinfo {pages} {1} (\bibinfo {year} {2015})}\BibitemShut {NoStop}%
\bibitem [{\citenamefont {Hartich}\ \emph {et~al.}(2016)\citenamefont
  {Hartich}, \citenamefont {Barato},\ and\ \citenamefont
  {Seifert}}]{hartich2016sensory}%
  \BibitemOpen
  \bibfield  {author} {\bibinfo {author} {\bibfnamefont {D.}~\bibnamefont
  {Hartich}}, \bibinfo {author} {\bibfnamefont {A.~C.}\ \bibnamefont
  {Barato}},\ and\ \bibinfo {author} {\bibfnamefont {U.}~\bibnamefont
  {Seifert}},\ }\bibfield  {title} {\bibinfo {title} {Sensory capacity: {A}n
  information theoretical measure of the performance of a sensor},\ }\href@noop
  {} {\bibfield  {journal} {\bibinfo  {journal} {Phys. Rev. E}\ }\textbf
  {\bibinfo {volume} {93}},\ \bibinfo {pages} {022116} (\bibinfo {year}
  {2016})}\BibitemShut {NoStop}%
\bibitem [{\citenamefont {Amano}\ \emph {et~al.}(2022)\citenamefont {Amano},
  \citenamefont {Esposito}, \citenamefont {Kreidt}, \citenamefont {Leigh},
  \citenamefont {Penocchio},\ and\ \citenamefont
  {Roberts}}]{amano2022insights}%
  \BibitemOpen
  \bibfield  {author} {\bibinfo {author} {\bibfnamefont {S.}~\bibnamefont
  {Amano}}, \bibinfo {author} {\bibfnamefont {M.}~\bibnamefont {Esposito}},
  \bibinfo {author} {\bibfnamefont {E.}~\bibnamefont {Kreidt}}, \bibinfo
  {author} {\bibfnamefont {D.~A.}\ \bibnamefont {Leigh}}, \bibinfo {author}
  {\bibfnamefont {E.}~\bibnamefont {Penocchio}},\ and\ \bibinfo {author}
  {\bibfnamefont {B.~M.~W.}\ \bibnamefont {Roberts}},\ }\bibfield  {title}
  {\bibinfo {title} {Insights from an information thermodynamics analysis of a
  synthetic molecular motor},\ }\href@noop {} {\bibfield  {journal} {\bibinfo
  {journal} {Nat. Chem.}\ }\textbf {\bibinfo {volume} {14}},\ \bibinfo {pages}
  {530} (\bibinfo {year} {2022})}\BibitemShut {NoStop}%
\bibitem [{\citenamefont {Penocchio}\ \emph {et~al.}(2022)\citenamefont
  {Penocchio}, \citenamefont {Avanzini},\ and\ \citenamefont
  {Esposito}}]{penocchio2022information}%
  \BibitemOpen
  \bibfield  {author} {\bibinfo {author} {\bibfnamefont {E.}~\bibnamefont
  {Penocchio}}, \bibinfo {author} {\bibfnamefont {F.}~\bibnamefont
  {Avanzini}},\ and\ \bibinfo {author} {\bibfnamefont {M.}~\bibnamefont
  {Esposito}},\ }\bibfield  {title} {\bibinfo {title} {Information
  thermodynamics for deterministic chemical reaction networks},\ }\href@noop {}
  {\bibfield  {journal} {\bibinfo  {journal} {J. Chem. Phys.}\ }\textbf
  {\bibinfo {volume} {157}},\ \bibinfo {pages} {034110} (\bibinfo {year}
  {2022})}\BibitemShut {NoStop}%
\bibitem [{\citenamefont {Landau}\ and\ \citenamefont
  {Lifshitz}(1959)}]{landau1959fluid}%
  \BibitemOpen
  \bibfield  {author} {\bibinfo {author} {\bibfnamefont {L.~D.}\ \bibnamefont
  {Landau}}\ and\ \bibinfo {author} {\bibfnamefont {E.~M.}\ \bibnamefont
  {Lifshitz}},\ }\href@noop {} {\emph {\bibinfo {title} {Fluid {M}echanics}}},\
  Vol.~\bibinfo {volume} {6}\ (\bibinfo  {publisher} {Addision-Wesley, Reading,
  MA},\ \bibinfo {year} {1959})\BibitemShut {NoStop}%
\bibitem [{\citenamefont {De~Zarate}\ and\ \citenamefont
  {Sengers}(2006)}]{de2006hydrodynamic}%
  \BibitemOpen
  \bibfield  {author} {\bibinfo {author} {\bibfnamefont {J.~M.~O.}\
  \bibnamefont {De~Zarate}}\ and\ \bibinfo {author} {\bibfnamefont {J.~V.}\
  \bibnamefont {Sengers}},\ }\href@noop {} {\emph {\bibinfo {title}
  {Hydrodynamic fluctuations in fluids and fluid mixtures}}}\ (\bibinfo
  {publisher} {Elsevier},\ \bibinfo {year} {2006})\BibitemShut {NoStop}%
\bibitem [{\citenamefont {Ruelle}(1979)}]{ruelle1979microscopic}%
  \BibitemOpen
  \bibfield  {author} {\bibinfo {author} {\bibfnamefont {D.}~\bibnamefont
  {Ruelle}},\ }\bibfield  {title} {\bibinfo {title} {Microscopic fluctuations
  and turbulence},\ }\href@noop {} {\bibfield  {journal} {\bibinfo  {journal}
  {Phys. Lett. A}\ }\textbf {\bibinfo {volume} {72}},\ \bibinfo {pages} {81}
  (\bibinfo {year} {1979})}\BibitemShut {NoStop}%
\bibitem [{\citenamefont {Komatsu}\ \emph {et~al.}(2014)\citenamefont
  {Komatsu}, \citenamefont {Matsumoto}, \citenamefont {Shimada},\ and\
  \citenamefont {Ito}}]{komatsu2014glimpse}%
  \BibitemOpen
  \bibfield  {author} {\bibinfo {author} {\bibfnamefont {T.~S.}\ \bibnamefont
  {Komatsu}}, \bibinfo {author} {\bibfnamefont {S.}~\bibnamefont {Matsumoto}},
  \bibinfo {author} {\bibfnamefont {T.}~\bibnamefont {Shimada}},\ and\ \bibinfo
  {author} {\bibfnamefont {N.}~\bibnamefont {Ito}},\ }\bibfield  {title}
  {\bibinfo {title} {A glimpse of fluid turbulence from the molecular scale},\
  }\href@noop {} {\bibfield  {journal} {\bibinfo  {journal} {Int. J. Mod. Phys.
  C}\ }\textbf {\bibinfo {volume} {25}},\ \bibinfo {pages} {1450034} (\bibinfo
  {year} {2014})}\BibitemShut {NoStop}%
\bibitem [{\citenamefont {Bandak}\ \emph {et~al.}(2021)\citenamefont {Bandak},
  \citenamefont {Eyink}, \citenamefont {Mailybaev},\ and\ \citenamefont
  {Goldenfeld}}]{bandak2021thermal}%
  \BibitemOpen
  \bibfield  {author} {\bibinfo {author} {\bibfnamefont {D.}~\bibnamefont
  {Bandak}}, \bibinfo {author} {\bibfnamefont {G.~L.}\ \bibnamefont {Eyink}},
  \bibinfo {author} {\bibfnamefont {A.}~\bibnamefont {Mailybaev}},\ and\
  \bibinfo {author} {\bibfnamefont {N.}~\bibnamefont {Goldenfeld}},\ }\bibfield
   {title} {\bibinfo {title} {Thermal noise competes with turbulent
  fluctuations below millimeter scales},\ }\href@noop {} {\bibfield  {journal}
  {\bibinfo  {journal} {arXiv preprint arXiv:2107.03184}\ } (\bibinfo {year}
  {2021})}\BibitemShut {NoStop}%
\bibitem [{\citenamefont {Bandak}\ \emph {et~al.}(2022)\citenamefont {Bandak},
  \citenamefont {Goldenfeld}, \citenamefont {Mailybaev},\ and\ \citenamefont
  {Eyink}}]{bandak2022dissipation}%
  \BibitemOpen
  \bibfield  {author} {\bibinfo {author} {\bibfnamefont {D.}~\bibnamefont
  {Bandak}}, \bibinfo {author} {\bibfnamefont {N.}~\bibnamefont {Goldenfeld}},
  \bibinfo {author} {\bibfnamefont {A.~A.}\ \bibnamefont {Mailybaev}},\ and\
  \bibinfo {author} {\bibfnamefont {G.}~\bibnamefont {Eyink}},\ }\bibfield
  {title} {\bibinfo {title} {Dissipation-range fluid turbulence and thermal
  noise},\ }\href@noop {} {\bibfield  {journal} {\bibinfo  {journal} {Phys.
  Rev. E}\ }\textbf {\bibinfo {volume} {105}},\ \bibinfo {pages} {065113}
  (\bibinfo {year} {2022})}\BibitemShut {NoStop}%
\bibitem [{\citenamefont {Eyink}\ and\ \citenamefont
  {Jafari}(2022)}]{eyink2022high}%
  \BibitemOpen
  \bibfield  {author} {\bibinfo {author} {\bibfnamefont {G.}~\bibnamefont
  {Eyink}}\ and\ \bibinfo {author} {\bibfnamefont {A.}~\bibnamefont {Jafari}},\
  }\bibfield  {title} {\bibinfo {title} {High schmidt-number turbulent
  advection and giant concentration fluctuations},\ }\href@noop {} {\bibfield
  {journal} {\bibinfo  {journal} {Phys. Rev. Research}\ }\textbf {\bibinfo
  {volume} {4}},\ \bibinfo {pages} {023246} (\bibinfo {year}
  {2022})}\BibitemShut {NoStop}%
\bibitem [{\citenamefont {McMullen}\ \emph {et~al.}(2022)\citenamefont
  {McMullen}, \citenamefont {Krygier}, \citenamefont {Torczynski},\ and\
  \citenamefont {Gallis}}]{mcmullen2022navier}%
  \BibitemOpen
  \bibfield  {author} {\bibinfo {author} {\bibfnamefont {R.~M.}\ \bibnamefont
  {McMullen}}, \bibinfo {author} {\bibfnamefont {M.~C.}\ \bibnamefont
  {Krygier}}, \bibinfo {author} {\bibfnamefont {J.~R.}\ \bibnamefont
  {Torczynski}},\ and\ \bibinfo {author} {\bibfnamefont {M.~A.}\ \bibnamefont
  {Gallis}},\ }\bibfield  {title} {\bibinfo {title} {Navier-{S}tokes
  {E}quations {D}o {N}ot {D}escribe the {S}mallest {S}cales of {T}urbulence in
  {G}ases},\ }\href@noop {} {\bibfield  {journal} {\bibinfo  {journal} {Phys.
  Rev. Lett.}\ }\textbf {\bibinfo {volume} {128}},\ \bibinfo {pages} {114501}
  (\bibinfo {year} {2022})}\BibitemShut {NoStop}%
\bibitem [{\citenamefont {Bell}\ \emph {et~al.}(2022)\citenamefont {Bell},
  \citenamefont {Nonaka}, \citenamefont {Garcia},\ and\ \citenamefont
  {Eyink}}]{bell2022thermal}%
  \BibitemOpen
  \bibfield  {author} {\bibinfo {author} {\bibfnamefont {J.~B.}\ \bibnamefont
  {Bell}}, \bibinfo {author} {\bibfnamefont {A.}~\bibnamefont {Nonaka}},
  \bibinfo {author} {\bibfnamefont {A.~L.}\ \bibnamefont {Garcia}},\ and\
  \bibinfo {author} {\bibfnamefont {G.}~\bibnamefont {Eyink}},\ }\bibfield
  {title} {\bibinfo {title} {Thermal fluctuations in the dissipation range of
  homogeneous isotropic turbulence},\ }\href@noop {} {\bibfield  {journal}
  {\bibinfo  {journal} {J. Fluid Mech.}\ }\textbf {\bibinfo {volume} {939}}
  (\bibinfo {year} {2022})}\BibitemShut {NoStop}%
\bibitem [{\citenamefont {Maes}(2021)}]{maes2021local}%
  \BibitemOpen
  \bibfield  {author} {\bibinfo {author} {\bibfnamefont {C.}~\bibnamefont
  {Maes}},\ }\bibfield  {title} {\bibinfo {title} {Local detailed balance},\
  }\href@noop {} {\bibfield  {journal} {\bibinfo  {journal} {SciPost Phys.
  Lect. Notes}\ }\textbf {\bibinfo {volume} {32}},\ \bibinfo {pages} {1}
  (\bibinfo {year} {2021})}\BibitemShut {NoStop}%
\bibitem [{\citenamefont
  {Tanogami}(2022{\natexlab{a}})}]{tanogami2022violation}%
  \BibitemOpen
  \bibfield  {author} {\bibinfo {author} {\bibfnamefont {T.}~\bibnamefont
  {Tanogami}},\ }\bibfield  {title} {\bibinfo {title} {Violation of the second
  fluctuation-dissipation relation and entropy production in nonequilibrium
  medium},\ }\href@noop {} {\bibfield  {journal} {\bibinfo  {journal} {J. Stat.
  Phys.}\ }\textbf {\bibinfo {volume} {187}},\ \bibinfo {pages} {1} (\bibinfo
  {year} {2022}{\natexlab{a}})}\BibitemShut {NoStop}%
\bibitem [{\citenamefont {Peliti}\ and\ \citenamefont
  {Pigolotti}(2021)}]{peliti2021stochastic}%
  \BibitemOpen
  \bibfield  {author} {\bibinfo {author} {\bibfnamefont {L.}~\bibnamefont
  {Peliti}}\ and\ \bibinfo {author} {\bibfnamefont {S.}~\bibnamefont
  {Pigolotti}},\ }\href@noop {} {\emph {\bibinfo {title} {Stochastic
  {T}hermodynamics: {A}n {I}ntroduction}}}\ (\bibinfo  {publisher} {Princeton
  University Press},\ \bibinfo {year} {2021})\BibitemShut {NoStop}%
\bibitem [{\citenamefont {Gardiner}(2009)}]{gardiner1985handbook}%
  \BibitemOpen
  \bibfield  {author} {\bibinfo {author} {\bibfnamefont {C.~W.}\ \bibnamefont
  {Gardiner}},\ }\href@noop {} {\emph {\bibinfo {title} {Handbook of
  {S}tochastic {M}ethods}}},\ \bibinfo {edition} {4th}\ ed.\ (\bibinfo
  {publisher} {Springer, Berlin},\ \bibinfo {year} {2009})\BibitemShut
  {NoStop}%
\bibitem [{\citenamefont {Cover}\ and\ \citenamefont
  {Thomas}(2006)}]{cover1999elements}%
  \BibitemOpen
  \bibfield  {author} {\bibinfo {author} {\bibfnamefont {T.~M.}\ \bibnamefont
  {Cover}}\ and\ \bibinfo {author} {\bibfnamefont {J.~A.}\ \bibnamefont
  {Thomas}},\ }\href@noop {} {\emph {\bibinfo {title} {Elements of
  {I}nformation {T}heory}}},\ \bibinfo {edition} {2nd}\ ed.\ (\bibinfo
  {publisher} {Wiley-Interscience, Hoboken, NJ},\ \bibinfo {year}
  {2006})\BibitemShut {NoStop}%
\bibitem [{\citenamefont {Hartich}\ \emph {et~al.}(2014)\citenamefont
  {Hartich}, \citenamefont {Barato},\ and\ \citenamefont
  {Seifert}}]{hartich2014stochastic}%
  \BibitemOpen
  \bibfield  {author} {\bibinfo {author} {\bibfnamefont {D.}~\bibnamefont
  {Hartich}}, \bibinfo {author} {\bibfnamefont {A.~C.}\ \bibnamefont
  {Barato}},\ and\ \bibinfo {author} {\bibfnamefont {U.}~\bibnamefont
  {Seifert}},\ }\bibfield  {title} {\bibinfo {title} {Stochastic thermodynamics
  of bipartite systems: transfer entropy inequalities and a {M}axwell’s demon
  interpretation},\ }\href@noop {} {\bibfield  {journal} {\bibinfo  {journal}
  {J. Stat. Mech.}\ }\textbf {\bibinfo {volume} {2014}},\ \bibinfo {pages}
  {P02016} (\bibinfo {year} {2014})}\BibitemShut {NoStop}%
\bibitem [{\citenamefont {Matsumoto}\ and\ \citenamefont
  {Sagawa}(2018)}]{matsumoto2018role}%
  \BibitemOpen
  \bibfield  {author} {\bibinfo {author} {\bibfnamefont {T.}~\bibnamefont
  {Matsumoto}}\ and\ \bibinfo {author} {\bibfnamefont {T.}~\bibnamefont
  {Sagawa}},\ }\bibfield  {title} {\bibinfo {title} {Role of sufficient
  statistics in stochastic thermodynamics and its implication to sensory
  adaptation},\ }\href@noop {} {\bibfield  {journal} {\bibinfo  {journal}
  {Phys. Rev. E}\ }\textbf {\bibinfo {volume} {97}},\ \bibinfo {pages} {042103}
  (\bibinfo {year} {2018})}\BibitemShut {NoStop}%
\bibitem [{\citenamefont {Horowitz}\ and\ \citenamefont
  {Esposito}(2014)}]{horowitz2014thermodynamics}%
  \BibitemOpen
  \bibfield  {author} {\bibinfo {author} {\bibfnamefont {J.~M.}\ \bibnamefont
  {Horowitz}}\ and\ \bibinfo {author} {\bibfnamefont {M.}~\bibnamefont
  {Esposito}},\ }\bibfield  {title} {\bibinfo {title} {Thermodynamics with
  continuous information flow},\ }\href@noop {} {\bibfield  {journal} {\bibinfo
   {journal} {Phys. Rev. X}\ }\textbf {\bibinfo {volume} {4}},\ \bibinfo
  {pages} {031015} (\bibinfo {year} {2014})}\BibitemShut {NoStop}%
\bibitem [{\citenamefont {Sekimoto}(2010)}]{sekimoto2010stochastic}%
  \BibitemOpen
  \bibfield  {author} {\bibinfo {author} {\bibfnamefont {K.}~\bibnamefont
  {Sekimoto}},\ }\href@noop {} {\emph {\bibinfo {title} {Stochastic
  Energetics}}}\ (\bibinfo  {publisher} {Springer, New York},\ \bibinfo {year}
  {2010})\BibitemShut {NoStop}%
\bibitem [{\citenamefont {Shiraishi}\ and\ \citenamefont
  {Sagawa}(2015)}]{shiraishi2015fluctuation}%
  \BibitemOpen
  \bibfield  {author} {\bibinfo {author} {\bibfnamefont {N.}~\bibnamefont
  {Shiraishi}}\ and\ \bibinfo {author} {\bibfnamefont {T.}~\bibnamefont
  {Sagawa}},\ }\bibfield  {title} {\bibinfo {title} {Fluctuation theorem for
  partially masked nonequilibrium dynamics},\ }\href@noop {} {\bibfield
  {journal} {\bibinfo  {journal} {Phys. Rev. E}\ }\textbf {\bibinfo {volume}
  {91}},\ \bibinfo {pages} {012130} (\bibinfo {year} {2015})}\BibitemShut
  {NoStop}%
\bibitem [{\citenamefont {L'vov}\ \emph {et~al.}(1998)\citenamefont {L'vov},
  \citenamefont {Podivilov}, \citenamefont {Pomyalov}, \citenamefont
  {Procaccia},\ and\ \citenamefont {Vandembroucq}}]{l1998improved}%
  \BibitemOpen
  \bibfield  {author} {\bibinfo {author} {\bibfnamefont {V.~S.}\ \bibnamefont
  {L'vov}}, \bibinfo {author} {\bibfnamefont {E.}~\bibnamefont {Podivilov}},
  \bibinfo {author} {\bibfnamefont {A.}~\bibnamefont {Pomyalov}}, \bibinfo
  {author} {\bibfnamefont {I.}~\bibnamefont {Procaccia}},\ and\ \bibinfo
  {author} {\bibfnamefont {D.}~\bibnamefont {Vandembroucq}},\ }\bibfield
  {title} {\bibinfo {title} {Improved shell model of turbulence},\ }\href@noop
  {} {\bibfield  {journal} {\bibinfo  {journal} {Phys. Rev. E}\ }\textbf
  {\bibinfo {volume} {58}},\ \bibinfo {pages} {1811} (\bibinfo {year}
  {1998})}\BibitemShut {NoStop}%
\bibitem [{\citenamefont {Biferale}(2003)}]{biferale2003shell}%
  \BibitemOpen
  \bibfield  {author} {\bibinfo {author} {\bibfnamefont {L.}~\bibnamefont
  {Biferale}},\ }\bibfield  {title} {\bibinfo {title} {Shell models of energy
  cascade in turbulence},\ }\href@noop {} {\bibfield  {journal} {\bibinfo
  {journal} {Annu. Rev. Fluid Mech.}\ }\textbf {\bibinfo {volume} {35}},\
  \bibinfo {pages} {441} (\bibinfo {year} {2003})}\BibitemShut {NoStop}%
\bibitem [{\citenamefont {Bohr}\ \emph {et~al.}(1998)\citenamefont {Bohr},
  \citenamefont {Jensen}, \citenamefont {Paladin},\ and\ \citenamefont
  {Vulpiani}}]{bohr1998dynamical}%
  \BibitemOpen
  \bibfield  {author} {\bibinfo {author} {\bibfnamefont {T.}~\bibnamefont
  {Bohr}}, \bibinfo {author} {\bibfnamefont {M.~H.}\ \bibnamefont {Jensen}},
  \bibinfo {author} {\bibfnamefont {G.}~\bibnamefont {Paladin}},\ and\ \bibinfo
  {author} {\bibfnamefont {A.}~\bibnamefont {Vulpiani}},\ }\href@noop {} {\emph
  {\bibinfo {title} {Dynamical systems approach to turbulence}}}\ (\bibinfo
  {publisher} {Cambridge university press},\ \bibinfo {year}
  {1998})\BibitemShut {NoStop}%
\bibitem [{\citenamefont {Kraskov}\ \emph {et~al.}(2004)\citenamefont
  {Kraskov}, \citenamefont {St{\"o}gbauer},\ and\ \citenamefont
  {Grassberger}}]{kraskov2004estimating}%
  \BibitemOpen
  \bibfield  {author} {\bibinfo {author} {\bibfnamefont {A.}~\bibnamefont
  {Kraskov}}, \bibinfo {author} {\bibfnamefont {H.}~\bibnamefont
  {St{\"o}gbauer}},\ and\ \bibinfo {author} {\bibfnamefont {P.}~\bibnamefont
  {Grassberger}},\ }\bibfield  {title} {\bibinfo {title} {Estimating mutual
  information},\ }\href@noop {} {\bibfield  {journal} {\bibinfo  {journal}
  {Phys. Rev. E}\ }\textbf {\bibinfo {volume} {69}},\ \bibinfo {pages} {066138}
  (\bibinfo {year} {2004})}\BibitemShut {NoStop}%
\bibitem [{\citenamefont {Khan}\ \emph {et~al.}(2007)\citenamefont {Khan},
  \citenamefont {Bandyopadhyay}, \citenamefont {Ganguly}, \citenamefont
  {Saigal}, \citenamefont {Erickson~III}, \citenamefont {Protopopescu},\ and\
  \citenamefont {Ostrouchov}}]{khan2007relative}%
  \BibitemOpen
  \bibfield  {author} {\bibinfo {author} {\bibfnamefont {S.}~\bibnamefont
  {Khan}}, \bibinfo {author} {\bibfnamefont {S.}~\bibnamefont {Bandyopadhyay}},
  \bibinfo {author} {\bibfnamefont {A.~R.}\ \bibnamefont {Ganguly}}, \bibinfo
  {author} {\bibfnamefont {S.}~\bibnamefont {Saigal}}, \bibinfo {author}
  {\bibfnamefont {D.~J.}\ \bibnamefont {Erickson~III}}, \bibinfo {author}
  {\bibfnamefont {V.}~\bibnamefont {Protopopescu}},\ and\ \bibinfo {author}
  {\bibfnamefont {G.}~\bibnamefont {Ostrouchov}},\ }\bibfield  {title}
  {\bibinfo {title} {Relative performance of mutual information estimation
  methods for quantifying the dependence among short and noisy data},\
  }\href@noop {} {\bibfield  {journal} {\bibinfo  {journal} {Phys. Rev. E}\
  }\textbf {\bibinfo {volume} {76}},\ \bibinfo {pages} {026209} (\bibinfo
  {year} {2007})}\BibitemShut {NoStop}%
\bibitem [{\citenamefont {Holmes}\ and\ \citenamefont
  {Nemenman}(2019)}]{holmes2019estimation}%
  \BibitemOpen
  \bibfield  {author} {\bibinfo {author} {\bibfnamefont {C.~M.}\ \bibnamefont
  {Holmes}}\ and\ \bibinfo {author} {\bibfnamefont {I.}~\bibnamefont
  {Nemenman}},\ }\bibfield  {title} {\bibinfo {title} {Estimation of mutual
  information for real-valued data with error bars and controlled bias},\
  }\href@noop {} {\bibfield  {journal} {\bibinfo  {journal} {Phys. Rev. E}\
  }\textbf {\bibinfo {volume} {100}},\ \bibinfo {pages} {022404} (\bibinfo
  {year} {2019})}\BibitemShut {NoStop}%
\bibitem [{\citenamefont {Carbone}\ and\ \citenamefont
  {Bragg}(2020)}]{Carbone2020_is_vortex}%
  \BibitemOpen
  \bibfield  {author} {\bibinfo {author} {\bibfnamefont {M.}~\bibnamefont
  {Carbone}}\ and\ \bibinfo {author} {\bibfnamefont {A.~D.}\ \bibnamefont
  {Bragg}},\ }\bibfield  {title} {\bibinfo {title} {Is vortex stretching the
  main cause of the turbulent energy cascade?},\ }\href@noop {} {\bibfield
  {journal} {\bibinfo  {journal} {J. Fluid Mech.}\ }\textbf {\bibinfo {volume}
  {883}},\ \bibinfo {pages} {R2} (\bibinfo {year} {2020})}\BibitemShut
  {NoStop}%
\bibitem [{\citenamefont {Wilson}(1975)}]{wilson1975renormalization}%
  \BibitemOpen
  \bibfield  {author} {\bibinfo {author} {\bibfnamefont {K.~G.}\ \bibnamefont
  {Wilson}},\ }\bibfield  {title} {\bibinfo {title} {The renormalization group:
  {C}ritical phenomena and the {K}ondo problem},\ }\href@noop {} {\bibfield
  {journal} {\bibinfo  {journal} {Rev. Mod. Phys.}\ }\textbf {\bibinfo {volume}
  {47}},\ \bibinfo {pages} {773} (\bibinfo {year} {1975})}\BibitemShut
  {NoStop}%
\bibitem [{\citenamefont {Tanogami}(2021)}]{tanogami2021theoretical}%
  \BibitemOpen
  \bibfield  {author} {\bibinfo {author} {\bibfnamefont {T.}~\bibnamefont
  {Tanogami}},\ }\bibfield  {title} {\bibinfo {title} {Theoretical analysis of
  quantum turbulence using the {O}nsager ideal turbulence theory},\ }\href@noop
  {} {\bibfield  {journal} {\bibinfo  {journal} {Phys. Rev. E}\ }\textbf
  {\bibinfo {volume} {103}},\ \bibinfo {pages} {023106} (\bibinfo {year}
  {2021})}\BibitemShut {NoStop}%
\bibitem [{\citenamefont {Tanogami}(2022{\natexlab{b}})}]{tanogami2022reply}%
  \BibitemOpen
  \bibfield  {author} {\bibinfo {author} {\bibfnamefont {T.}~\bibnamefont
  {Tanogami}},\ }\bibfield  {title} {\bibinfo {title} {Reply to ``{C}omment on
  `{T}heoretical analysis of quantum turbulence using the {O}nsager ideal
  turbulence theory' ''},\ }\href@noop {} {\bibfield  {journal} {\bibinfo
  {journal} {Phys. Rev. E}\ }\textbf {\bibinfo {volume} {105}},\ \bibinfo
  {pages} {027102} (\bibinfo {year} {2022}{\natexlab{b}})}\BibitemShut
  {NoStop}%
\bibitem [{\citenamefont {Krstulovic}\ \emph {et~al.}(2022)\citenamefont
  {Krstulovic}, \citenamefont {L'vov},\ and\ \citenamefont
  {Nazarenko}}]{krstulovic2022comment}%
  \BibitemOpen
  \bibfield  {author} {\bibinfo {author} {\bibfnamefont {G.}~\bibnamefont
  {Krstulovic}}, \bibinfo {author} {\bibfnamefont {V.}~\bibnamefont {L'vov}},\
  and\ \bibinfo {author} {\bibfnamefont {S.}~\bibnamefont {Nazarenko}},\
  }\bibfield  {title} {\bibinfo {title} {Comment on ``{T}heoretical analysis of
  quantum turbulence using the {O}nsager ideal turbulence theory''},\
  }\href@noop {} {\bibfield  {journal} {\bibinfo  {journal} {Phys. Rev. E}\
  }\textbf {\bibinfo {volume} {105}},\ \bibinfo {pages} {027101} (\bibinfo
  {year} {2022})}\BibitemShut {NoStop}%
\bibitem [{\citenamefont {Skrbek}\ \emph {et~al.}(2021)\citenamefont {Skrbek},
  \citenamefont {Schmoranzer}, \citenamefont {Midlik},\ and\ \citenamefont
  {Sreenivasan}}]{skrbek2021phenomenology}%
  \BibitemOpen
  \bibfield  {author} {\bibinfo {author} {\bibfnamefont {L.}~\bibnamefont
  {Skrbek}}, \bibinfo {author} {\bibfnamefont {D.}~\bibnamefont {Schmoranzer}},
  \bibinfo {author} {\bibfnamefont {{\v{S}}.}~\bibnamefont {Midlik}},\ and\
  \bibinfo {author} {\bibfnamefont {K.~R.}\ \bibnamefont {Sreenivasan}},\
  }\bibfield  {title} {\bibinfo {title} {Phenomenology of quantum turbulence in
  superfluid helium},\ }\href@noop {} {\bibfield  {journal} {\bibinfo
  {journal} {Proc. Natl. Acad. Sci. U. S. A.}\ }\textbf {\bibinfo {volume}
  {118}} (\bibinfo {year} {2021})}\BibitemShut {NoStop}%
\bibitem [{\citenamefont {Tanogami}\ and\ \citenamefont
  {Sasa}(2021)}]{tanogami2021van}%
  \BibitemOpen
  \bibfield  {author} {\bibinfo {author} {\bibfnamefont {T.}~\bibnamefont
  {Tanogami}}\ and\ \bibinfo {author} {\bibfnamefont {S.-i.}\ \bibnamefont
  {Sasa}},\ }\bibfield  {title} {\bibinfo {title} {Van der {W}aals cascade in
  supercritical turbulence near a critical point},\ }\href
  {https://doi.org/10.1103/PhysRevResearch.3.L032027} {\bibfield  {journal}
  {\bibinfo  {journal} {Phys. Rev. Research}\ }\textbf {\bibinfo {volume}
  {3}},\ \bibinfo {pages} {L032027} (\bibinfo {year} {2021})}\BibitemShut
  {NoStop}%
\bibitem [{\citenamefont {Nazarenko}(2011)}]{Nazarenko_2011}%
  \BibitemOpen
  \bibfield  {author} {\bibinfo {author} {\bibfnamefont {S.}~\bibnamefont
  {Nazarenko}},\ }\href@noop {} {\emph {\bibinfo {title} {Wave
  {T}urbulence}}},\ Vol.\ \bibinfo {volume} {825}\ (\bibinfo  {publisher}
  {Springer Science \& Business Media},\ \bibinfo {year} {2011})\BibitemShut
  {NoStop}%
\bibitem [{\citenamefont {Zakharov}\ \emph {et~al.}(1992)\citenamefont
  {Zakharov}, \citenamefont {L'vov},\ and\ \citenamefont
  {Falkovich}}]{zakharov1992kolmogorov}%
  \BibitemOpen
  \bibfield  {author} {\bibinfo {author} {\bibfnamefont {V.~E.}\ \bibnamefont
  {Zakharov}}, \bibinfo {author} {\bibfnamefont {V.~S.}\ \bibnamefont
  {L'vov}},\ and\ \bibinfo {author} {\bibfnamefont {G.}~\bibnamefont
  {Falkovich}},\ }\href@noop {} {\emph {\bibinfo {title} {Kolmogorov spectra of
  turbulence {I}: {W}ave turbulence}}}\ (\bibinfo  {publisher} {Springer,
  Berlin},\ \bibinfo {year} {1992})\BibitemShut {NoStop}%
\bibitem [{\citenamefont {Tanogami}\ and\ \citenamefont
  {Sasa}(2022)}]{tanogami2022xy}%
  \BibitemOpen
  \bibfield  {author} {\bibinfo {author} {\bibfnamefont {T.}~\bibnamefont
  {Tanogami}}\ and\ \bibinfo {author} {\bibfnamefont {S.-i.}\ \bibnamefont
  {Sasa}},\ }\bibfield  {title} {\bibinfo {title} {{XY} model for cascade
  transfer},\ }\href@noop {} {\bibfield  {journal} {\bibinfo  {journal} {Phys.
  Rev. Research}\ }\textbf {\bibinfo {volume} {4}},\ \bibinfo {pages} {L022015}
  (\bibinfo {year} {2022})}\BibitemShut {NoStop}%
\bibitem [{\citenamefont {Tsubota}\ \emph {et~al.}(2013)\citenamefont
  {Tsubota}, \citenamefont {Aoki},\ and\ \citenamefont
  {Fujimoto}}]{tsubota2013spin}%
  \BibitemOpen
  \bibfield  {author} {\bibinfo {author} {\bibfnamefont {M.}~\bibnamefont
  {Tsubota}}, \bibinfo {author} {\bibfnamefont {Y.}~\bibnamefont {Aoki}},\ and\
  \bibinfo {author} {\bibfnamefont {K.}~\bibnamefont {Fujimoto}},\ }\bibfield
  {title} {\bibinfo {title} {Spin-glass-like behavior in the spin turbulence of
  spinor {B}ose-{E}instein condensates},\ }\href@noop {} {\bibfield  {journal}
  {\bibinfo  {journal} {Phys. Rev. A}\ }\textbf {\bibinfo {volume} {88}},\
  \bibinfo {pages} {061601(R)} (\bibinfo {year} {2013})}\BibitemShut {NoStop}%
\bibitem [{\citenamefont {Rodriguez-Nieva}(2021)}]{rodriguez2021turbulent}%
  \BibitemOpen
  \bibfield  {author} {\bibinfo {author} {\bibfnamefont {J.~F.}\ \bibnamefont
  {Rodriguez-Nieva}},\ }\bibfield  {title} {\bibinfo {title} {Turbulent
  relaxation after a quench in the {H}eisenberg model},\ }\href@noop {}
  {\bibfield  {journal} {\bibinfo  {journal} {Phys. Rev. B}\ }\textbf {\bibinfo
  {volume} {104}},\ \bibinfo {pages} {L060302} (\bibinfo {year}
  {2021})}\BibitemShut {NoStop}%
\bibitem [{\citenamefont {Harada}\ and\ \citenamefont
  {Sasa}(2006)}]{harada2006energy}%
  \BibitemOpen
  \bibfield  {author} {\bibinfo {author} {\bibfnamefont {T.}~\bibnamefont
  {Harada}}\ and\ \bibinfo {author} {\bibfnamefont {S.-i.}\ \bibnamefont
  {Sasa}},\ }\bibfield  {title} {\bibinfo {title} {Energy dissipation and
  violation of the fluctuation-response relation in nonequilibrium {L}angevin
  systems},\ }\href@noop {} {\bibfield  {journal} {\bibinfo  {journal} {Phys.
  Rev. E}\ }\textbf {\bibinfo {volume} {73}},\ \bibinfo {pages} {026131}
  (\bibinfo {year} {2006})}\BibitemShut {NoStop}%
\bibitem [{\citenamefont {Risken}(1996)}]{risken1996fokker}%
  \BibitemOpen
  \bibfield  {author} {\bibinfo {author} {\bibfnamefont {H.}~\bibnamefont
  {Risken}},\ }\bibfield  {title} {\bibinfo {title} {The {F}okker-{P}lanck
  {E}quation}\ }(\bibinfo  {publisher} {Springer},\ \bibinfo {year}
  {1996})\BibitemShut {NoStop}%
\bibitem [{\citenamefont {Lord}\ and\ \citenamefont
  {Rougemont}(2004)}]{lord2004numerical}%
  \BibitemOpen
  \bibfield  {author} {\bibinfo {author} {\bibfnamefont {G.~J.}\ \bibnamefont
  {Lord}}\ and\ \bibinfo {author} {\bibfnamefont {J.}~\bibnamefont
  {Rougemont}},\ }\bibfield  {title} {\bibinfo {title} {A numerical scheme for
  stochastic {PDE}s with {G}evrey regularity},\ }\href@noop {} {\bibfield
  {journal} {\bibinfo  {journal} {IMA J. Numer. Anal.}\ }\textbf {\bibinfo
  {volume} {24}},\ \bibinfo {pages} {587} (\bibinfo {year} {2004})}\BibitemShut
  {NoStop}%
\end{thebibliography}%

\end{document}